%% file: Review2_HuertasRoldan_UFs_IC418_clean.tex
\documentclass{aa}  

\usepackage{lscape}
\usepackage{graphicx,float}
\usepackage{txfonts}

\usepackage{array,multirow}
\usepackage{soul}
\usepackage{xspace}
\usepackage{subcaption}
\usepackage[usenames,dvipsnames]{color}
\usepackage{easyReview}
\usepackage{amssymb,amsmath}
\usepackage[breaklinks=true,colorlinks=true,linkcolor=blue,citecolor=blue,urlcolor=blue]{hyperref}

\usepackage{natbib}
\bibpunct{(}{)}{;}{a}{}{,}             

\renewcommand{\ga}{$\alpha$\xspace}

\renewcommand{\gg}{$\gamma$\xspace}
\newcommand{\gd}{$\delta$\xspace}
\renewcommand{\ge}{$\epsilon$\xspace}
\newcommand{\gz}{$\zeta$\xspace}
\newcommand{\gh}{$\eta$\xspace}
\newcommand{\gq}{$\theta$\xspace}
\newcommand{\gi}{$\iota$\xspace}

\newcommand{\hii}{\mbox{\ion{H}{ii}}\xspace}
\newcommand{\hei}{\mbox{\ion{He}{i}}\xspace}
\newcommand{\trihei}{\mbox{\ion{\element[][3]{He}}{i}}\xspace}

\newcommand{\triheii}{\mbox{\ion{\element[][3]{He}}{ii}}\xspace}
\newcommand{\fourheii}{\mbox{\ion{\element[][4]{He}}{ii}}\xspace}
\newcommand{\ci}{\mbox{\ion{C}{i}}\xspace}
\newcommand{\cii}{\mbox{\ion{C}{ii}}\xspace}
\newcommand{\ciii}{\mbox{\ion{C}{iii}}\xspace}
\newcommand{\oi}{\mbox{\ion{O}{i}}\xspace}
\newcommand{\oii}{\mbox{\ion{O}{ii}}\xspace}
\newcommand{\oiii}{\mbox{\ion{O}{iii}}\xspace}

\let\oldAA\AA
\renewcommand*{\AA}{\,\oldAA\xspace}

\newcommand{\mm}{\,mm\xspace}
\newcommand{\mhz}{\,MHz\xspace}
\newcommand{\ghz}{\,GHz\xspace}
\newcommand{\kms}{\,\mbox{km\,s$^{-1}$}\xspace}

\newcommand{\kel}{\,K\xspace}
\newcommand{\mkel}{\,mK\xspace}

\newcommand{\ic}{\mbox{IC 418}\xspace}
\newcommand{\ngc}{\mbox{NGC 7027}\xspace}

\begin{document}

\title{Detection of unidentified molecular pure rotational lines in C-rich PNe I. The fullerene-containing PN IC 418}

\titlerunning{Detection of UFs pure rotational lines in C-rich PNe I. The PN IC 418}


   \author{T. Huertas-Roldán\inst{1}\fnmsep\inst{2} 
   	\and J. P. Fonfría\inst{3}
   	\and J. Alcolea\inst{4}
   	\and D. A. García-Hernández\inst{1}\fnmsep\inst{2}
   	\and S. Mato\inst{5}
   	\and J. J. Díaz-Luis\inst{4}
   	\and R. Barzaga\inst{6}
   	\and A. Manchado\inst{7}
   	\and V. Bujarrabal\inst{4}
   	\and M. A. Gómez-Muñoz\inst{8}\fnmsep\inst{9}
   }
   
   \institute{Instituto de Astrofísica de Canarias (IAC),
   	C/ Vía Láctea s/n, 38205 La Laguna, Spain\\
   	\email{thuertas@iac.es}
   	\and 
   	Departamento de Astrofísica, Universidad de La Laguna (ULL),
   	38206 La Laguna, Spain
   	\and
   	Departamento de Astrofísica Molecular, Instituto de Física Fundamental (IFF, CSIC),
   	C/ Serrano 123, 28006 Madrid, Spain
   	\and
   	Observatorio Astronómico Nacional (OAN, IGN/CNIG),
   	C/ Alfonso XII 3, 28014 Madrid, Spain
   	\and
   	Grupo de Espectrocopía Molecular (GEM), Edificio Quifima, Laboratorios de Espectroscopía y Bioespectroscopía, Unidad Asociada CSIC, Parque Científico UVa, Universidad de Valladolid, 47011 Valladolid, Spain
   	\and
   	Departamento de Física and IUdEA, Universidad de La Laguna (ULL), 38200 Tenerife, Spain
   	\and
   	Consejo Superior de Investigaciones Científicas (CSIC), Spain
   	\and
   	Departament de Fïsica Quàntica i Astrofísica (FQA), Universitat de Barcelona (UB),
   	C/ Martí i Franquès 1, 08028 Barcelona, Spain
   	\and
   	Institut de Ciències del Cosmos (ICCUB), Universitat de Barcelona (UB),
   	C/ Martí i Franquès 1, 08028 Barcelona, Spain.
   	}

   \date{Received 00 Month, 2026 / Accepted 00 Month, 2026}

 
\abstract
{Molecular emission is observed in a wide variety of astrophysical environments, yet a substantial fraction of spectral features detected at mm wavelengths remains unidentified. Identifying these features is essential for constraining the inventory of interstellar and circumstellar molecules and for understanding the chemical pathways operating in evolved stars, including those leading to fullerene formation in planetary nebulae (PNe).}
{In this study, we investigate a set of weak unidentified molecular features detected in the C-rich and fullerene containing PN \ic. We aim to constraint the nature of their molecular carriers and assessing their possible connection to fullerene-related chemistry.}
{High-sensitivity observations at 2, 3, and 7\mm were carried out using the RT40m and IRAM 30m radio telescopes. The spectral features found in these data sets were compared with public molecular spectroscopic databases. Spectral-pattern searches and line fits were performed under linear, symmetric, and asymmetric rotor approximations to estimate rotational constants of potential molecular carriers.}
{We report the detection of 20 weak (SNR $\sim 2 - 13$) unidentified features (UFs), none of which can be attributed to mm radio recombination lines, instrumental artifacts, or any known molecular species. The observed features are inconsistent with the regular line spacing expected from linear molecules or symmetric rotors. A recurrent doublet-like pattern is identified, although no other spectral patterns are found. The estimated rotational constant ($B \sim 2\,500-3\,660$\mhz) suggests carriers with $4-13$ atoms. The rotational constants estimated for the spectral lines are incompatible with those expected from C$_{60}$ derivatives ($B \sim 100$\mhz), disfavoring fullerene-related species as the dominant carriers of the detected UFs.}
{The detected UFs point to the presence of molecular species not yet identified in space within the fullerene-rich PN \ic. Non-planar carbonaceous molecules produced during the destruction/processing of hydrogenated amorphous carbon (HAC) grains are suggested as promising candidates. The catalog of UFs presented here is made publicly available to facilitate future comparisons with laboratory measurements and theoretical calculations, which will be essential for identifying the molecular carriers and understanding the chemical pathways leading to fullerene formation in PNe.}


\maketitle


\section{Introduction}

Molecular species are widespread throughout the Universe and have been detected in a wide range of astrophysical environments, from dense molecular clouds to evolved stars  in the Milky Way and other galaxies. The methylidyne radical (CH) was the first molecule identified in the interstellar medium (ISM) \citep{Dunham1937, Swings1937, McKellar1940}. Since then, molecular discoveries have expanded, reaching 325 known species by 2025 \citep[][]{McGuire_astromol}, with new detections continuing to emerge \citep[see e.g.][and references therein, for some of the most recent new detections]{Cernicharo2023a, Cernicharo2024b, Pardo2023a, Gupta2024a, Remijan2024a, Remijan2025a, SanzNovo2025, Wenzel2025a}.

Despite the large number of molecules known up to date, a significant number of features in spectral spectral observations remains unidentified, indicating that their molecular carriers are still unknown. The advent of high-resolution and high-sensitivity instruments, particularly the JWST and ALMA, has led to the detection of an increasing number of unidentified spectral features, both in the infrared (IR) and radio domains, in the form of ro-vibrational and rotational transitions, respectively \citep[see e.g.,][]{Appleton2023, Bianchi2023, Leisawitz2023, GonzalezAlfonso2024, Nickerson2025}. These unidentified features (UFs) constitute a major challenge for astrochemistry, specially in evolved star circumstellar environments such as planetary nebulae (PNe) and in the ISM, where gas-phase and grain-surface reactions yield to increasingly complex molecules \citep[e.g.,][]{Tielens2005a, Tielens2005b}.

Identifying the molecular carriers responsible for specific spectral lines is a complex process. Molecular transitions are typically classified into electronic, vibrational, and rotational types. Vibrational transitions, typically observed in the IR range, have been used to identify complex species such as fullerenes C$_{60}$ and C$_{70}$ in young PNe \citep{Cami2010, GarciaHernandez2010} and C$_{60}^+$ in the ISM \citep{Berne2013, Campbell2015}. In contrast, molecular families, such as polycyclic aromatic hydrocarbons (PAHs), exhibit very similar IR signatures even at the high spectral resolution of the JWST \citep[see e.g.,][]{SmithPerez2025}, complicating the unambiguous identification of individual species. Pure rotational transitions occur when molecules with a permanent dipole moment change their rotational energy states. These transitions provide unique spectral fingerprints that make them specially powerful for molecular identification. These lines are typically observed at millimeter (mm), sub-millimeter (sub-mm), and centimeter (cm) wavelengths, using radio telescopes. However, as molecules increase in size and structural complexity, their rotational spectra become more complex and their individual lines weaker, and harder to predict theoretically or obtain experimentally, complicating their astronomical identification.

Radio observations have played a fundamental role in the discovery of new molecules in space, particularly toward evolved stars. The high spectral resolution and sensitivity of modern radio facilities enable the detection of pure rotational lines that serve as distinctive molecular fingerprints, making this technique the most powerful tool for molecular identification \citep[see e.g.,][]{Wilson1970}. Surveys of low- and intermediate-mass evolved stars, such as asymptotic giant branch (AGB) stars and proto-PNe, have historically been among the richest sources of new molecular detections, as their dense, molecule-rich envelopes host complex gas-phase and grain surface chemistry under diverse physical conditions \citep[see e.g.][]{Cernicharo2000, Pardo2007, Pardo2022, VelillaPrieto2017, Zhang2013}. Extending such studies to PNe provides crucial insights into how circumstellar molecules survive, evolve, or reform after photodissociation and photoionization \citep[e.g.][]{Edwards2014, Schmidt2016, Gold2024}, and helps trace the chemical transformation of stellar ejecta into the ISM. Continued radio spectroscopy of these objects, enabled by facilities like the Green Bank Telescope (GBT), ALMA, and the IRAM 30m and Yebes 40m (RT40m) telescopes, is therefore essential for uncovering new molecular species and constraining the pathways of molecular evolution in the final stages of stellar evolution.

The detection of exotic carbonaceous species like fullerenes (e.g. C$_{60}$, C$_{70}$) in young PNe has revealed that these environments host complex dust and molecular processing under UV irradiation and/or shock conditions \citep[e.g.][]{GarciaHernandez2010, GarciaHernandez2011b, GarciaHernandez2012, DiazLuis2016}. Objects like \ic are particularly valuable laboratories for these studies. \ic, a carbon-rich elliptical PN, exhibits bands of both aliphatic and aromatic molecular compounds along with IR emission features attributed to C$_{60}$ \citep{Otsuka2014}, making it a prime target for exploring the formation and survival of fullerenes and their derivatives in H-rich circumstellar gas. Observations of this object therefore probe not only the formation pathways and stability of fullerenes, but also their potential role as tracers of dust chemistry, molecular evolution, and the chemical feedback of evolved stars to the ISM. Consequently, high-sensitivity radio and mm/sub-mm spectral surveys of PNe such as \ic are key to advance our understanding of chemical complexity in evolved stellar environments.

Also, the detection of new radio spectral features provides critical input for molecular astrophysics, as it directly links observational astronomy with laboratory and theoretical chemistry. High-sensitivity spectral surveys reveal candidate lines of previously unknown species, motivating laboratory measurements and ab initio calculations of their rotational spectra and dipole moments. Such coordinated efforts are essential to confirm molecular identifications and to expand spectral databases for complex molecules, including radicals, ions, and fullerene-related species. In this context, radio detections act as the primary driver for cross-disciplinary progress in characterizing the molecular inventory of evolved stars and the ISM.

In this paper, we present for first time the detection of 20 weak pure rotational lines with a signal to noise ratio (SNR) $\sim 2-13$ in high spectral resolution observations of the PN \ic ranging from 30 to 140\ghz and obtained with the RT40m and IRAM 30m telescopes \citep{HuertasRoldan2025}, aimed at detecting new UFs and identifying potential molecular carriers associated with them. In Sec.~\ref{sec:obs_datared} we describe the data acquisition and reduction process. In Sec.~\ref{sec:results} we detail the set of UFs and the analyses performed to learn about their molecular origin and characterize their possible carriers. Finally, in Secs.~\ref{sec:discussion} and \ref{sec:conclusions} we discuss the results and state our conclusions, respectively.

\begin{table}
	\centering
	\caption{Frequencies of the UFs found in \ic.}
	\label{table:ufs_frequencies}
	\addtolength{\tabcolsep}{-0.3em}
	\begin{tabular}{l c c l}
		\hline\hline\noalign{\smallskip}
		Species & $\nu$ (MHz) & SNR & Comments \\
		\hline\noalign{\smallskip}
		\multicolumn{4}{c}{Detections $^a$} \\
		U35300 & $35300.1 \pm 0.5$ & 3.9 & Isolated \\
		U36344 & $36344.9 \pm 0.5$ & 4.2 & Isolated \\
		U37403 & $37403.7 \pm 0.5$ & 5.3 & Isolated \\
		U37410 & $37410.3 \pm 0.5$ & 5.0 & Isolated \\
		U85582 & $85582.1 \pm 1.0$ & 3.3 & Isolated \\
		U87129 & $87129.5 \pm 1.0$ & 4.0 & Isolated \\
		U87336 & $87336.7 \pm 1.0$ & 4.0 & Isolated \\
		U131474 & $131474.4 \pm 1.0$ & 5.2 & Isolated \\
		Abs136891 & $136891.5 \pm 1.0$ & 6.8 & Absorption, isolated \\
		U138455 & $138455.1 \pm 1.0$ & 3.8 & Isolated \\
		U139054 & $139054.9 \pm 1.0$ & 13.3 & Isolated \\
		\hline\noalign{\smallskip}
		\multicolumn{4}{c}{Tentative detections $^b$} \\
		U33180 & $33180.0 \pm 0.5$ & $3.4 \pm 0.9$ & Blended with H103\gz \\
		U35519 & $35519.3 \pm 0.5$ & $3.9 \pm 1.1$ & Blended with H95\ge \\
		U35593 & $35593.9 \pm 0.5$ & $2.0 \pm 1.2$ & Blended with H110\gq \\
		U39019 & $39019.4 \pm 0.5$ & $2.4 \pm 0.8$ & Blended with H92\ge \\
		U42876 & $42876.6 \pm 0.5$ & $3.5 \pm 1.0$ & Blended with H83\gd \\
		U42987 & $42987.9 \pm 0.5$ & $3.6 \pm 4.0$ & Blended with H89\ge \\
		U47746 & $47746.0 \pm 0.5$ & $2.4 \pm 0.9$ & Blended with H73\gg \\
		U85342 & $85342.2 \pm 1.0$ & $3.3 \pm 0.9$ & Blended with H78\gh \\
		U133126 & $133126.5 \pm 1.0$ & $4.2 \pm 1.3$ & Blended with H72\gi \\
		\hline
	\end{tabular}
	\tablefoot{$a$: we consider detection of isolated lines with SNR $>3$ as well as blended features with SNR $>5$. $b$: we consider tentative detections of isolated and blended spectral features with SNR $>2$.}
\end{table}


\section{Observations and Data Reduction}
\label{sec:obs_datared}

\begin{figure*}[h!]
	\captionsetup[sub]{skip=0mm, belowskip=0pt}
	\centering
	\begin{subfigure}[]{0.34\linewidth}
		\includegraphics[width=\linewidth]{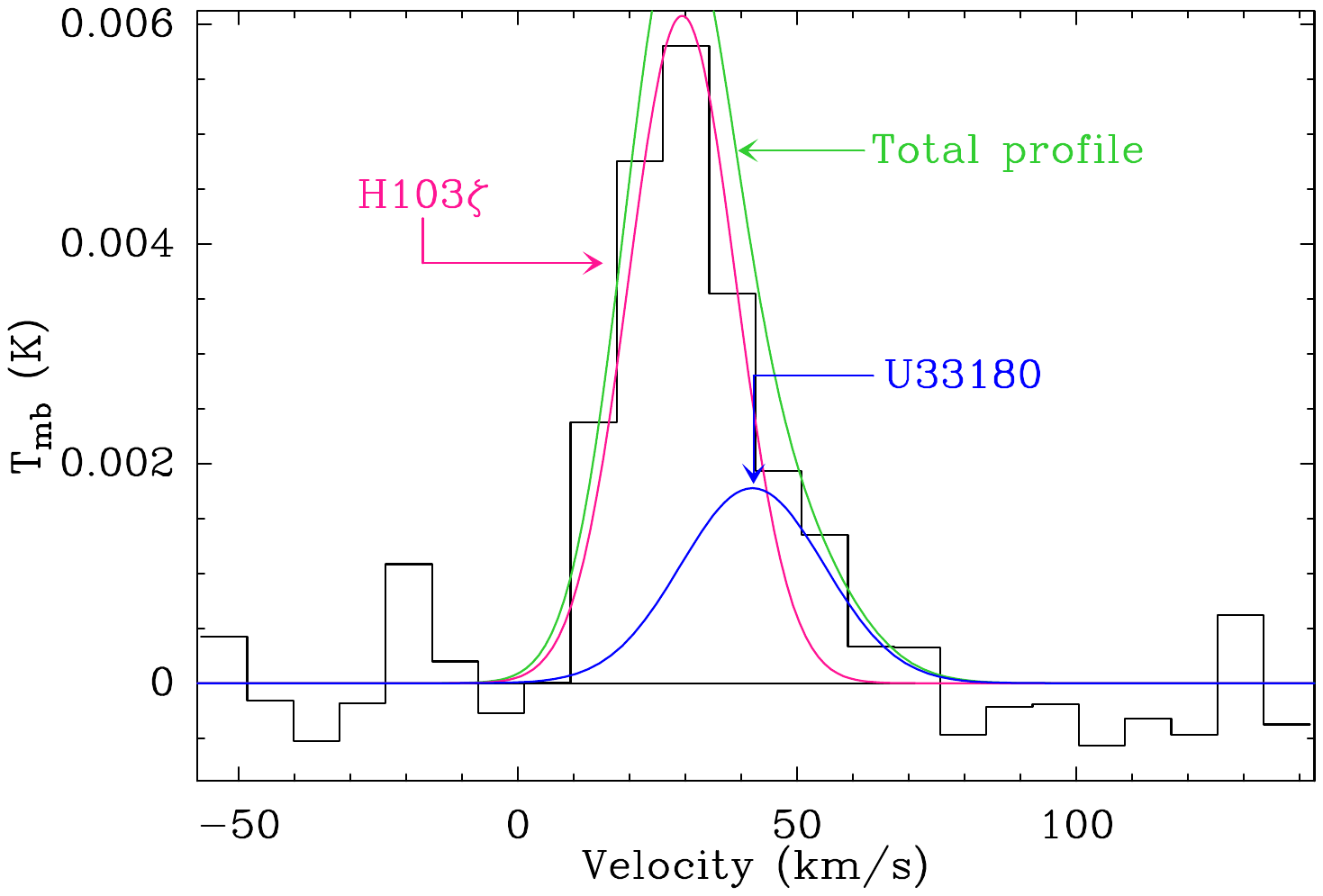}
		\caption{Tentative}\label{fig:ufs_U33180}
	\end{subfigure}
	\hspace{-6mm}
	\begin{subfigure}[]{0.34\linewidth}
		\includegraphics[width=\linewidth]{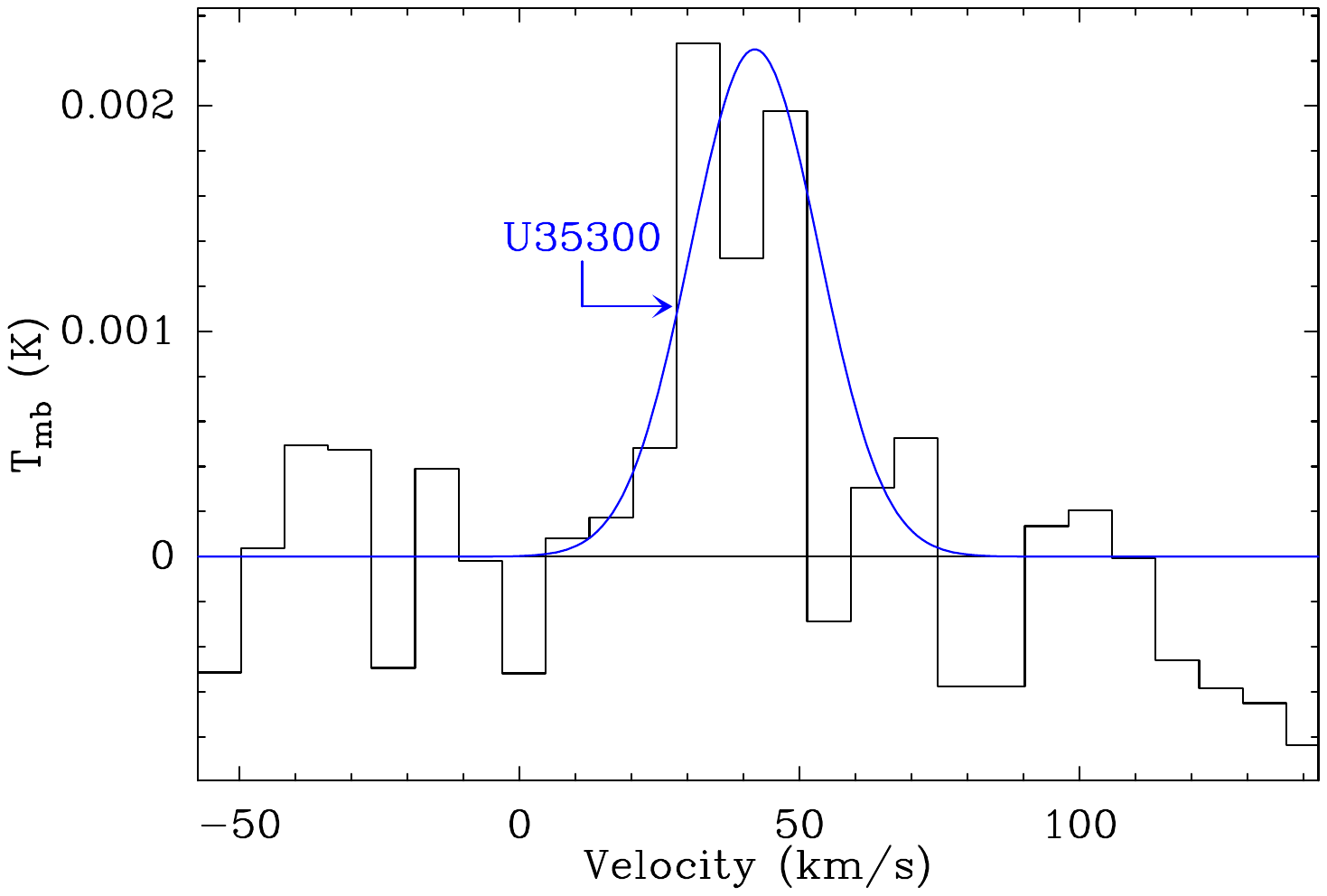}
		\caption{Detection}\label{fig:ufs_U35300}
	\end{subfigure}
	\hspace{-6mm}
	\begin{subfigure}[]{0.34\linewidth}
		\includegraphics[width=\linewidth]{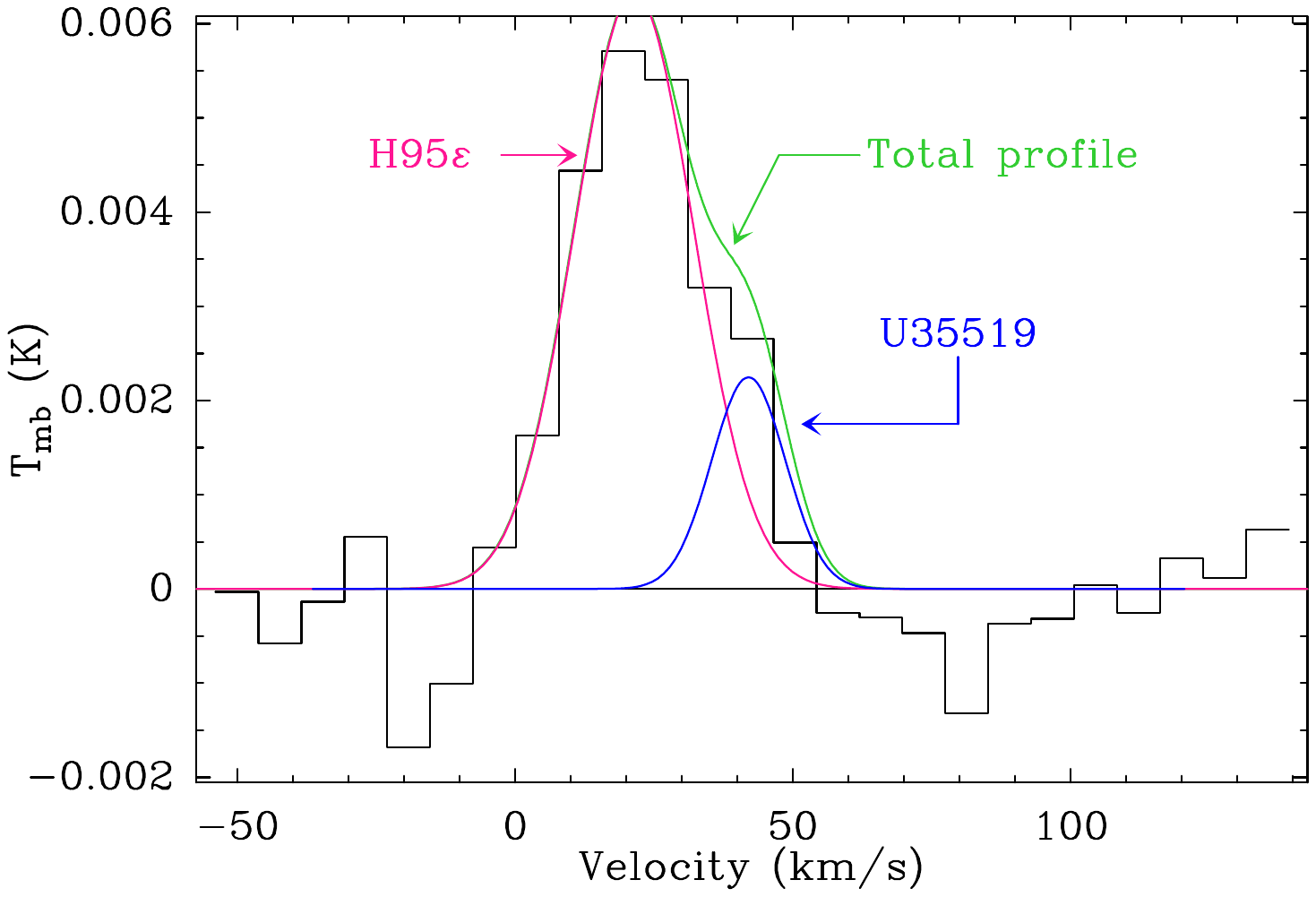}
		\caption{Tentative}\label{fig:ufs_U35519}
	\end{subfigure}
	\hspace{-6mm}
	\begin{subfigure}[]{0.34\linewidth}
		\includegraphics[width=\linewidth]{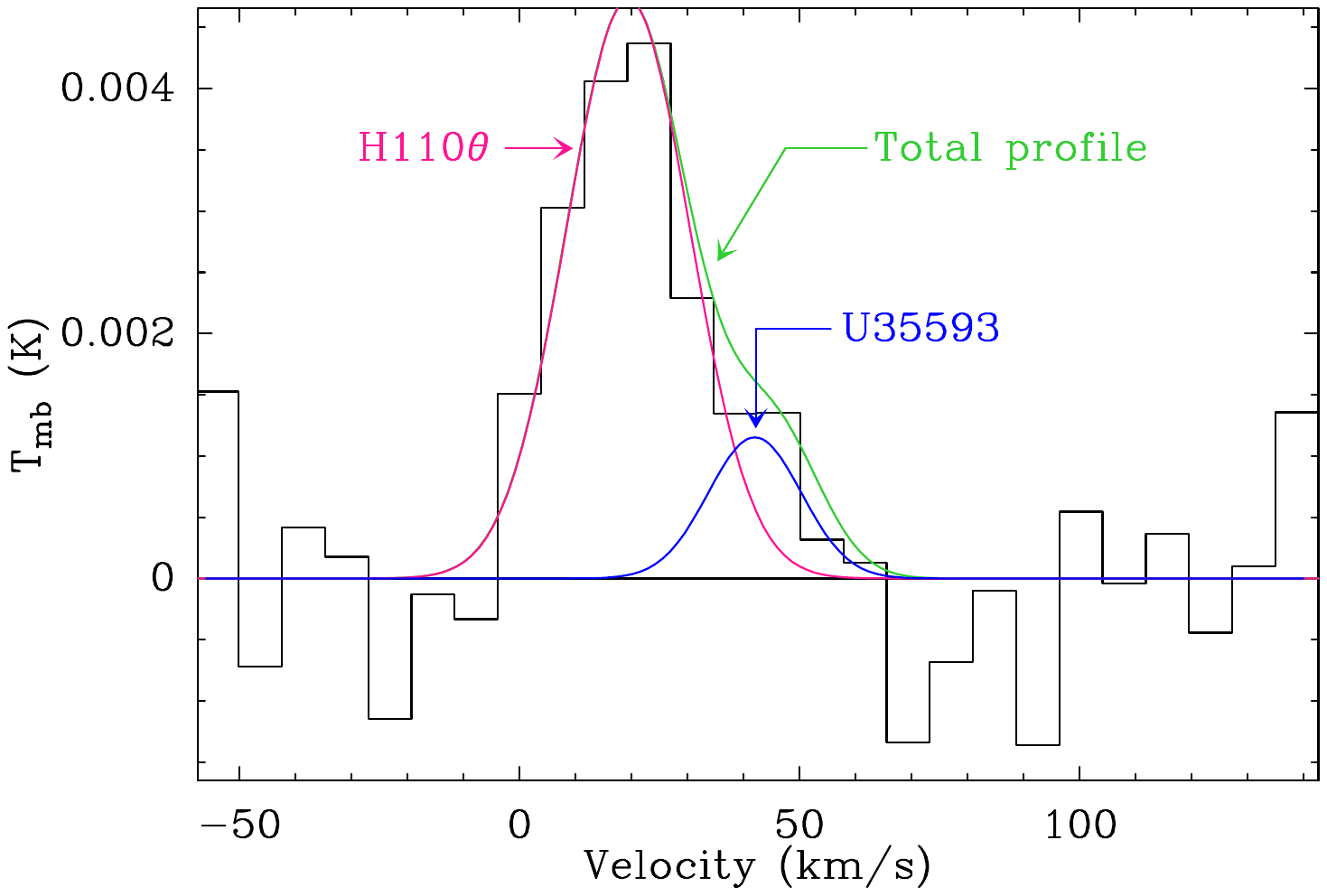}
		\caption{Tentative}\label{fig:ufs_U35593}
	\end{subfigure}
	\hspace{-6mm}
	\begin{subfigure}[]{0.34\linewidth}
		\includegraphics[width=\linewidth]{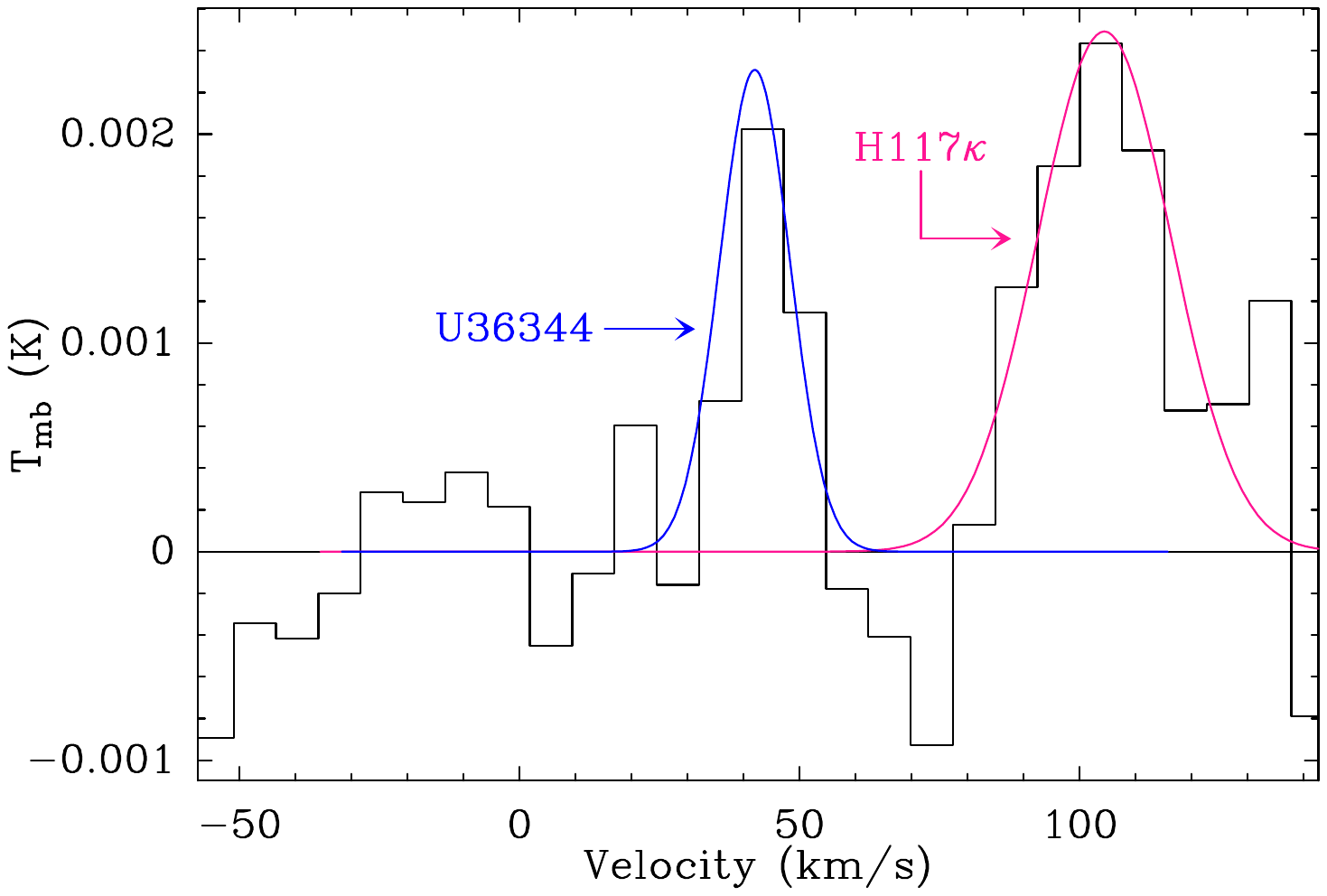}
		\caption{Detection}\label{fig:ufs_U36344}
	\end{subfigure}
	\hspace{-6mm}
	\begin{subfigure}[]{0.34\linewidth}
		\includegraphics[width=\linewidth]{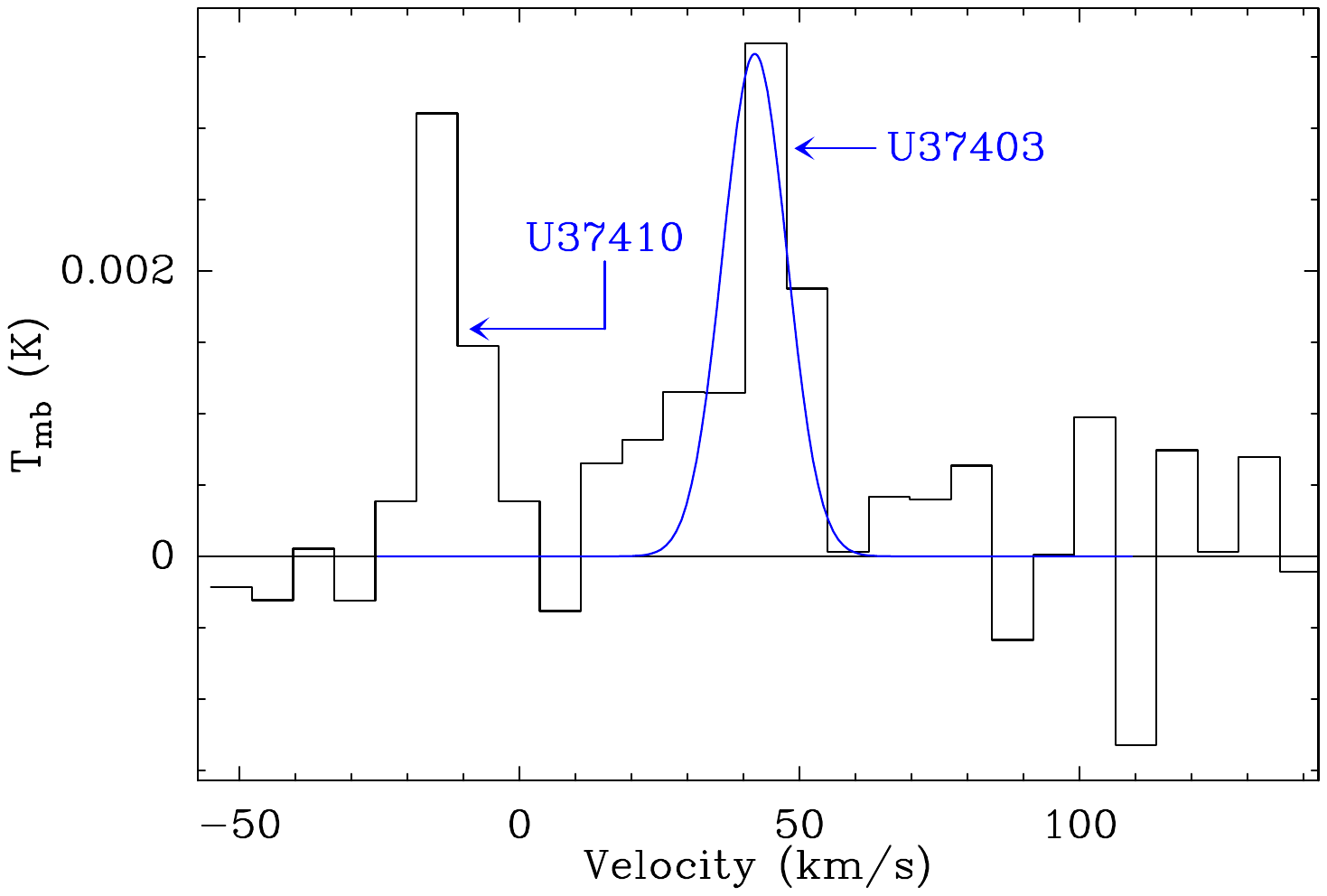}
		\caption{Detection}\label{fig:ufs_U37403}
	\end{subfigure}
	\hspace{-6mm}
	\begin{subfigure}[]{0.34\linewidth}
		\includegraphics[width=\linewidth]{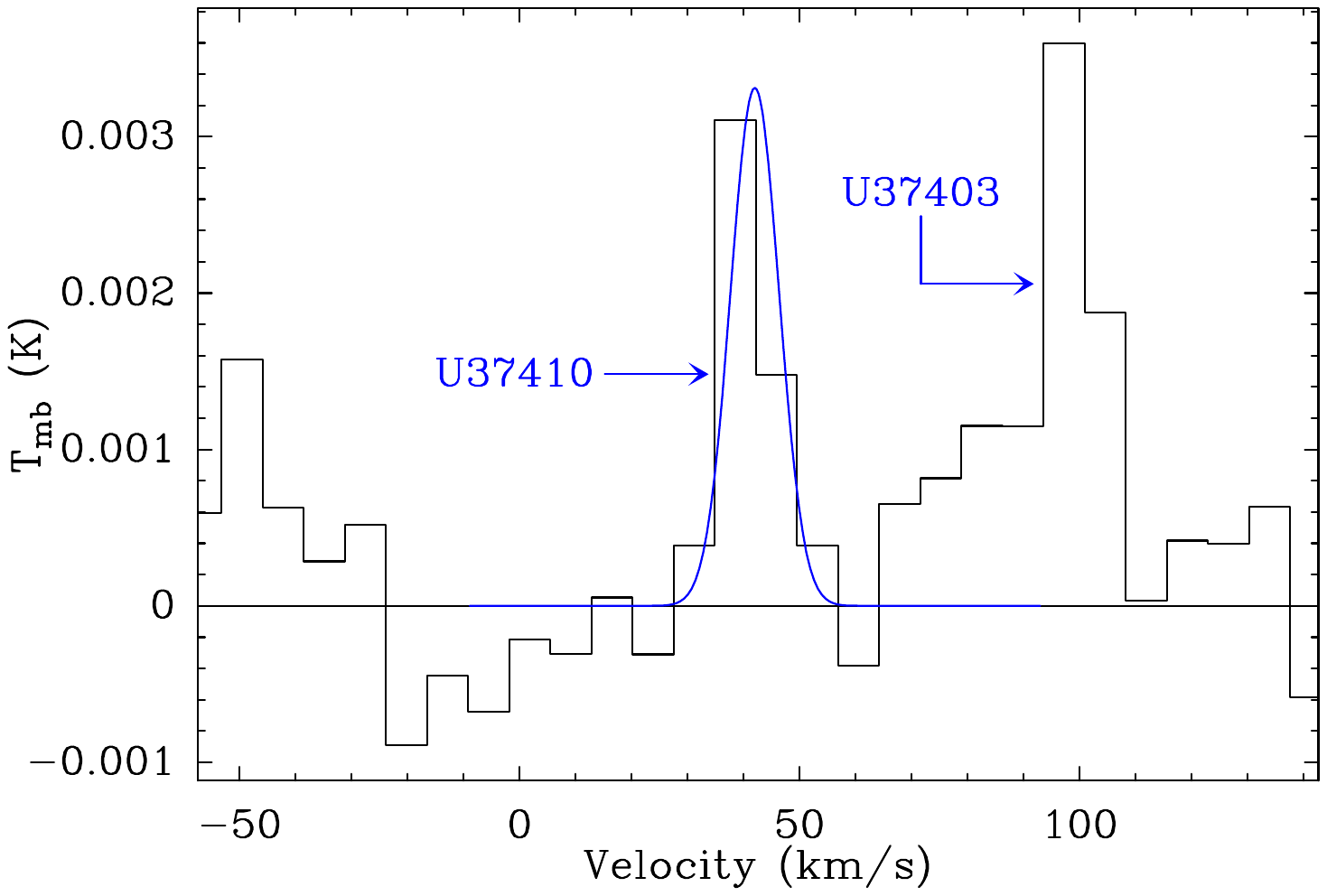}
		\caption{Detection}\label{fig:ufs_U37410}
	\end{subfigure}
	\hspace{-6mm}
	\begin{subfigure}[]{0.34\linewidth}
		\includegraphics[width=\linewidth]{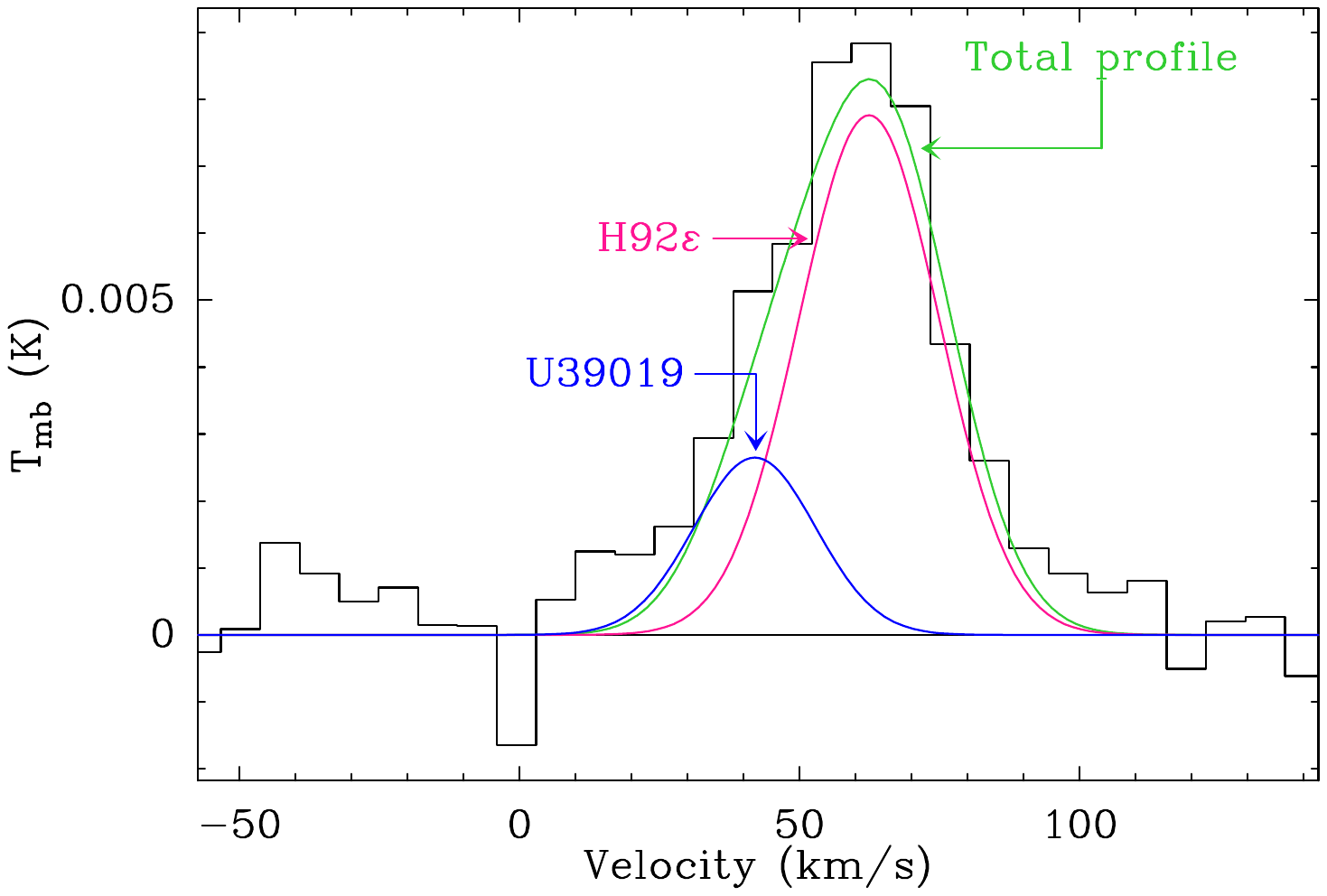}
		\caption{Tentative}\label{fig:ufs_U39019}
	\end{subfigure}
	\hspace{-6mm}
	\begin{subfigure}[]{0.34\linewidth}
		\includegraphics[width=\linewidth]{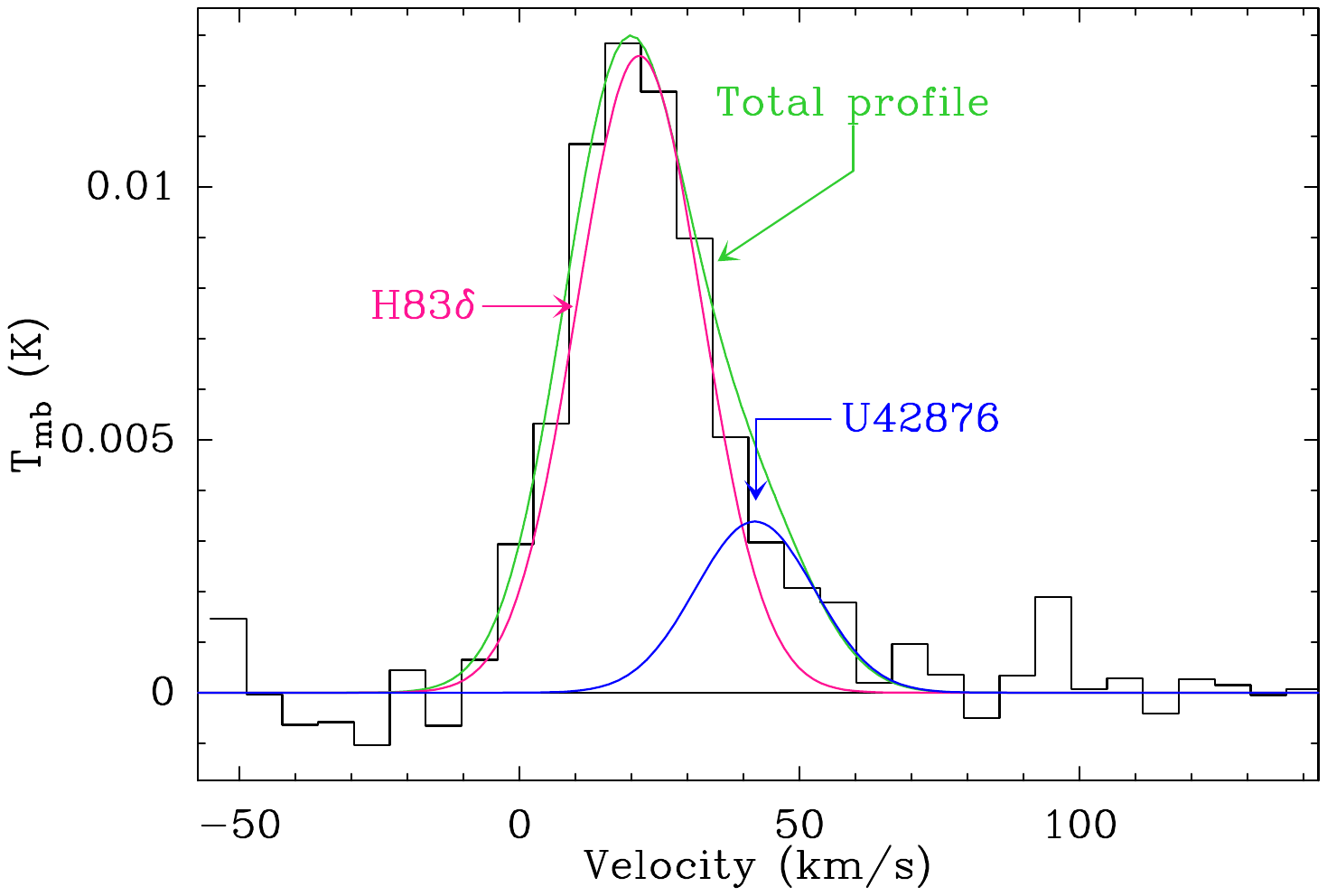}
		\caption{Tentative}\label{fig:ufs_U42876}
	\end{subfigure}
	\hspace{-6mm}
	\begin{subfigure}[]{0.34\linewidth}
		\includegraphics[width=\linewidth]{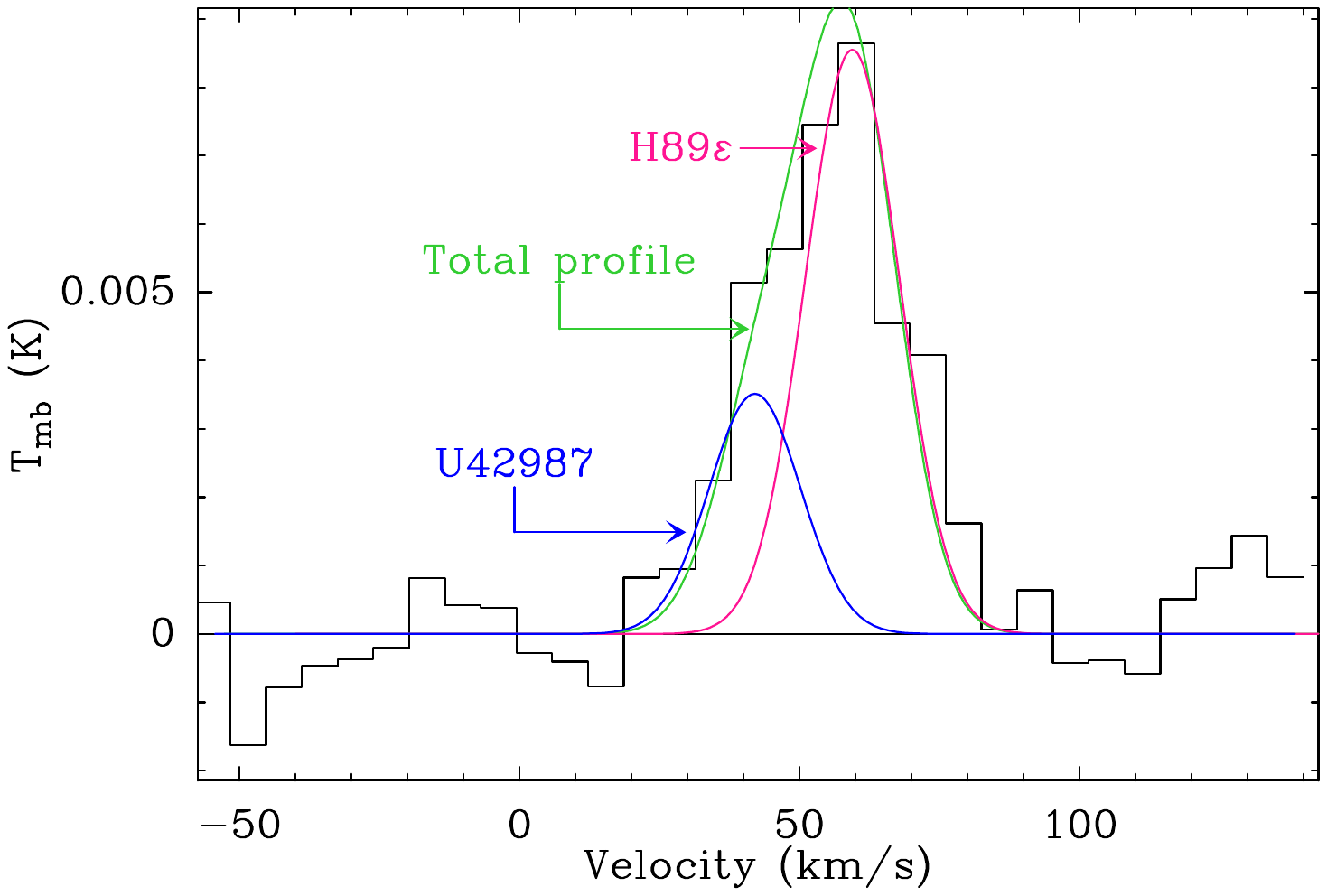}
		\caption{Tentative}\label{fig:ufs_U42987}
	\end{subfigure}
	\hspace{-6mm}
	\begin{subfigure}[]{0.34\linewidth}
		\includegraphics[width=\linewidth]{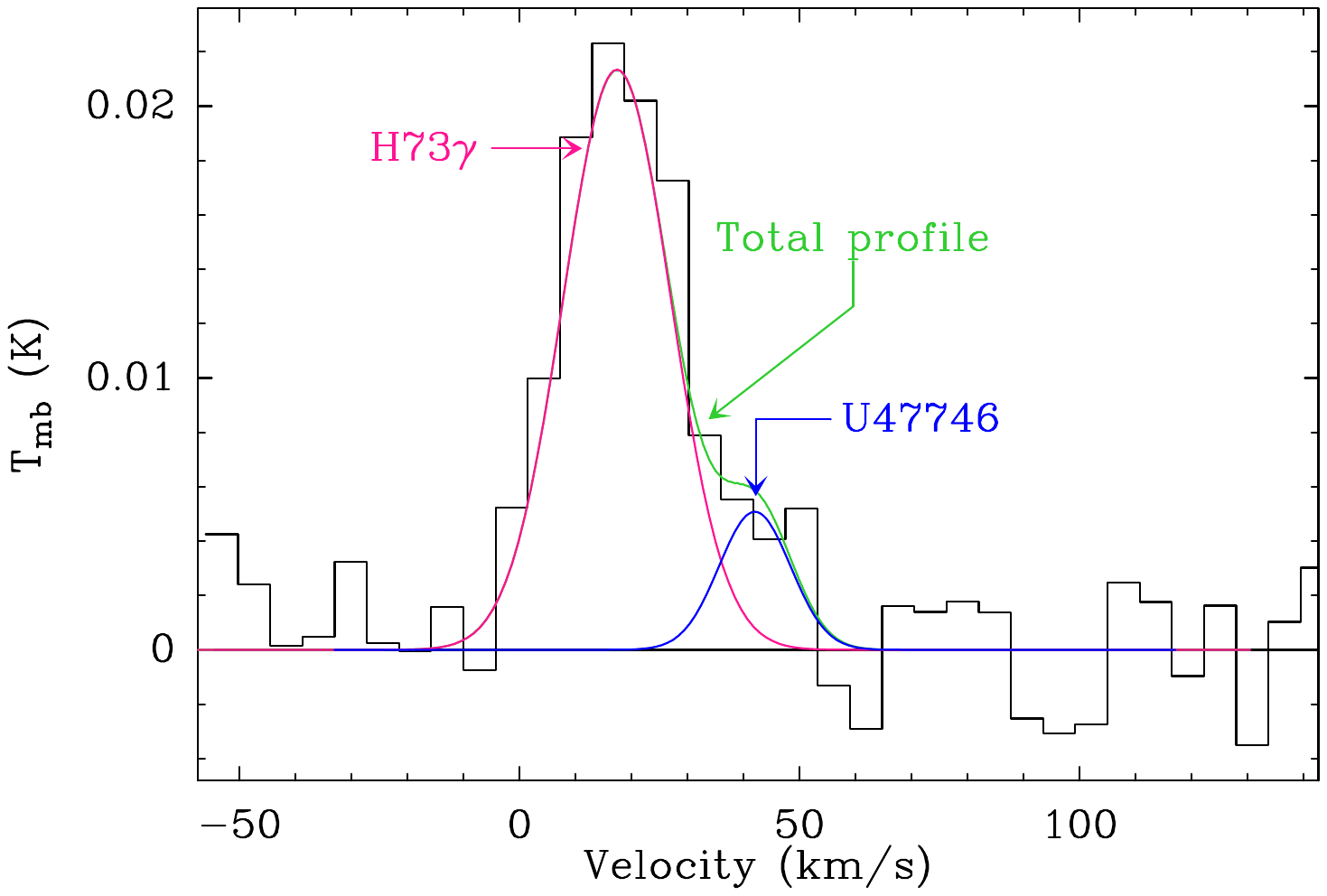}
		\caption{Tentative}\label{fig:ufs_U47746}
	\end{subfigure}
	\caption{UFs detected and tentatively detected across the spectra of \ic in the Q-band. Pink curves show the Gaussian fits of some RRLs, while blue curves display the Gaussian fits of UFs. The total profiles (RRL+UF) are shown in green (when applicable).}
	\label{fig:ufs_Yebes}
\end{figure*}

\begin{figure*}[htbp]
	\captionsetup[sub]{skip=0mm, belowskip=0pt}
	\centering
	\begin{subfigure}[]{0.34\linewidth}
		\includegraphics[width=\linewidth]{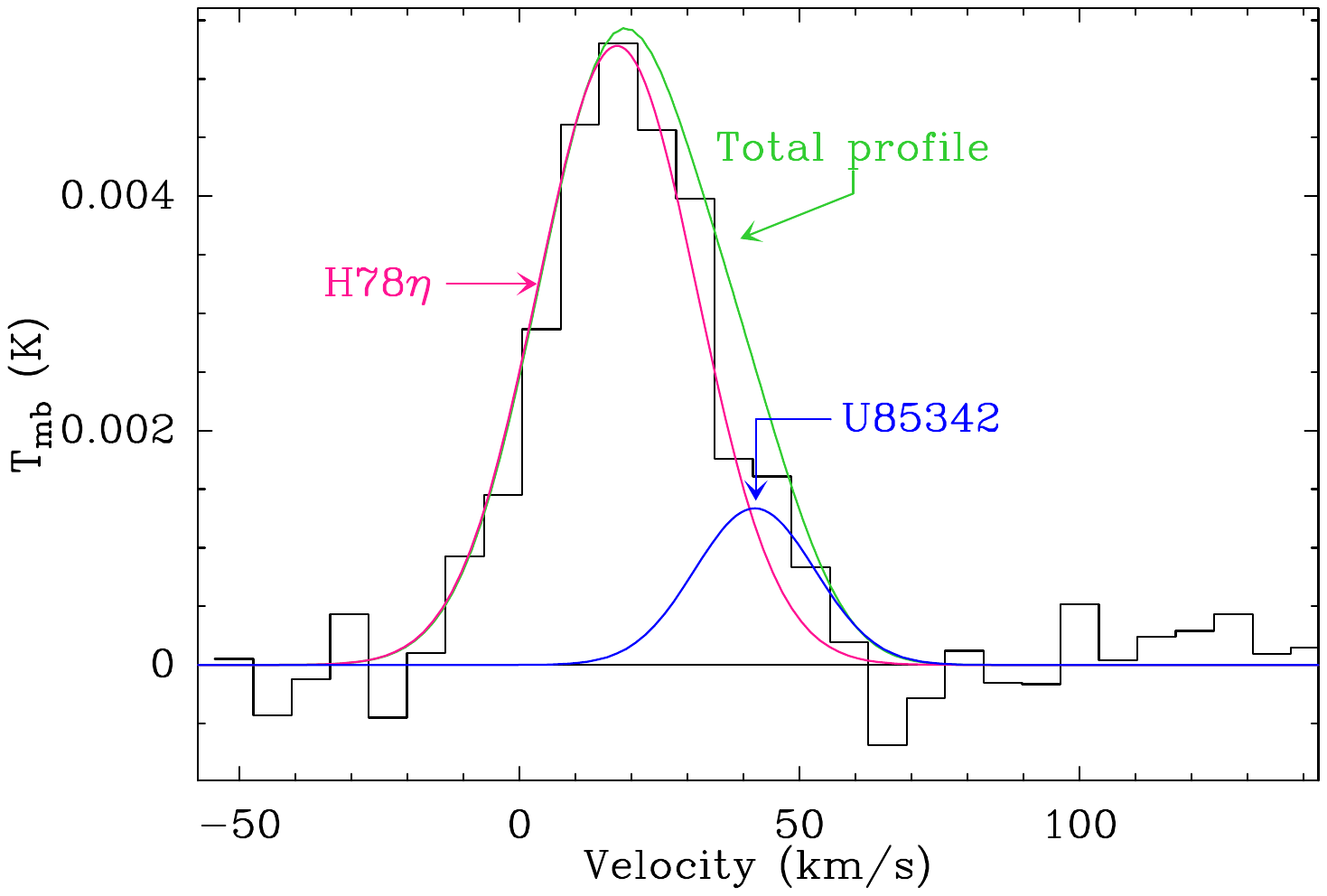}
		\caption{Tentative}\label{fig:ufs_U85342}
	\end{subfigure}
	\hspace{-6mm}
	\begin{subfigure}[]{0.34\linewidth}
		\includegraphics[width=\linewidth]{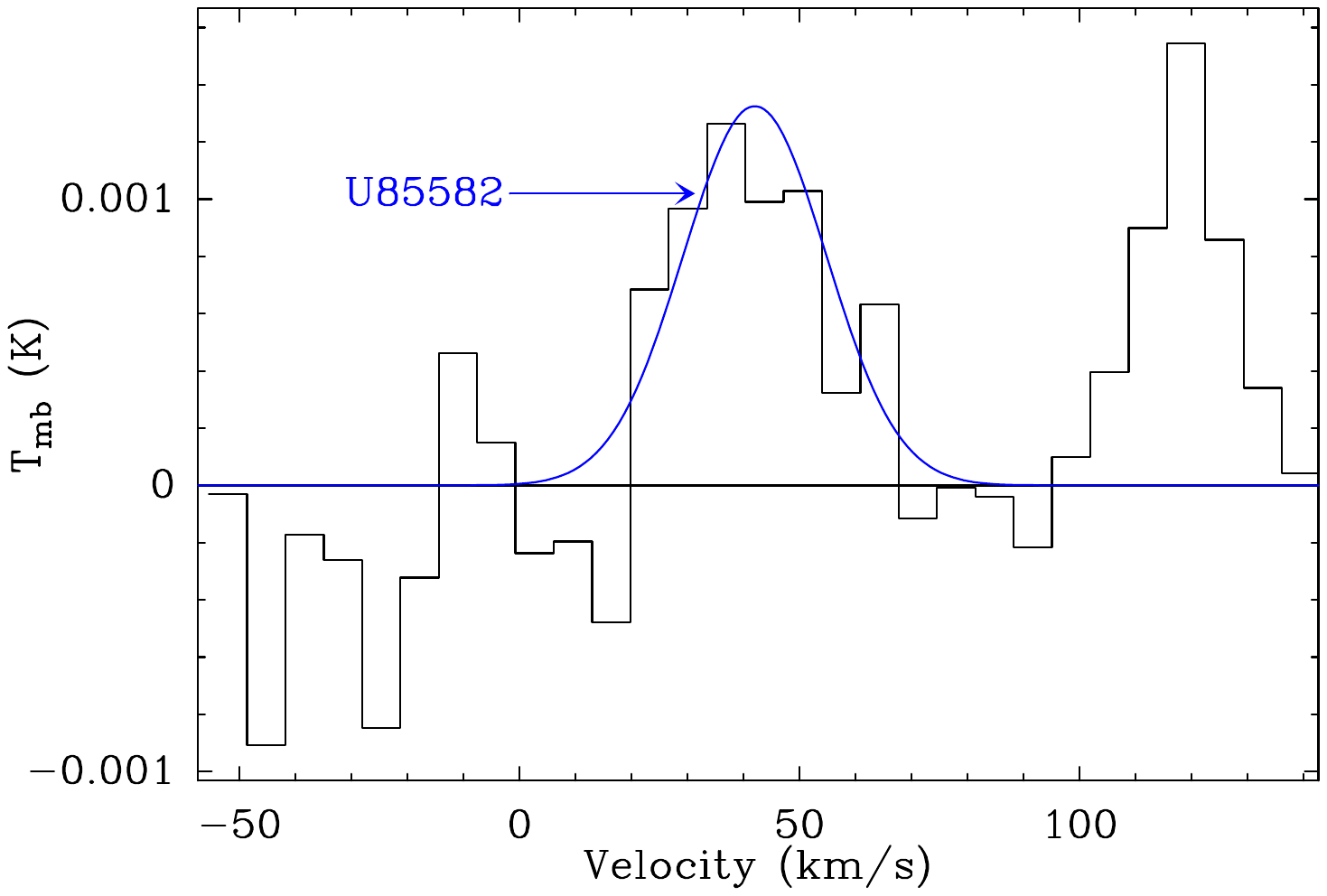}
		\caption{Detection}\label{fig:ufs_U85582}
	\end{subfigure}
	\hspace{-6mm}
	\begin{subfigure}[]{0.34\linewidth}
		\includegraphics[width=\linewidth]{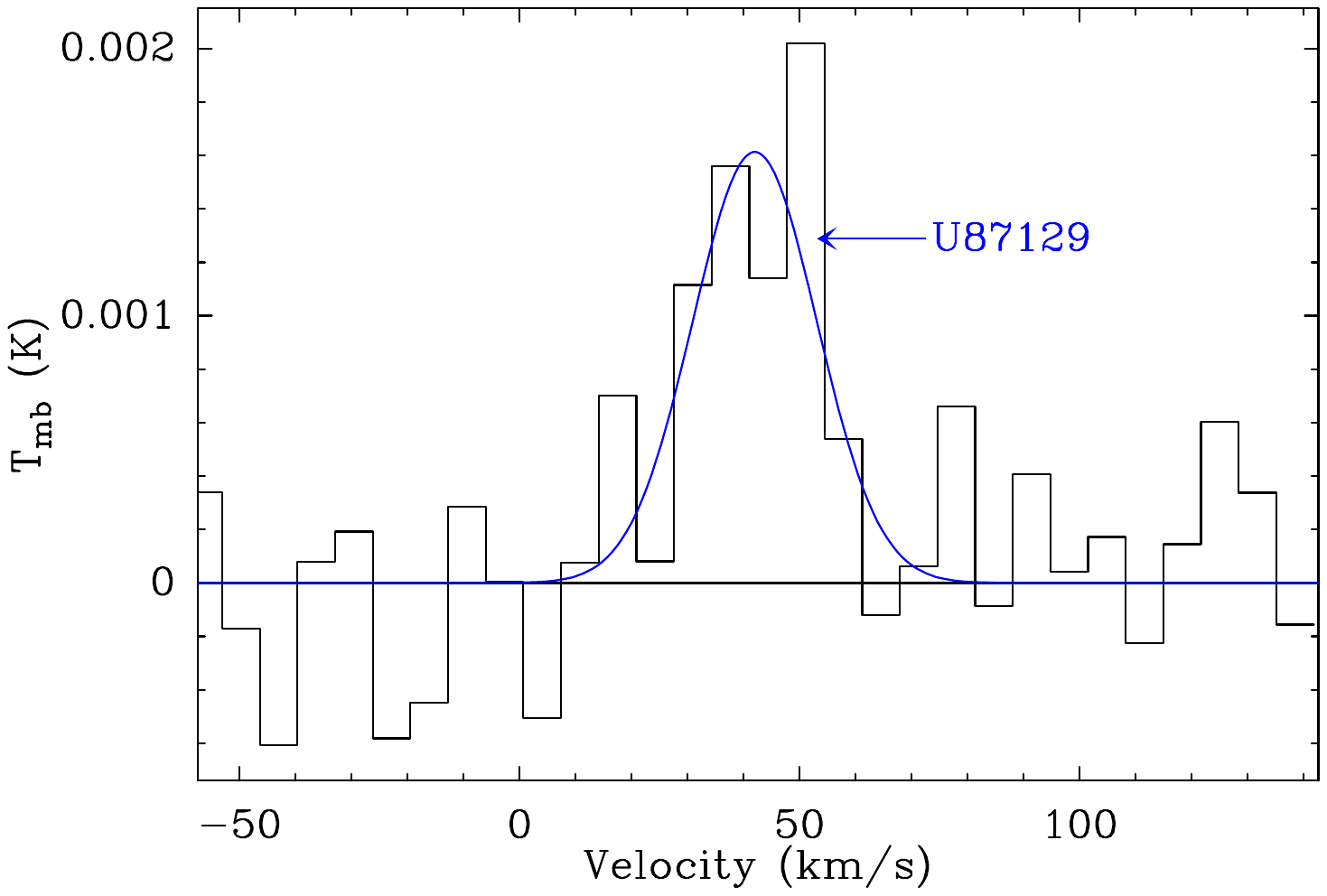}
		\caption{Detection}\label{fig:ufs_U87129}
	\end{subfigure}
	\hspace{-6mm}
	\begin{subfigure}[]{0.34\linewidth}
		\includegraphics[width=\linewidth]{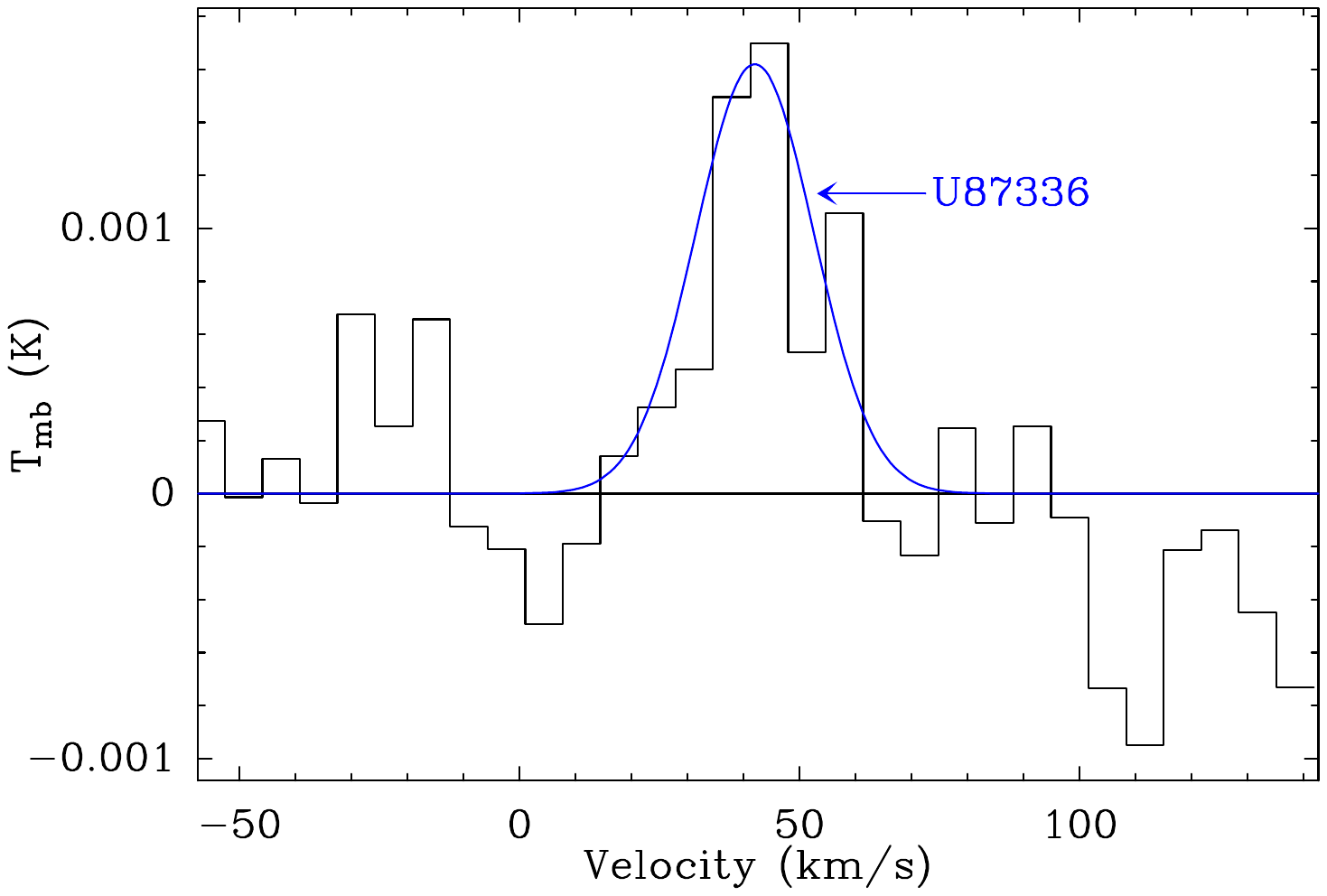}
		\caption{Detection}\label{fig:ufs_U87336}
	\end{subfigure}
	\hspace{-6mm}
	\begin{subfigure}[]{0.34\linewidth}
		\includegraphics[width=\linewidth]{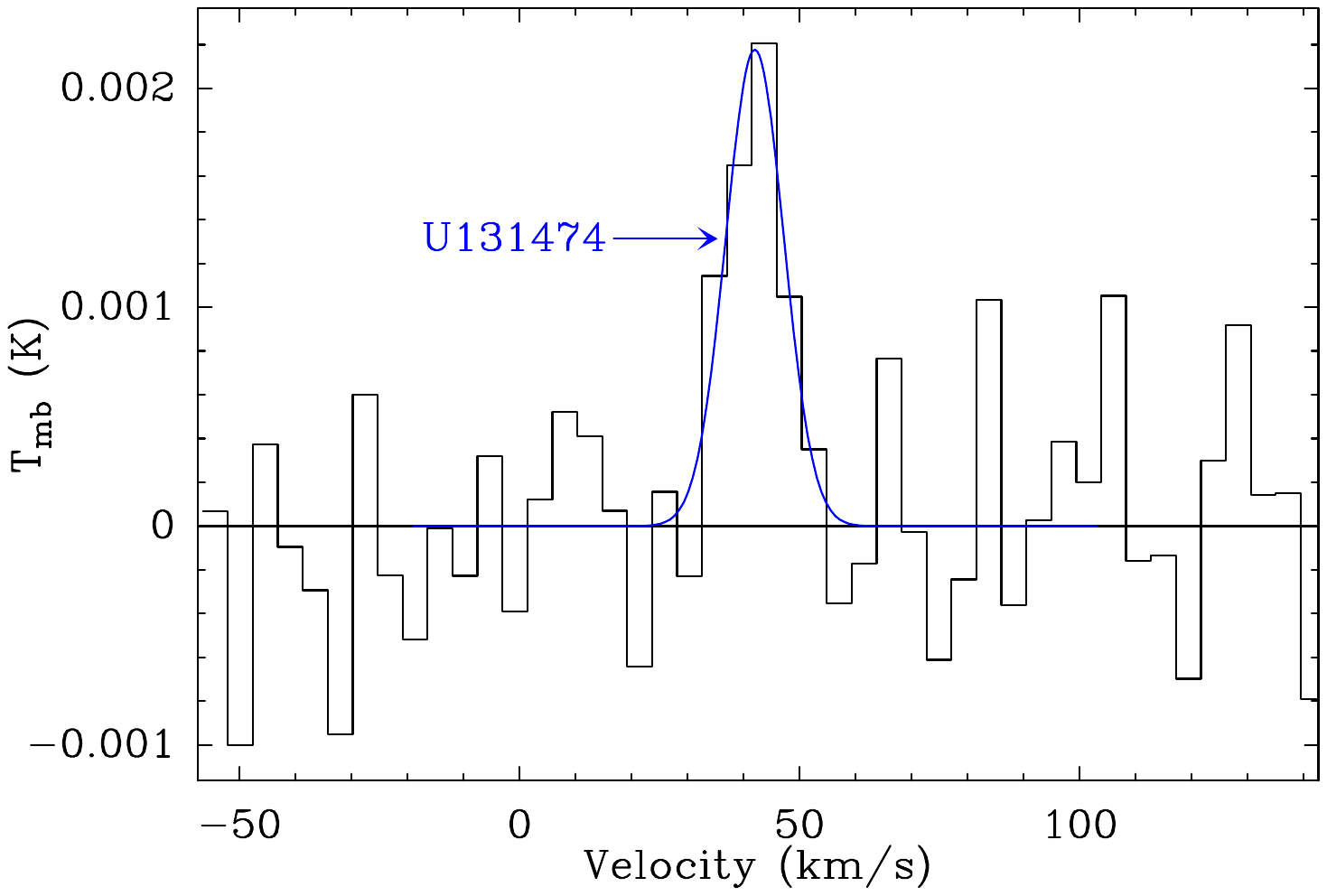}
		\caption{Detection}\label{fig:ufs_U131474}
	\end{subfigure}
	\hspace{-6mm}
	\begin{subfigure}[]{0.34\linewidth}
		\includegraphics[width=\linewidth]{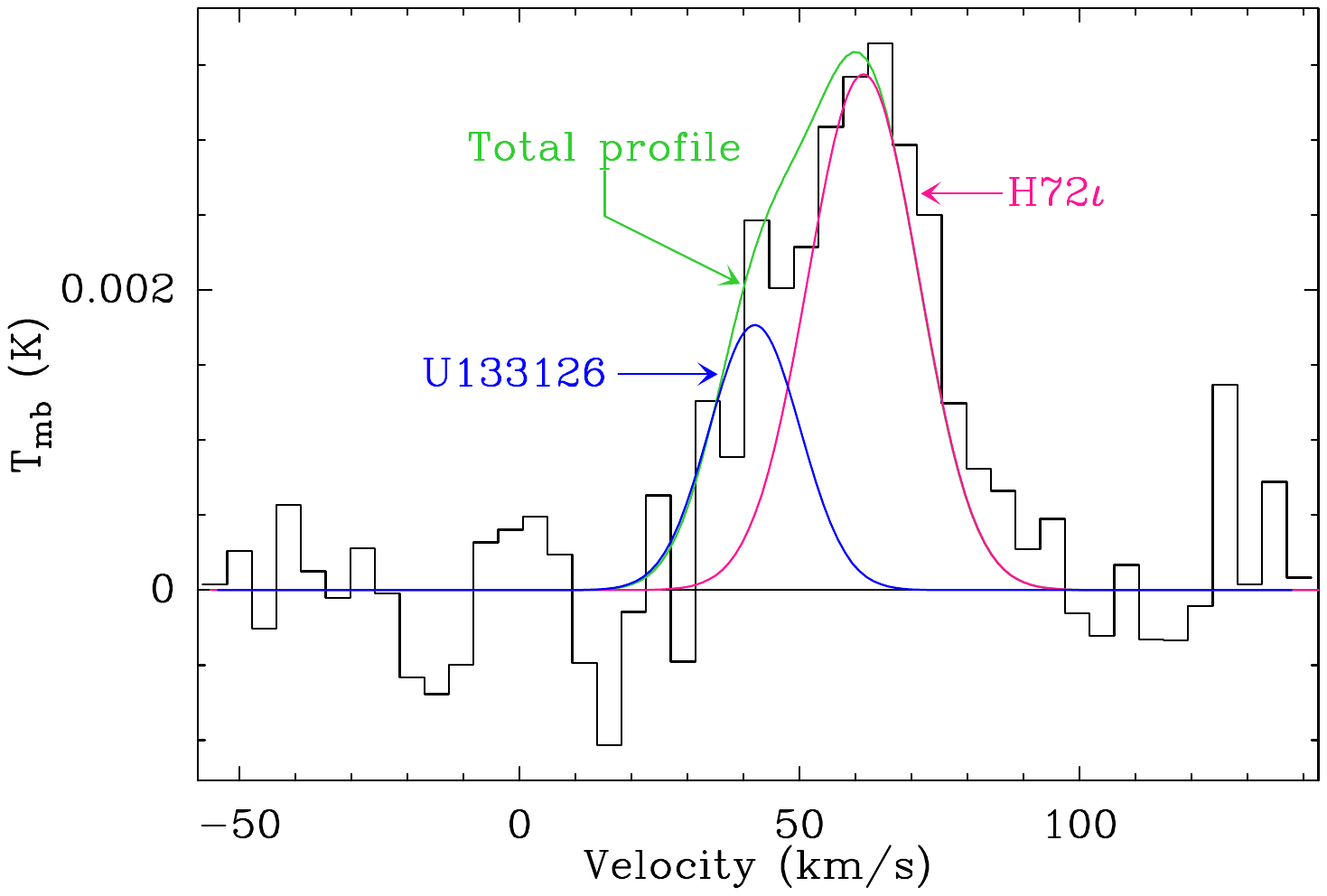}
		\caption{Tentative}\label{fig:ufs_U133126}
	\end{subfigure}
	\hspace{-6mm}
	\begin{subfigure}[]{0.34\linewidth}
		\includegraphics[width=\linewidth]{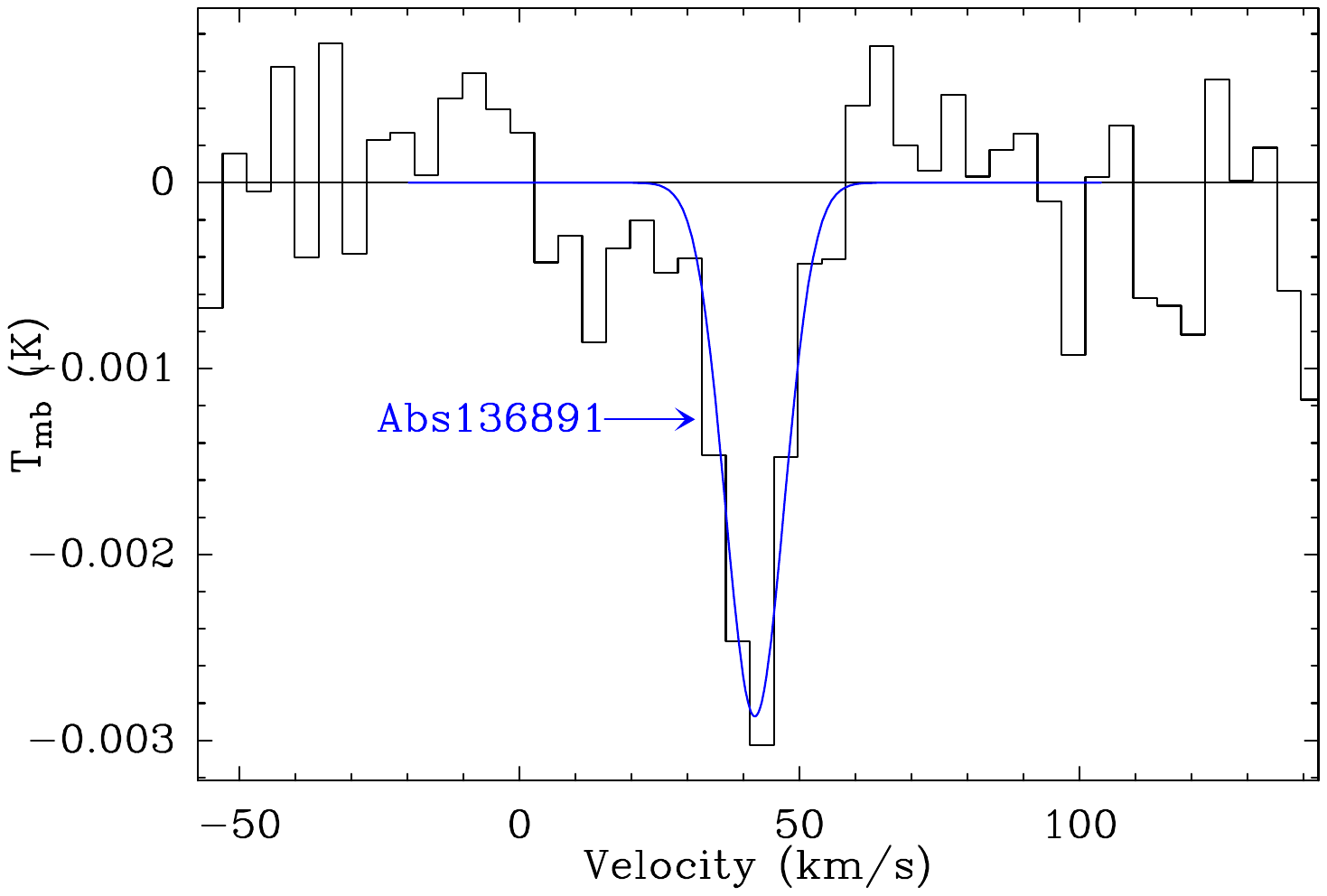}
		\caption{Detection}\label{fig:ufs_U136891}
	\end{subfigure}
	\hspace{-6mm}
	\begin{subfigure}[]{0.34\linewidth}
		\includegraphics[width=\linewidth]{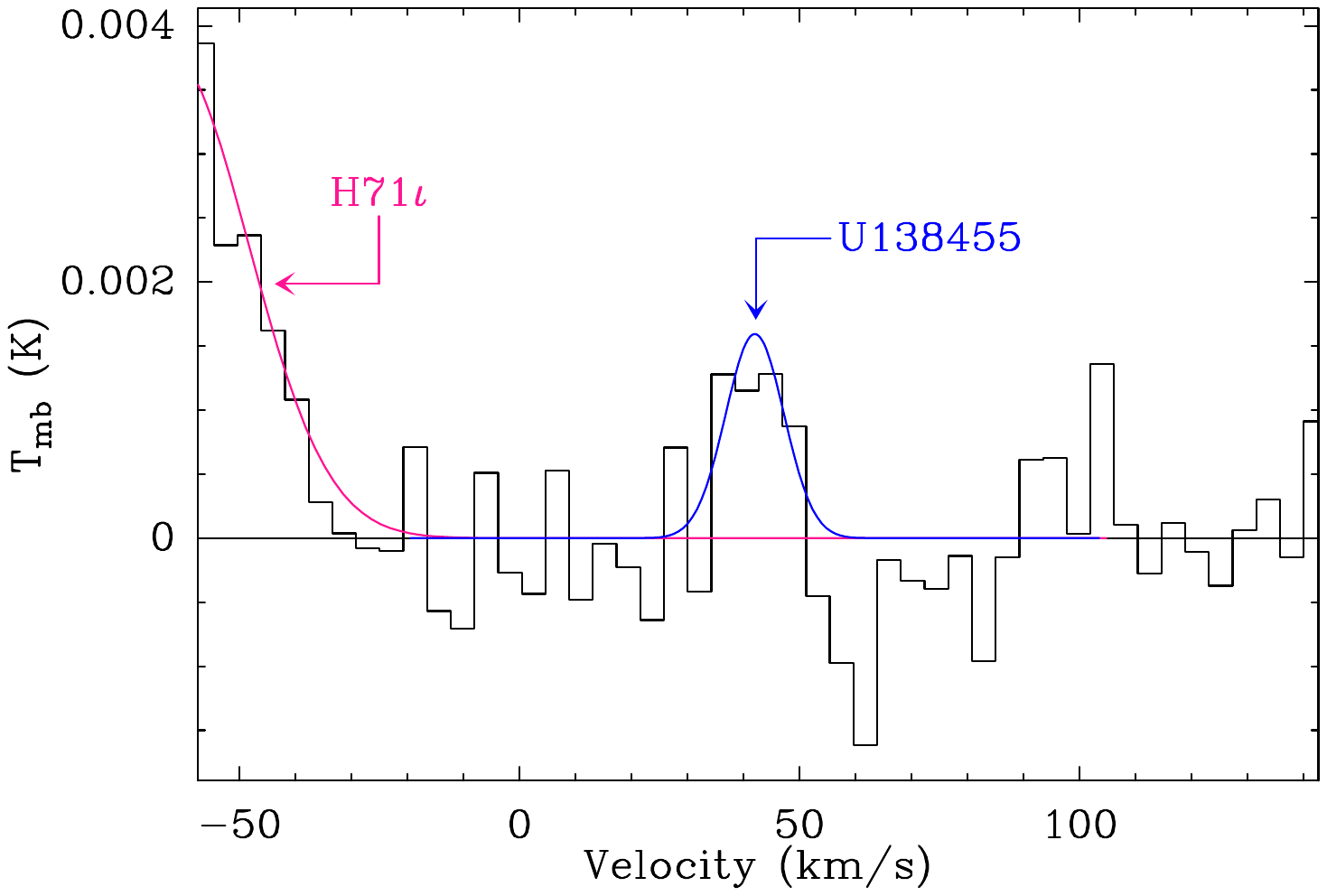}
		\caption{Detection}\label{fig:ufs_U138455}
	\end{subfigure}
	\hspace{-6mm}
	\begin{subfigure}[]{0.34\linewidth}
		\includegraphics[width=\linewidth]{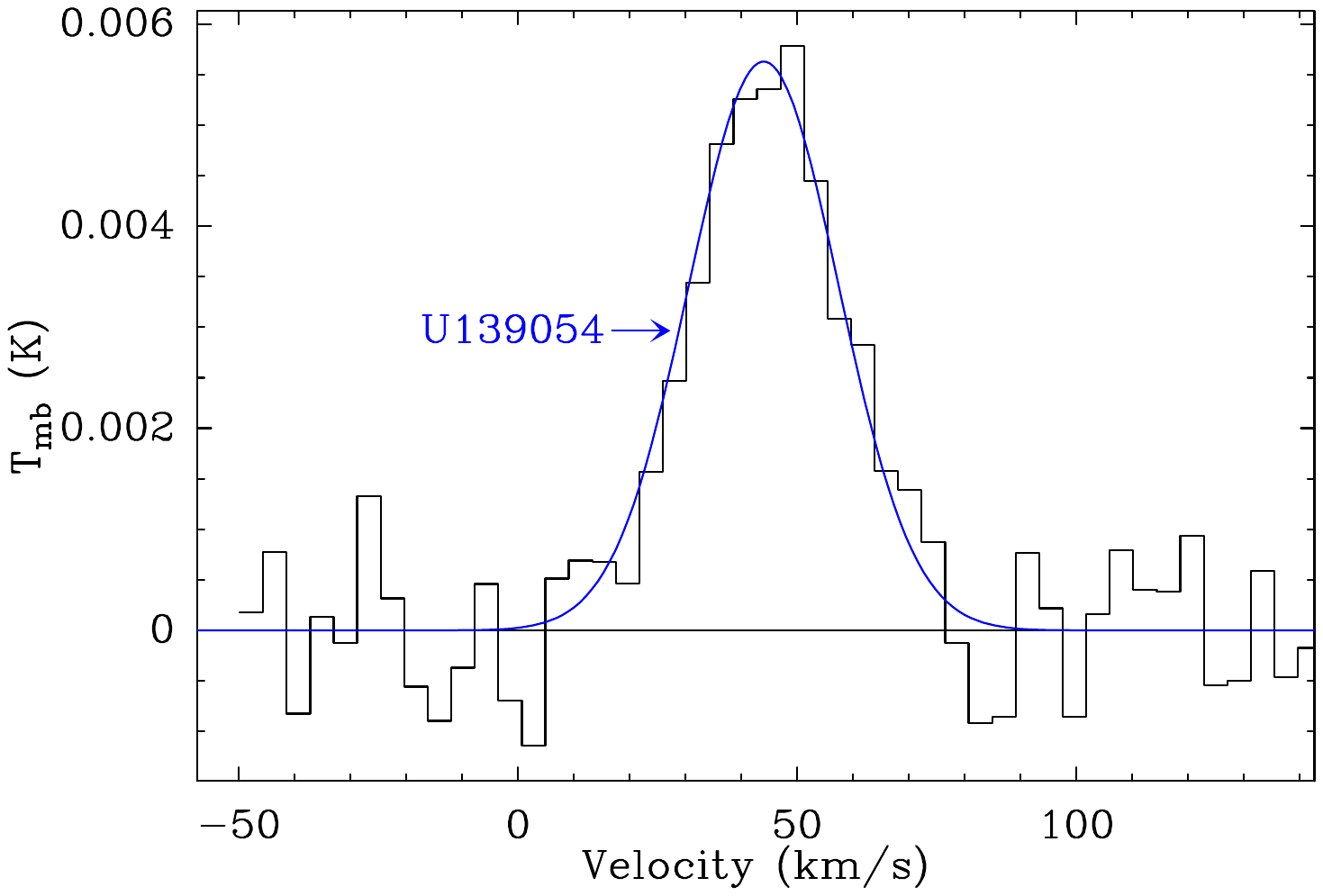}
		\caption{Detection}\label{fig:ufs_U139054}
	\end{subfigure}
	\caption{UFs detected and tentatively detected across the spectra of \ic in the Q-band. Pink curves show the Gaussian fits of some RRLs, while blue curves display the Gaussian fits of UFs. The total profiles (RRL+UF) are shown in green (when applicable).}
	\label{fig:ufs_IRAM}
\end{figure*}

The observations of the PN \ic were carried out during the winter semester of 2021. These observations are part of the IRAM 30m telescope project 158-21 (PI: D. A. García-Hernández) and RT40m project 22A011 (PI: D. A. García-Hernández).

Project 22A011 aimed to observe the whole Q-band (7\mm, i.e., \mbox{30.5 - 50.0\ghz}) with the RT40m, a 40\,m diameter such dish located in the Centro Astronómico de Yebes (Guadalajara, Spain) and operated by the Spanish Instituto Geográfico Nacional (IGN)\footnote{\href{https://rt40m.oan.es/rt40m_en.php}{https://rt40m.oan.es/rt40m\_en.php}}. These observations were executed in dual polarization using the position switching mode (PSW). The 2 and 3\mm (i.e., 131.5 - 139\ghz and 81.5 - 89\ghz, respectively) data were acquired with the IRAM 30m telescope located on Pico Veleta in Sierra Nevada (Granada, Spain) and operated by the Institut de Radioastronomie Millimétrique (IRAM). We observed the targets (see below) in single sideband and dual polarization with the E090 and E130 bands of the Eight MIxer super-heterodyne Receiver  \citep[EMIR\footnote{\href{https://publicwiki.iram.es/EmirforAstronomers}{https://publicwiki.iram.es/EmirforAstronomers}},][]{Carter2012} using the wobbler switching mode \citep[WSW; see][for more details]{HuertasRoldan2025}.

Together with the C$_{60}$-rich PN \ic, the projects included the observation of the young PN \ngc. This object is the archetypal PAH-rich PN and IR observations clearly show that it does not contain fullerenes \citep[][]{Otsuka2014, SmithPerez2025}. We note that a detailed analysis of the molecular rotational lines in NGC 7027 will be presented in a forthcoming paper (Huertas-Roldán, \textit{in prep}.) but see Sec.~\ref{sec:comparison_IC418-NGC7027} for a qualitative comparison between \ngc and \ic. All datasets were previously reduced using the Continuum and Line Analysis Single-dish Software (CLASS) program of the GILDAS \footnote{For more information, see \href{https://www.iram.fr/IRAMFR/GILDAS}{https://www.iram.fr/IRAMFR/GILDAS} (accessed on 24 September 2024).} astronomical software package \citep[sept22a version; see][]{HuertasRoldan2025} to identify and analyze the radio recombination lines (RRLs)\footnote{The RRLs are emission lines in the radio regime produced in a plasma when a recently captured electron de-excites to a lower Rydberg level.} present in the spectra of both sources. Some weak lines remained unidentified in these spectra. For \ic, we reduced the rms noise by resampling the spectra to a final resolution of 0.916\mhz (8.4\kms - 5.7\kms) at 7\mm, 1.954\mhz (6.7\kms) at 3\mm, and 1.954\mhz (4.3\kms) at 2\mm. At these resolutions, the rms noise level is $0.5 - 1.4$\mkel at 7\mm [$T_\mathrm{mb}$] and 0.4\mkel [$T_\mathrm{mb}$] at 2 and 3\mm.

Weak emission lines can be mistaken with instrumental features, radio frequency interferences (RFI), or resonant signals due to strong lines in the image side band. To confirm the real emission lines and remove the parasites, we first identified any known RFI with the help of the IRAM and Yebes Observatory staff, then compared the spectra acquired with different local oscillator (LO) frequencies, and finally we inspected the polarizations of the observations. The comparison of data taken using different LO frequencies reveals a displacement in frequency of the artifacts different to that expected for real signals coming from the source. By comparing the V and H polarizations, genuine emission (sub)thermal molecular lines are expected to exhibit similar shapes and intensities. Otherwise, they are considered artifacts. We compared, line by line, the total average of scans at each polarization and removed from the list of candidates to real molecular lines all features that clearly showed displacement on a polarization with respect to the other or emission just in one polarization. With the remaining candidates, we compared the average of each individual day in both polarizations. Those lines which showed significantly different emission profiles or intensities on each day were also removed from the list. In this analysis we also confirmed that there are not signal from the image side band of the receiver related to strong RRLs contaminating the data \citep[see][]{HuertasRoldan2025}. No exceedingly narrow emission typical of masers has been detected in the spectra.

\section{Results}
\label{sec:results}

\begin{figure*}[ht]
	\centering
	\includegraphics[width=0.9\linewidth]{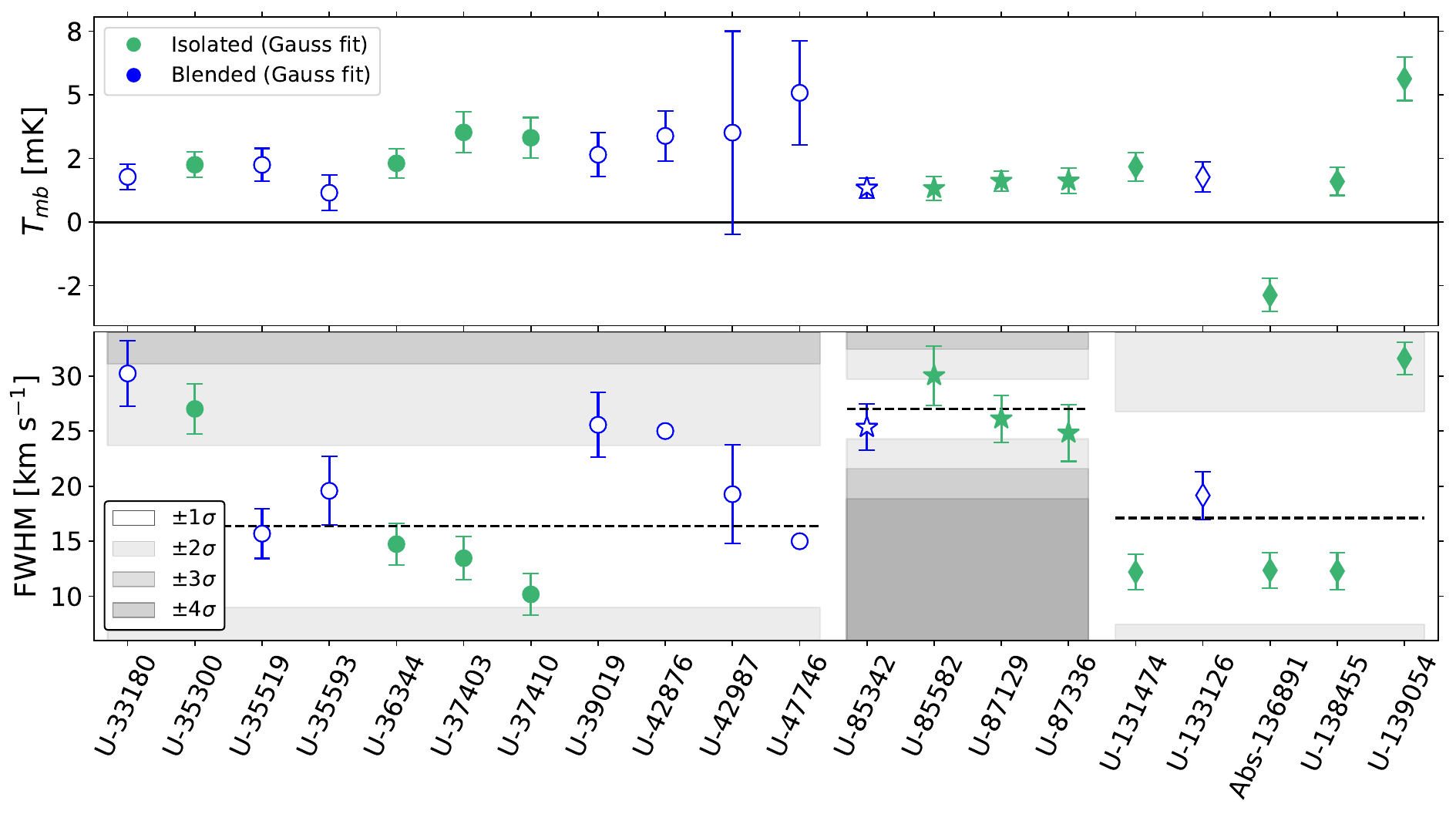}
	\caption{Peak temperature (K, upper panel) and FWHM (\kms, lower panel) of the UFs found in \ic. Dots, stars, and diamonds represent the 7, 3, and 2\mm band data, respectively. Open blue symbols correspond to tentatively detected lines because they are blended with RRLs and their SNR is $\sim2-4$ (see Table \ref{table:ufs_frequencies}). The gray area in the lower panel represent the $\pm 1\sigma$, $\pm 2\sigma$, $\pm 3\sigma$, and $\pm 4\sigma$ values around the mean.}
	\label{fig:UFs_Tmb-FWHM}
\end{figure*}

Among the approximately 160 RRLs that populate the radio spectrum of \ic \citep[see][]{HuertasRoldan2025}, we have found 20 UFs with SNR $>2$ (see Table \ref{table:ufs_frequencies} and Figs. \ref{fig:ufs_Yebes} and \ref{fig:ufs_IRAM}). Out of the total, 11 UFs are observed above the detection level and they are all isolated lines (SNR $>3$ for isolated lines), this is, not-blended with any RRL. The isolated lines show intensities varying from 1 to 5\,mK (see Fig.~\ref{fig:UFs_Tmb-FWHM}) and Gaussian-like profiles. At 7\mm, the four isolated lines display quite similar peak intensities. Their FWHMs are compatible with the $\pm1\sigma$ region around the mean (see Fig.~\ref{fig:UFs_Tmb-FWHM}),  with the exception of U35300, which is the broadest UF. At 3\mm, the lines display also similar peak intensities and FWHMs statistically compatible. At 2\mm, the peak intensities of the lines are more diverse. We find in this range the only line in absorption from the whole data set (Abs136891 Fig. \ref{fig:ufs_U136891}). The FWHMs are compatible with their average value at $1\sigma$ level, except for U139054, which doubles the mean FWHM value.

The 9 tentatively detected UFs are all partially blended with RRLs (see Figs.~\ref{fig:ufs_Yebes} and \ref{fig:ufs_IRAM}). We have assumed Gaussian shapes to fit and analyze them. In the 7\mm band, 7 lines display a wide range of intensities and FWHMs broader than the values of the clearly detected and isolated lines. However, this latter parameter is not well determined due to the uncertainties imposed by the partial blending on the fitting process. The tentative line at 3\mm has an intensity and FWHM similar to the values of the clearly detected lines. The same is found for the tentative lines at 2\mm.

Tentative UFs can be weak real lines produced by molecules, heavy chemical elements, and inhomogeneities in the envelope of \ic, or unusually strong and broad noise structures difficult to identify without observational ratification. All these possibilities are particularly important in the current work given that all tentative UFs are blended with RRLs. We have performed several sanity checks in order to keep only the molecular features, but possible noise structures could not be completely removed from the linelist \textit{a priori}, with the expectation that the data analyses presented in Section \ref{sec:discussion} will identify any spurious features.

\subsection{Analyzing coincidences of UFs with RRLs}
\label{sec:UFs-RRLs}

Given the high number of RRLs detected in this source, it can be argued that the UFs partially blended with the hydrogen RRLs could actually be unassigned RRLs of heavier atoms. Comparing the predictions with the observed spectra, we found all H and \hei RRLs that were expected to appear on each spectral range in \ic \citep[see][]{HuertasRoldan2025} and we conclude that the UFs are not RRLs of these atoms. We also checked RRL frequencies of other neutral chemical elements or their ionized species (D, \trihei, \triheii, \fourheii, \ci, \cii, \ciii, \oi, \oii, and \oiii) even though their low abundances in this environment result in peak flux intensities below the noise level of our observations \citep[for example, for \ci lines the intensities are $\sim 10^{-6}$\kel; see e.g.,][to find the abundances of the elements]{Pottasch2004}. Despite there are a few coincidences between some of the UFs and RRLs of chemical elements other than H and hei, they have to be considered accidental. Any real contamination to RRLs associated to particularly dense gas inhonomogeneities in the \hii region should affect all lines of the same series systematically, this is, all $\alpha$ lines ($\Delta n=1$), all $\beta$ lines ($\Delta n=2$), and so on. This behavior is not found, so the coincidences are considered accidental.

Furthermore, in order to check if the UFs blended with RRLs were in fact features associated to possible denser clumps in the ionized wind of \ic, we tested if RRLs profiles display systematic asymmetries. The emission of each clump would produce a noticeable bump in the overall emission that could be shifted due to a possible distinct expansion velocity. So, any systematic asymmetry in the line profiles of the RRLs can suggest the presence of such clumps, but such variations have to be found in all the RRLs simultaneously and, in particular, in the H\ga lines. We stacked all the H\ga RRLs to get an average, high quality line profile and found no bump that could suggest the presence of such clumps.

\subsection{Searching for coincidences with known molecules in space} \label{sec:UFs-molecules}

We have compared these UFs frequencies with those of the rotational transitions of the 1300 molecules available at the Cologne Database for Molecular Spectroscopy (CDMS)\footnote{\href{https://cdms.astro.uni-koeln.de/classic/catalog}{https://cdms.astro.uni-koeln.de/classic/catalog}}, the $\sim 400$ molecules at the JPL Molecular Spectroscopy Catalog\footnote{\href{https://spec.jpl.nasa.gov/ftp/pub/catalog/catdir.html}{https://spec.jpl.nasa.gov/ftp/pub/catalog/catdir.html}}, and with recent detections towards sources as \mbox{TMC-1} since 2021 \citep{Agundez2021a, Agundez2021b, Agundez2022a, Agundez2023a, Agundez2023b, Agundez2024a, Cabezas2021d, Cabezas2021a, Cabezas2021c, Cabezas2021b, Cabezas2022a, Cabezas2022b, Cabezas2022c, Cabezas2023b, Cabezas2024a, Cabezas2024b, Cernicharo2021c, Cernicharo2021f, Cernicharo2021h, Cernicharo2021i, Cernicharo2021d, Cernicharo2021e, Cernicharo2021b, Cernicharo2021g, Cernicharo2021a, Cernicharo2022a, Cernicharo2023b, Cernicharo2023c, Cernicharo2024a, Cernicharo2024b, Cernicharo2024c, Cernicharo2024d, Cooke2023a, Fried2025, Fuentetaja2022a, Fuentetaja2023a, Fuentetaja2024a, Fuentetaja2024b, Fuentetaja2025, Loru2023a, Marcelino2021, Marcelino2023a, Remijan2023a, Remijan2024a, Remijan2025a, Silva2023a, Tennis2023a, Wenzel2024, Wenzel2025a, Wenzel2025b} and IRC+10216 \citep{Cabezas2023a, Cernicharo2023a, Gupta2024a, Pardo2022, Pardo2023a, VelillaPrieto2023a}. The signals do not appear to correspond to any known interstellar or circumstellar molecule, species with rotational spectra available in the latter public databases, or more than 100 other species with unpublished or uncataloged data (including numerous heterocyclic species; B. A. McGuire 2025, priv. comm.).

Knowing that \ic is C$_{60}$-rich and displays a variety of hydrocarbon species, it may be possible that complex molecules are present in its circumstellar envelope (CSE), but their rotational lines are intrinsically weak. We have tried the line stacking method \citep{Loomis2021} with the frequencies of complex molecules, mainly the PAHs detected in TMC-1, to search for accumulated features above the detection level. This method takes advantage of the noise reduction by stacking the spectrum at the frequencies corresponding to the rotational transitions of a given molecule. With the stacked frequencies showing an emission above the noise level, constraints on the abundance of the molecule can be set. Besides the line stacking of the spectrum of \ic at the frequencies of the PAHs discovered in TMC-1, no feature has been detected over the noise level. Therefore, these tests do not provide any additional information about the molecular content of \ic above the noise level.

\subsection{Comparing the spectrum of \ic with \ngc}
\label{sec:comparison_IC418-NGC7027}

The purpose of observing \ic and \ngc was to compare their different fullerene chemistry to extract information about the molecular content of \ic. The radio spectrum of \ic does not contain features of any detected molecule in space, and even CO is undetected \citep[see][]{Dayal1996}. In the IR, only C$_{60}$ bands have been detected \citep{Otsuka2014}. In contrast, \ngc displays several carbon molecules in the mm region \citep[see e.g.][]{Zhang2008}, and very diverse PAHs in the IR domain \citep[see e.g.,][]{SmithPerez2025}; but no hint of C$_{60}$ or other fullerenes in the IR. Therefore, any spectral feature present only in the spectrum of \ic and not in the spectrum of \ngc could be potentially related with fullerene-based species.

The comparison of both spectra shows that only two UFs of \ic may be also observed in \ngc. These are U39019 (U39013 in \ngc) and U139054. In both sources, these UFs show Gaussian-like profiles. U39019 (Fig.~\ref{fig:ufs_U39019},  with SNR $2.4 \pm 0.8$) is partially blended with a RRL, while U139054 (Fig.~\ref{fig:ufs_U139054}, with SNR $4.2 \pm 1.3$) is isolated. We also explored whether these two UFs could be related by applying the same linear fitting procedure used for the doublets discussed above. However, no consistent pattern was found, and thus no evidence supports a common carrier.

The fact that two UFs of \ic are also detected in \ngc could suggest that the same molecule or molecules producing these rotational lines are present in both C$_{60}$-rich and C$_{60}$-poor circumstellar environments.

\section{Discussion}
\label{sec:discussion}

There are three main types of molecular rotors: linear, symmetric, and asymmetric\footnote{Spherical rotors do not display a permanent electric dipole moment, so they are not considered in this paper.}. Each type can be associated to characteristic patterns of rotational transitions in the spectrum \citep[see e.g.][]{Cooke2013}, which can be even more complex if internal quantum effects such as couplings between molecular rotation and other angular momenta (e.g., orbital momentum or electronic spin) are present. We have searched for these patterns in the set of 20 UFs presented in Table \ref{table:ufs_frequencies}, regardless they are clear detections or not. Identifying any of these line patterns in the observed spectrum can provide key clues to the identity of the emitting molecule.

Within this set, some features appear grouped in pairs of closely spaced lines. These pairs, hereafter referred to as ``doublets'', were identified based on their proximity in frequency and their appearance in the spectra across the observed bands (2, 3, and 7\mm). Ten UFs (five clearly detected and five tentatively detected) are involved in this set of doublets (e.g. U35519 \& U35593 in Table \ref{table:ufs_doublets} comprise UD35556 in Table \ref{table:ufs_frequencies}), while the remaining UFs do not show any clear grouping or systematic pattern (examples of Loomis-Woods diagrams used to search for correlations among UFs can be found in the Appendix \ref{sec:LW_diagrams}). The doublets are plotted in Fig.~\ref{fig:ufs_doublets} and their frequencies are listed in Table \ref{table:ufs_doublets}.

\begin{table*}[h]
	\centering
	\caption{Frequencies of the doublets of UFs observed on \ic.}
	\label{table:ufs_doublets}
	\addtolength{\tabcolsep}{-0.3em}
	\begin{tabular}{l c c c c}
		\hline\hline\noalign{\smallskip}
		Doublet $^a$ & $\nu_\mathrm{low}$ (MHz) & $\nu_\mathrm{high}$ (MHz) & $\nu_c$ (MHz) $^b$ & Splitting (MHz) \\
		\hline\noalign{\smallskip}
		UD35556 & $35519.3 \pm 0.5$ & $35593.9 \pm 0.5$ & $35556.6 \pm 0.5$ & $74.6 \pm 0.5$ \\
		UD42932 & $42876.6 \pm 0.5$ & $42987.9 \pm 0.5$ & $42932.3 \pm 0.5$ & $111.3 \pm 0.5$ \\
		UD85462 & $85342.2 \pm 1.0$ & $85582.1 \pm 1.0$ & $85462.1 \pm 1.0$ & $239.9 \pm 1.0$ \\
		UD87233 & $87129.5 \pm 1.0$ & $87336.7 \pm 1.0$ & $87233.1 \pm 1.0$ & $207.2 \pm 1.0$ \\
		UD138755 & $138455.1 \pm 1.0$ & $139054.9 \pm 1.0$ & $138755.0 \pm 1.0$ & $599.8 \pm 1.0$ \\
		\hline
	\end{tabular}
	\tablefoot{$a$: the letters ``UD'' mean Unidentified Doublet. $b$: central frequency of the doublet.}
\end{table*}

\subsection{Analysis of doublets of UFs}
\subsubsection{Fitting under the linear rotor approximation}
\label{sec:linear_rotor_model}

\begin{figure*}[ht]
	\captionsetup[sub]{skip=0mm, belowskip=0pt}
	\centering
	\begin{subfigure}[]{0.49\linewidth}
		\includegraphics[width=0.95\linewidth]{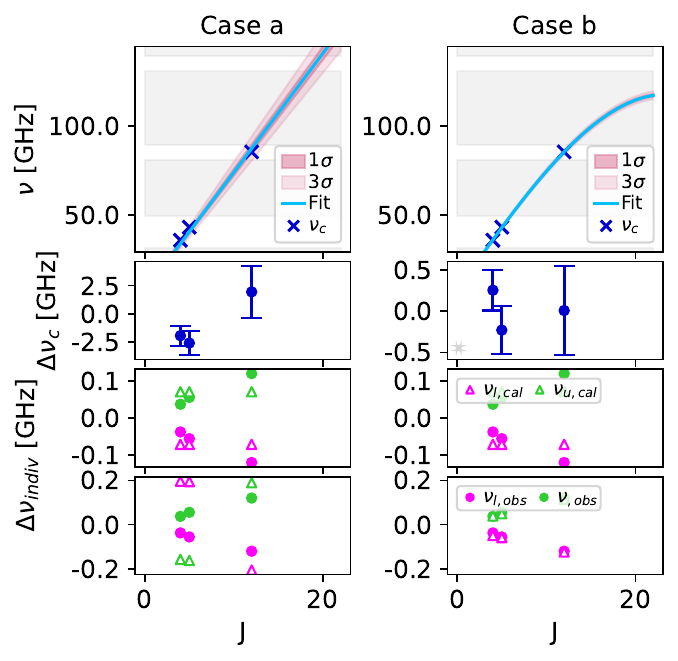}
		\caption{UD35556, UD42932, and UD85462 (Pattern 1), $J_\mathrm{init} = 4$} \label{fig:fit_UD35UD42UD85_J0-4}
	\end{subfigure}
	\begin{subfigure}[]{0.49\linewidth}
		\includegraphics[width=0.95\linewidth]{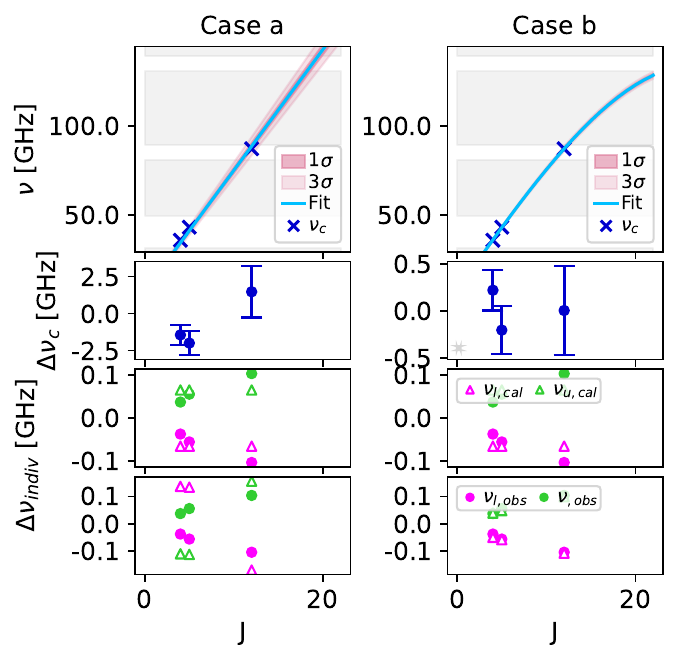}
		\caption{UD35556, UD42932, and UD87233 (Pattern 2), $J_\mathrm{init} = 4$} \label{fig:fit_UD35UD42UD87_J0-4}
	\end{subfigure}
	\begin{subfigure}[]{0.49\linewidth}
		\includegraphics[width=0.95\linewidth]{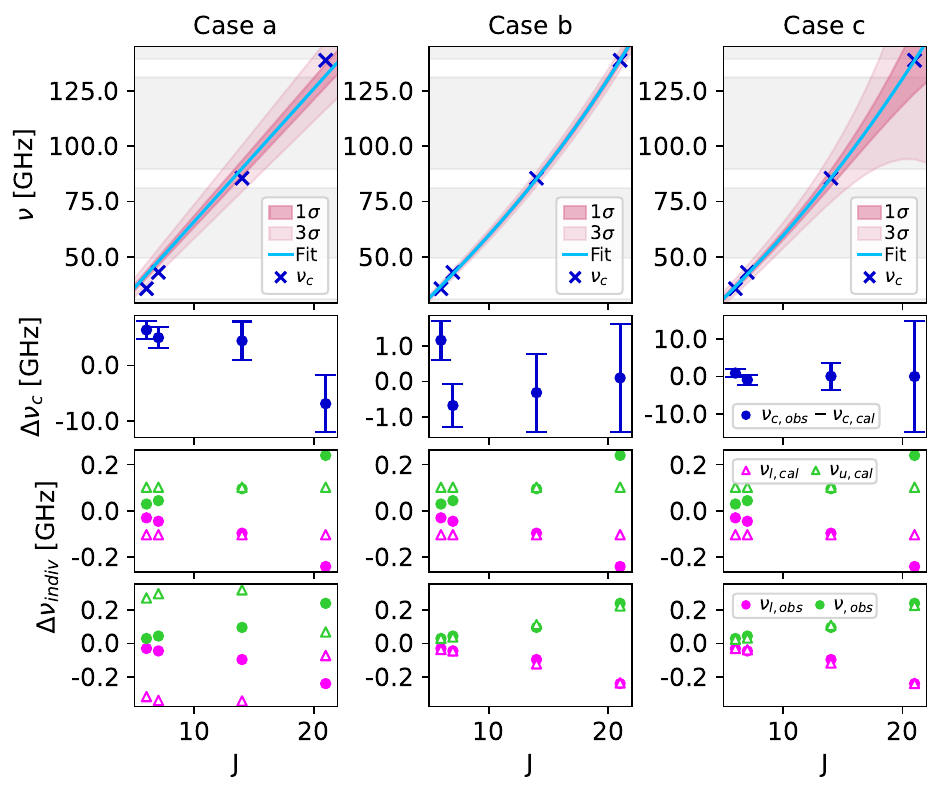}
		\caption{UD35556, UD42932, UD85462, and UD138755 (Pattern 3), $J_\mathrm{init} = 6$} \label{fig:fit_UD35UD42UD85UD138_J0-6}
	\end{subfigure}
	\begin{subfigure}[]{0.49\linewidth}
		\includegraphics[width=0.95\linewidth]{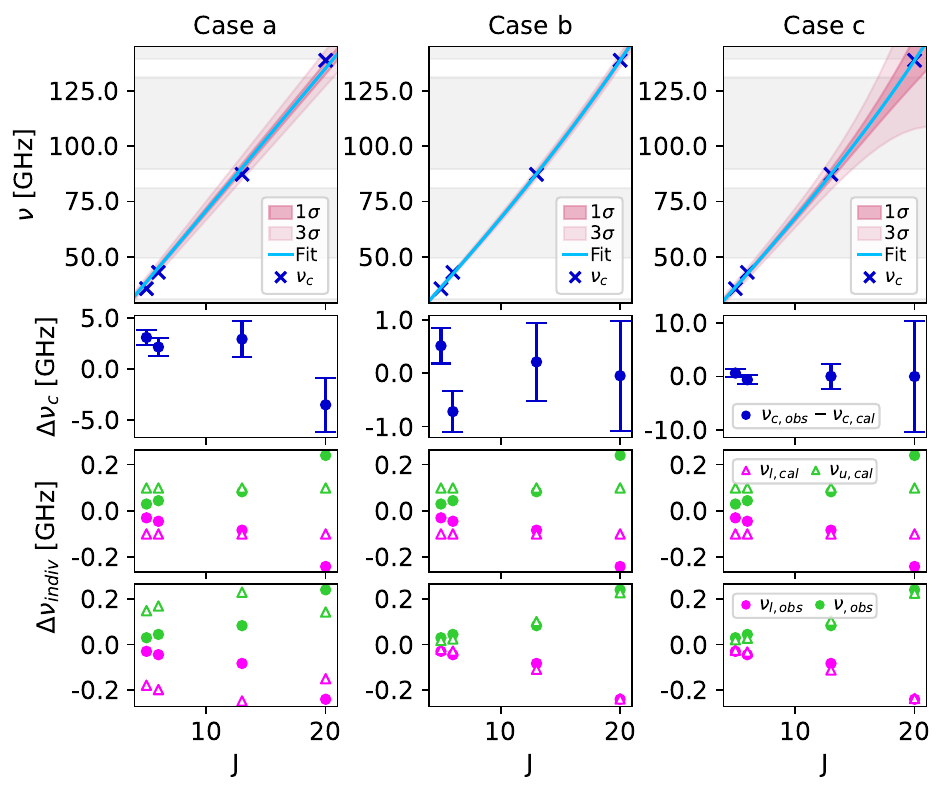}
		\caption{UD35556, UD42932, UD87233, and UD138755 (Pattern 4), $J_\mathrm{init} = 5$} \label{fig:fit_UD35UD42UD87UD138_J0-5}
	\end{subfigure}
	\caption{Fits of four different combinations of doublets. For each dataset, three cases have been considered: \textit{a}, \textit{b}, and \textit{c}. Each case is described using different panels. The top panel represents the overall fit results. Blue crosses show the central frequency of the observational pairs. The middle upper panel plots with blue dots the difference between the observed central frequency and the predicted value. Middle bottom and bottom panels represent the modeling of the doublet splitting using Eqs.~\ref{eq:freq_RotTrans_G} (middle bottom) and \ref{eq:freq_RotTrans_JG} (bottom). The green and pink markers show the difference between individual lines and the central frequency of the doublet. The light gray areas in all plots indicate the unobserved ranges.}
	\label{fig:doublets_freq_fit}
\end{figure*}

The central frequencies of the doublets observed in the spectrum (Table \ref{table:ufs_doublets}) seem to follow a linear dependence with a rotational quantum number $J$. We have fitted them assuming a linear molecule with doublets using the equation
\begin{equation} \label{eq:freq_RotTrans}
	\nu = 2B(J+1) -4D(J+1)^3 + H(J+1)^3[(J+2)^3 - J^3] \,,
\end{equation}
where $B$ is the rotational constant, $D$, and $H$ the centrifugal distortion constants of the molecule, and $J$ the rotational quantum number of the lower level of each rotational transition. Higher order rotational constants have been neglected by default. In order to inspect the effect of each constant to the fit, we have considered three cases: case \textit{a} has no distortions (just the term with $B$), case \textit{b} considers quartic centrifugal distortions (terms with $B$ and $D$), and case \textit{c} adds sextic distortions (terms with $B$, $D$, and $H$). To assign the $J$ of each doublet, we assumed that UD35556 and UD42932 are consecutive. Their frequency separation is relatively large, which suggests a pattern more consistent with species exhibiting sparse and regular spectra, as opposed to the dense line structure typically found in larger molecules. No additional lines are detected between these features. While the nature of the carrier remains unknown, this interpretation provides a reasonable working hypothesis in the absence of further constraints. They allowed us to derive a first estimate of the B constant and their rotational numbers following Eq.~\ref{eq:freq_RotTrans}. However, since we are fitting doublets, the position of the individual components with respect to the central frequency is in first approximation
\begin{equation} \label{eq:freq_RotTrans_G}
	\nu = 2B(J+1) -4D(J+1)^3 + H(J+1)^3[(J+2)^3 - J^3] \pm \gamma \,,
\end{equation}
where $\gamma$ is the splitting constant that applies to the lower frequency component of the doublet ($-\gamma$) and to the higher frequency component of the doublet ($+\gamma$).

We have fitted four different combinations of three or four doublets or patterns (labeled from 1 to 4) using Eq.~\ref{eq:freq_RotTrans_G}. The most accurate fits for each pattern are plotted in Fig.~\ref{fig:doublets_freq_fit} (see all fits in the Appendix \ref{appen:fit_plots_doublets} and Tables \ref{tab:pattern1_params}, \ref{tab:pattern2_params}, \ref{tab:pattern3_params}, and \ref{tab:pattern4_params}). Comparing the fits with three doublets (see Figs.~\ref{fig:fit_UD35UD42UD85_J0-4} and \ref{fig:fit_UD35UD42UD87_J0-4}), a larger dispersion with respect to the observational frequencies is found in case \textit{a} than in case \textit{b}. The statistical uncertainties are also reduced when $D$ is included (case \textit{b}). Focusing on the fits with four doublets (see Figs.~\ref{fig:fit_UD35UD42UD85UD138_J0-6} and \ref{fig:fit_UD35UD42UD87UD138_J0-5}) we observe that case \textit{a} provides the highest value of constant $B$, while the other two cases display quite similar values within the errors. The dispersion of the predicted central frequencies is higher for case \textit{a}. The difference between observed and predicted central frequencies is smaller for cases \textit{b} and \textit{c} and the smallest differences are found in case \textit{b}. This means that including $H$ does not improve the results. In fact, case \textit{c} results show a slightly higher dispersion and larger uncertainties. Therefore, the case \textit{b} provides smaller differences between the central frequencies and more accurate values of constants $B$ and $D$. Despite the large errors that are found in all cases, pattern 1 and 2 provide less accurate frequency predictions and larger $B$ and $D$ constants than patterns 3 and 4.

The bottom panels in the plots clearly show that the doublet splitting increases with frequency. However, the predictions from the fits with Eq.~\ref{eq:freq_RotTrans_G} remain constant. This disagreement may point out that the splitting depends on the $J$ value and the parameter $\gamma$ in Eq.~\ref{eq:freq_RotTrans_G} should be modified as
\begin{align} \label{eq:freq_RotTrans_JG}
	\nu & = 2B(J+1) -4D(J+1)^3 + H(J+1)^3[(J+2)^3 - J^3] \notag \\
	& \quad \pm [\gamma + \gamma_J (J + 1)] F(J) \,,
\end{align}
where $F(J)$ is a function that encodes the value for the higher frequency line ($J/2$) and the lower frequency line ($-(J  + 1)/2$), and $\gamma_J$ is a second order constant related to the doublet splitting. The inclusion of the last term in Eq.~\ref{eq:freq_RotTrans_JG} clearly improves the description of the higher and lower frequency components of the doublets. Besides this improvement, the statistical uncertainties are not sufficiently small to clearly discard any pattern to be real.

However, we could constrain the number of atoms of the possible carrier, after a comparison with public data. The results plotted in Fig.~\ref{fig:doublets_freq_fit} display the best fits for each different combination of doublets and, in every case, the fit that does not consider the $H$ constant provides the smallest errors. The fits using three doublets provide $B = 3\,632 \pm 27$\mhz (using UD85462) and  $B = 3\,617 \pm 24$\mhz (using UD87233), which set a range of $B = 3\,590 - 3\,660$\mhz (see Figs.~\ref{fig:fit_UD35UD42UD85_J0-4} and \ref{fig:fit_UD35UD42UD87_J0-4}). These values are compatible with molecules with $6 - 12$ atoms that have been already detected in space. The fits with four doublets give $B = 2\,562 \pm 41$\mhz (using UD85462) and  $B = 2\,979 \pm 29$\mhz (using UD87233), so a  range of $B =2\,520 - 3\,010$\mhz (see Figs.~\ref{fig:fit_UD35UD42UD85UD138_J0-6} and \ref{fig:fit_UD35UD42UD87UD138_J0-5}) can be considered. In this case, by considering a total range $B =2\,520 - 3\,660$\mhz, species of $4 - 13$ atoms could be considered as possible carriers of the observed doublets.

\subsubsection{Fitting under the asymmetric rotor approximation}

The fits described in Sec.~\ref{sec:linear_rotor_model} using a linear rotor approximation were found to be unsatisfactory, which was further confirmed using the standard SPFIT/SPCAT \citep{Pickett1991} spectroscopic software package and the JB95\footnote{\href{https://www.nist.gov/services-resources/software/jb95-spectral-fitting-program}{https://www.nist.gov/services-resources/software/jb95-spectral-fitting-program}} graphical interface program to visualize the spectrum and the SPFIT/SPCAT results. To explore the nature of the potential carriers, we expanded our analysis to include various coupling terms, such as Coriolis coupling, spin-rotation coupling, and nuclear quadrupole coupling constants. However, these additional parameters did not improve the convergence of the fits. The errors on the constants and the predicted frequencies are not comparable with the precision of our data. The series of doublets could actually be an apparent doublet-like pattern, in which the components of the doublets are actually lines with different K (K$_\mathrm{a}$, K$_\mathrm{c}$) subbands of a symmetric (asymmetric) rotor. Yet, the line splitting for symmetric rotors depending on K is expected to be exceedingly small (i.e., of the order of the centrifugal distortion constants) and an asymmetric rotor is more likely responsible for the detected apparent doublets.

We made use of the standard SPFIT/SPCAT \citep{Pickett1991} spectroscopic software package to analyze and predict our sets of doublets using the asymmetric rotor approximation. A distinct characteristic of them is that their internal frequency splitting is not constant; rather, it systematically increases with growing frequency. This behavior is potentially suggestive of a pair of \textit{b-type} transitions ($K_a = 2 \leftarrow1$) from an asymmetric rigid rotor \citep[see e.g.,][]{Cooke2013}, or splitting arising from large-amplitude motion \citep[LAM; see e.g.][]{Nguyen2020} between double equivalent potential energy surface minima. Despite comprehensive attempts to model this doublet progression, including efforts to account for such splitting mechanisms, a definitive assignment could not be achieved. The derived rotational constants were therefore useless to predict additional transitions consistent with the observed data, nor to extend the series beyond the detected features.

A systematic search revealed an additional promising progression of single UFs, distinct from the doublet-like pattern discussed above, in the $30-50$\ghz range, characterized by a quasi-regular frequency separation of approximately 4300\mhz. Standard Hamiltonian models for linear, symmetric, and asymmetric rotors were rigorously tested using SPFIT/SPCAT. Numerous combinations of trial rotational constants and quantum number assignments ($J$, $K_a$, $K_c$) were explored for this 4300\mhz additional pattern. Unfortunately, these efforts proved unsuccessful and no tested combination converged to a physically plausible solution or produced a fit with acceptable residuals (root-mean-square error). Consequently, both the $\sim4300$\mhz progression and the doublet-like pattern remain unassigned, highlighting the significant challenge posed by the unidentified spectral line inventory in this source.

\subsection{Can UFs be produced by fullerene derivatives?}

The C$_{60}$ IR detection in \ic \citep{Otsuka2014} suggests that C$_{60}$-based species could be present in this PN and potentially detectable in the radio domain. The C$_{60}$ molecule does not display a permanent dipole moment and, therefore, it does not produce a rotational spectrum to be observed at radio frequencies. However,  some C$_{60}$ derivatives, such as C$_{60}$H$_2$, C$_{60}$H$_4$, or C$_{60}$H$_{18}$, do \citep[see][who report values in the $0.57 - 8.35\,$D range]{Sabirov2022}. Also, some exohedral metallofullerenes, like C$_{60}$Mg,  C$_{60}$Na,  C$_{60}$K,  C$_{60}$Fe$^+$, display very high dipole moment values (4.9\,D, 11.8\,D, 15.8\,D, and 11.8\,D, respectively; R. Barzaga 2025, \textit{priv. comm.}). These species are thus potentially detectable at the frequencies covered by our observations, provided that their abundances are sufficiently high and the rotational temperature sufficiently low. Given the interest in detecting large complex molecules in space, which may provide key clues to unveil the formation pathways of fullerenes in astronomical environments, we discuss below whether our UFs could be produced by fullerene related species.

Theoretical simulations of pure rotational spectra using the rigid rotor approximation of some C$_{60}$ derivatives such as C$_{60}$NH$_2$ and C$_{60}$CN (R. Barzaga 2023, priv. comm.), C$_{60}$CN$^\pm$, C$_{60}$Fe, and C$_{60}$Fe$^+$ (GL. Hou et al. 2023, priv. comm.), or C$_{70}^+$ \citep[][and references therein]{Nemes2024}, give rotational constants of $A,\,B,\,C < 100$\mhz. This implies that the transition patterns, within the same branch, are separated by $< 200$\mhz, something that it is not observed in our \ic radio spectra. In addition, the doublets detected in our spectrum are compatible with values of $B\sim 2520-3660$\mhz (see Sec.~\ref{sec:linear_rotor_model}), which implies distances between consecutive features a factor of $\sim 20$ higher than for those of fullerene derivatives. The large rotational line separations observed are thus inconsistent with theoretical simulations of C$_{60}$-based species, and we conclude that very likely fullerene-derivatives are not the carriers of the UFs.

In contrast, the planar C$_{24}$ molecule has been tentatively detected in the mid-IR spectra of some PNe containing fullerenes \citep{GarciaHernandez2011b, GarciaHernandez2012} and its closed fullerene isomer could be the smallest fullerene possibly present in space \citep[see e.g.,][]{Bernstein2017}. The planar and cage isomers of C$_{24}$ have not a permanent dipole moment. In contrast, some other C$_{24}$ isomers, such as bowl and ring structures (neutrals, cations, and anions), have a significant dipole moment \citep[in the $0.25 - 12.63\,$D range; see][]{Karri2023} and could be possible carriers of the UFs. Their rotational constants are not known to date. They are potential molecular targets to be studied theoretically and in the laboratory for their rotational spectra.

In short, the spectral behavior of fullerene derivatives is inconsistent with the properties of the observed UFs, indicating that, even if present, they cannot account for the detected radio features, given our detection limit level. The lack of radio detections of rotational transitions attributable to fullerene derivatives may be related to several reasons both intrinsic and observational effects. On the one hand, the abundance of these species may be too low to produce detectable emission. On the other hand, the sensitivity of current \mbox{IRAM 30m} and RT40m observations may be still insufficient to detect such weak features. In addition, their optical depths are expected to be small, leading to intrinsically faint spectral lines. The partition function is also a critical factor. The large number of atoms in C$_{60}$ derivatives implies that their partition function is expected to be quite large at the temperatures prevailing in the molecular regions where fullerenes survive \citep[$T_\mathrm{exc} \sim 300$\kel on average; see Table 3 in][]{GarciaHernandez2012}. The rotational level significantly populated should have a rotational energy of the order of $2 - 3 T_\mathrm{rot}$. Assuming a rotational constant $B = 100$\mhz for C$_{60}$ and its derivatives and taking $T_\mathrm{rot} \sim 50$\kel, the populated rotational level with the highest energy value\footnote{We have used $E_\mathrm{rot} = 2 T_\mathrm{rot}$.} should have $J \sim 150$. Increasing the value of $T_\mathrm{rot}$ would increase the value of $J$. Moreover, at these temperatures the number of vibrational modes is, in general $3N - 6$, in particular, 174 for the C$_{60}$. Populating several tens of levels is experimentally observed as very weak emission or absorption spectral lines. Our conclusion is based on the results of the simulations of fullerene derivatives rotational spectra. We note that, to our best knowledge, there are no laboratory microwave spectra of such fullerene species available in the literature.

Finally, after analyzing the set of doublets (see Sec.~\ref{sec:linear_rotor_model}), we find that any combination of them cannot be explained by currently known molecules in space or C$_{60}$ derivatives. According to the different fitting results and the values of the rotational constants obtained ( Sec.~\ref{sec:linear_rotor_model}), the set of doublets is consistent with molecular species displaying $B\sim 2\,500-3\,660$\mhz. Rotational constants in this range are characteristic of molecules much smaller than fullerenes. By comparing this range of values with the data of molecular rotational constants available in the CDMS and JPL databases, the most plausible carriers are molecules containing approximately $4-13$ atoms, this is, clearly distinct from fullerene‐based species and fully consistent with the frequency separations observed for the UFs.

\subsection{Possible carrier species depending on the fullerene formation process}

The formation mechanism of fullerenes in space is a fundamental open question in astrochemistry.  Two main mechanisms (bottom-up vs top-down) have been proposed to explain their presence in space. The bottom-up schema involves small C-based species, like small hydrocarbons or small clusters, that progressively grow into larger structures. Fullerenes would form through sequential reactions building larger carbonaceous species that would eventually close into stable cage structures \citep{Kroto1985, Heath1992}. This mechanism is efficient in high temperature and high-density environments. However, in the low-density interstellar and circumstellar media, the growth reactions would be extremely slow and the formation of fullerenes is expected to be very inefficient \citep{Micelotta2012}. The top-down scenario proposes the photochemical processing or destruction of larger C-based species to finally form fullerenes. Two main options are considered that differ in the initial carbonaceous compound: PAHs \citep{Berne2012} or hydrogenated amorphous carbon (HACs) \citep{GarciaHernandez2010, GarciaHernandez2011a}.

\cite{Berne2012} proposed that fullerenes in the ISM form from the photo-fragmentation of large PAHs. The strong UV radiation from the star would produce the complete dehydrogenation of the PAHs, turning them into planar graphene sheets, which then would loose C$_2$ units from the border. These border defects in the structure may force the formation of pentagons that curve the planar structure until it definitely closes and forms cages like C$_{60}$ and C$_{70}$. Some experimental results support the photochemical processing of large PAH cations ($\geq 60$ C-atoms) as a possible top-down route towards C$_{60}$, among other species \citep{Zhen2014}. Some byproducts of these photochemical processing, like small graphene flakes, rings, and small cages, could be detectable in the mm range/through rotational emission if they: (i) display a permanent dipole moment; (ii) their abundance is high enough, and (iii) they are highly stable against UV radiation. \cite{Berne2015} point out that the formation of C$_{60}$ from PAHs is efficient only for species with $60 - 66$ C atoms and the process is extremely fast. Therefore, most intermediate transition species are not expected to reach sufficiently high abundances.

However, \cite{Micelotta2012} demonstrate that the large PAHs scenario for fullerene formation proposed by \cite{Berne2012} requires a very specific tuning of the dissociation parameters, being very unlikely to happen in space; at least under the physical conditions of young PNe with fullerenes like PN \ic. Fullerenes could form in young PNe due to the destruction/processing (e.g., by UV radiation) of dust grains composed by a mixture of aromatic and aliphatic species, such as HACs, coal, etc. \citep{GarciaHernandez2010, GarciaHernandez2011a} \citep[see also][and references therein]{GomezMunoz2024}. Fullerenes are detected in conjunction with HAC-like grains in young PNe \citep[e.g.][]{GarciaHernandez2011a, GarciaHernandez2012, BernardSalas2012} and laboratory experiments found fullerenes among the dissociation products of UV-irradiated HACs \citep{Scott1997}; i.e. small alkanes, unsaturated carbon-chain radicals, and small dehydrogenated PAH-like species are released first followed by proto-graphitic, aromatic clusters with $\simeq 40$ C atoms similar to the locally aromatic polycyclic hydrocarbons (LAPHs) proposed by \cite{Petrie2003}. Indeed, \cite{Micelotta2012} argue how the UV-induced dehydrogenation of HAC grains naturally provides the conditions for fullerene formation in young PNe; both H removal and pentagonal ring formation, which force the evolving structure to curl up. They suggest that the HACs dehydrogenation would produce LAPH-like species, which they call ``arophatic'' structures, mainly characterized by aliphatic bridging groups linking the aromatic clusters. The ``arophatic'' molecules proposed by \cite{Micelotta2012} and their expected fragments are curled-up structures that are expected to exhibit significant dipole moments, potentially enhanced by the incorporation of atoms such as N, O, or S. We note that up to date no molecules containing these elements are detected in our source. Such complex structures (neutral/anion/cation forms), if being stable and abundant enough, may be present in C$_{60}$-rich PNe and are strong potential candidate carriers of the UFs detected in the mm spectrum of \ic. Apart from that, very recently \cite{Jones2025} shows that the end products of the UV-induced processing of HACs may be non-planar and highly conjugated cyclic species, such as 1,3- and 1,4-cyclohexadiene and cyclooctatetraene. Such non-planar species are highly stable and emit the strong $3.3\,\mu$m emission seen in the ISM and evolved PNe (i.e. usually hotter than \ic\footnote{Note that \ic is one of the C$_{60}$-rich PNe also showing PAH-like emission at 3.3, 6.2, 7.7, and 11.3$\,\mu$m \citep[see e.g,][]{Otsuka2014}.}, where even fullerenes have been already destroyed) and largely attributed to planar PAHs\footnote{During the evolution from the AGB to the PN phase, there is a shift from aliphatic IR features (e.g., at 3.4, 6.9, 7.3$\,\mu$m) in proto-PNe to classic PAH bands (at 3.3, 6.2, 7.7$\,\mu$m) in evolved PNe, possibly reflecting a progressive dehydrogenation (e.g. UV-induced ) of initially aliphatic-rich (e.g. HAC-like) carbonaceous dust \citep[e.g.,][]{Sloan1997, Kwok2011}.}. Unfortunately, to our best knowledge, there is no laboratory and/or theoretical rotational microwave spectra of such non-planar species, both arophatic and highly conjugated molecules, available in the literature.

Interestingly, the molecular size of the non-planar highly conjugated cyclic species recently suggested to be products of the HACs processing \citep{Jones2025} agree quite well with the possible molecule/s ($\sim 4-13$ atoms) producing the doublet UFs detected in \ic. Highly conjugated species in such range like 1,3-cyclohexadiene (C$_6$H$_8$), 1,4-cyclohexadiene (C$_6$H$_8$), and cyclooctatetraene (C$_8$H$_8$), among possibly others, are thus strong potential candidate UF carriers to search for in our \ic radio spectra. The cyclooctatetraene does not have permanent dipole moment, but the 1,3-cyclohexadiene and the 1,4-cyclohexadiene have permanent dipole moments of $|\mu| = 0.437$\,D \citep{Butcher1965} and $|\mu| \simeq 0.3$\,D \citep{Scharpen1968}, respectively. The latter species could be thus identified through their pure rotational spectrum if their abundances toward PN \ic are high enough. For example, indene, with a dipole moment of $\sim 0.6$\,D was detected toward \mbox{TMC-1} \citep{Cernicharo2021f} because of its unexpected large abundance. Moreover, this kind of species could react with other atoms or small radicals, forming new molecules with significantly higher dipole moments; e.g., 1,4-cyclohexanedione (1,4-C$_6$H$_8$O$_2$) displays a permanent dipole moment of $\sim 1.4$\,D \citep{Rogers1961}.

We note that from the processing/destruction of HACs, other carbonaceous species could be produced, like small chain radicals, alkanes, and others. More recent experiments simulating the processing/destruction of HAC grains by UV radiation (or cosmic rays) at ISM conditions generally find that small hydrocarbon species (e.g., C$_x$H$_y$) are efficiently released into the gas phase \citep[see][]{Alata2015, Dartois2017}. Thus, the UV irradiation of HACs around PNe is expected to produce a wide variety of small carbonaceous molecules that could eventually grow up and form PAHs. If the newly formed PAHs are exposed to the UV radiation of the central star, they can get ionized and undergo radical isomerization prior to dehydrogenation \citep{Patch2025}, efficiently forming PAH$^+$ isomers containing pentagons. This process would produce a set of highly symmetric and compact PAH$^+$ isomers with significant dipole moment and potentially detectable in the radio domain by their pure rotational spectra\footnote{Note that their neutral counterparts may also easily form via neutralization from collisions with an electron \citep{Patch2025}, being detectable at radio wavelengths. For example, azulene, a pentagon-containing isomer of naphthalene, has a significant permanent dipole moment of $\sim 0.8$\,D \citep{Huber2005}.}. If the lifetime and abundance of these PAH$^+$ isomers are sufficiently large, then they are also potential candidate carriers of the UFs detected in our \ic radio spectra.

In summary, current understanding indicate that fullerene formation in young C$_{60}$-rich PNe is most likely driven by the processing/destruction of HAC-like grains \citep{GarciaHernandez2010, GarciaHernandez2011a} than by the fragmentation of large PAHs \citep{Berne2012}. For example, the UV-induced dehydrogenation of HACs would generate curved, defect-rich carbonaceous structures (e.g., the so-called arophatic species) that provide chemically realistic precursors to closed-cage fullerenes and represent a diverse population of non-planar, dipole-bearing molecules displaying a pure rotational spectrum. Such UV processing could be also expected to act on newly formed PAHs, which could undergo radical isomerization, producing pentagon-containing PAH isomers with significant dipole moments. All these species, previously undetected in the radio range, represent promising candidate carriers for the UFs observed in the mm spectrum of \ic. In particular, if the doublet UFs detected in \ic pertain to the same molecule, we suggest small ($\sim 4 - 13$ atoms) non-planar highly conjugated cyclic molecules, like those recently suggested by \cite{Jones2025}, to be strong candidate carriers. Laboratory measurements and theoretical simulations of their rotational spectra will be essential to confirm or discard their viability as carriers and to advance in our understanding of the chemical pathways leading from complex carbonaceous solids to fullerenes in astrophysical environments.

\section{Conclusions}
\label{sec:conclusions}

We report the presence of molecular emission in the high-sensitivity radio spectrum of the fullerene-containing PN \ic based on the detection of 20 weak (at SNR $\sim 2-13$) unidentified molecular pure rotational lines at 2, 3, and 7\mm. These unidentified features (UFs) cannot be attributed neither to the numerous radio recombination lines observed in this source \citep{HuertasRoldan2025}, not to instrumental artifacts, radio frequency interferences, or any known molecular species listed in the literature and public spectroscopic databases.

A search for spectral patterns was performed to estimate the rotational constants of the possible molecular carriers. Apart from a recurrent doublet-like structure, no clear or systematic pattern was identified. The detected UFs are unlikely to arise from linear molecules, as the characteristic regular line spacing of $2B$ is not observed. However, if real, this may indicate that an internal mechanism in the linear molecule is producing them. The same conclusion applies to symmetric and asymmetric rotors, which would exhibit similarly regular spacing in each $K$-dependent branch. The rotational constant $B$ for the apparent doublet pattern using different combinations of the observed doublets has been estimated using the linear rotor approximation ($B \sim 2\,500-3\,660$\mhz) and comparison with published data, could likely correspond to a molecule containing approximately $4-13$ atoms. Given the difficulty in identifying clear patterns, a machine learning-based analysis could be explored in the future to systematically search for connections among the detected UFs.

Fullerene derivatives were considered as potential UF carriers due to the IR detection of C$_{60}$ in \ic, but we conclude that most likely fullerene-derivatives are not the carriers of the UFs. Therefore, we have discussed possible carrier species depending on the fullerene formation process. In this context, diverse non-planar molecular species not previously detected with radio astronomy, and expected to be produced from the processing/destruction of HAC-like grains, are suggested as promising candidates for the carriers of these UFs. However, substantial theoretical and laboratory efforts are still required to obtain reliable rotational spectra for comparison with observations. Such data would not only help explain some of the UFs detected in \ic, but also provide valuable constraints on the chemical pathways leading to fullerene formation in PNe.

Here, the catalog of UFs detected towards the fullerene-rich environment around \ic is made publicly available to the radio astronomical community, for comparisons with future laboratory and theoretical data that could eventually lead to the detection of new molecules in space.


\begin{acknowledgements}

The authors acknowledge Miguel Santander-García from the OAN for carrying out the observations at the Yebes RT40m and the IRAM staff for help provided during the observations. The authors acknowledge B. A. McGuire for comparing our data with their private database and GL. Hou and his group for sharing the synthetic rotational spectra of some fullerene-based species. THR, DAGH, AMT, and RB acknowledge the support from the State Research Agency (AEI) of the Ministry of Science, Innovation and Universities (MICIU) of the Government of Spain, and the European Regional Development Fund (ERDF), under grant PID2023-147325NB-I00/AEI/10.13039/501100011033. THR acknowledges support from grant PID2020-115758GB-I00/PRE2021-100042 financed by MCIN/AEI/10.13039/501100011033 and the European Social Fund Plus (ESF+). JPF acknowledges support from grants PID2023-147545NB-I00 and PID2023-146056NB-C22. S. Mato has been funded by the call for UVa 2023 predoctoral contracts, cofunded by Banco Santander and by the predoctoral contract of Junta de Castilla y León 2023, cofunded by European Social Fund (FSE+). JA, JJDL, and VB acknowledge support from the project CRISPNESS, grant  PID2023-146056NB-C21, funded by MICIU/AEI/10.13039/501100011033 and by ERDF/EU. MAGM acknowledge to be funded by the European Union (ERC, CET-3PO, 101042610). Views and opinions expressed are however those of the author(s) only and do not necessarily reflect those of the European Union or the European Research Council Executive Agency. Neither the European Union nor the granting authority can be held responsible for them. MAGM also acknowledge financial support from grant CEX2024-001451-M funded by MICIU/AEI/10.13039/501100011033. This work is based on observations carried out under project number 158-21 with the IRAM 30m telescope. IRAM is supported by INSU/CNRS (France), MPG (Germany) and IGN (Spain). Based on observations carried out with the Yebes 40 m telescope (22A011). The 40 m radio telescope at Yebes Observatory is operated by the Spanish Geographic Institute (IGN; Ministerio de Transportes y Movilidad Sostenible). This publication is based upon work from COST Action CA21126 - Carbon molecular nanostructures in space (NanoSpace), supported by COST (European Cooperation in Science and Technology).

\end{acknowledgements}

\typeout{}
\bibliographystyle{aa} 
\bibliography{PaperUFs} 

\begin{appendix}


\section{Plots of the doublets}
\label{appen:spec_plots_doublets}

\begin{figure*}[htbp]
	\captionsetup[sub]{skip=0mm, belowskip=0pt}
	\centering
	\begin{subfigure}[]{0.49\linewidth}
		\includegraphics[width=\linewidth]{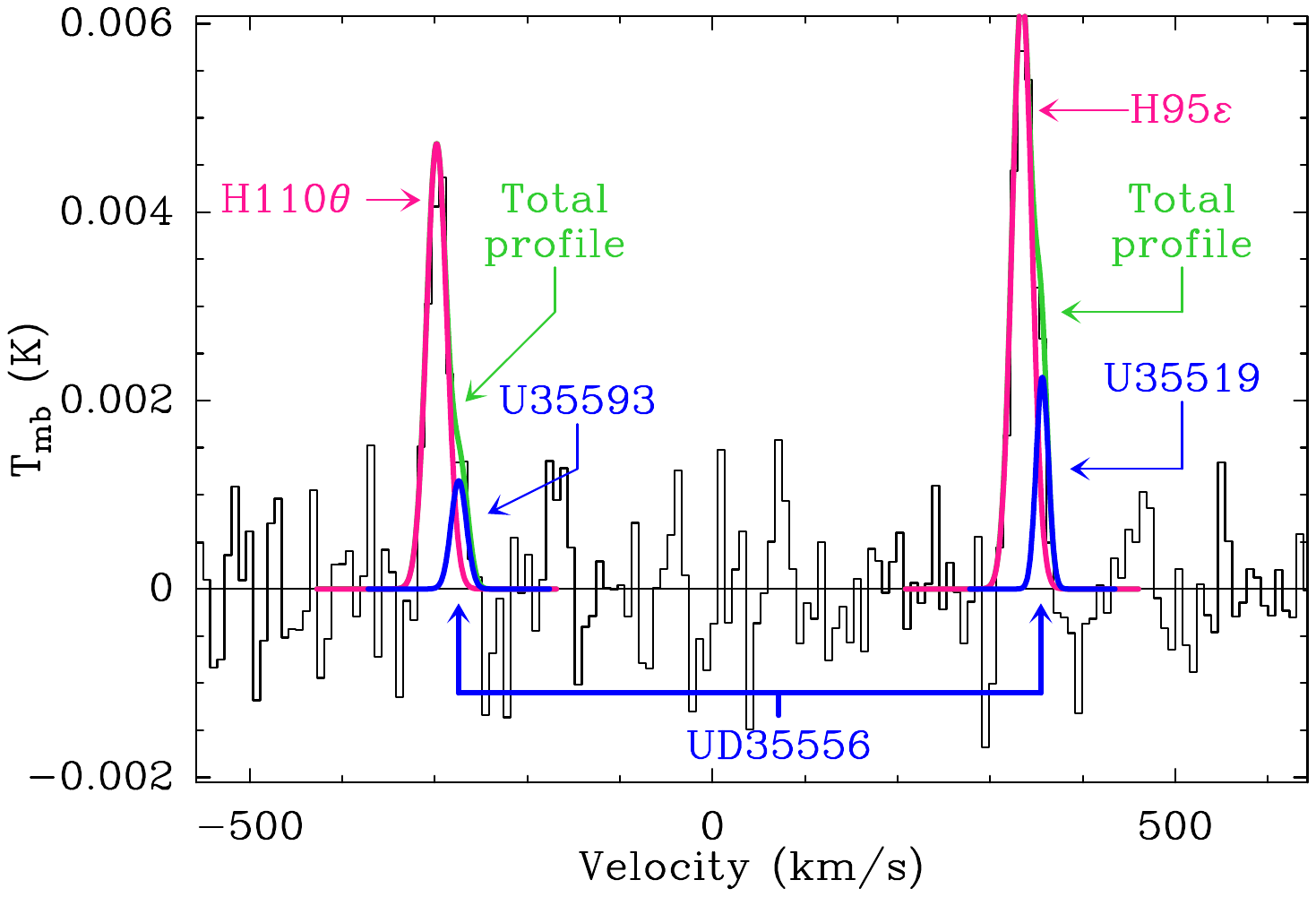}
		\caption{}\label{fig:ufs_UD35556}
	\end{subfigure}
	\hspace{-6mm}
	\begin{subfigure}[]{0.49\linewidth}
		\includegraphics[width=\linewidth]{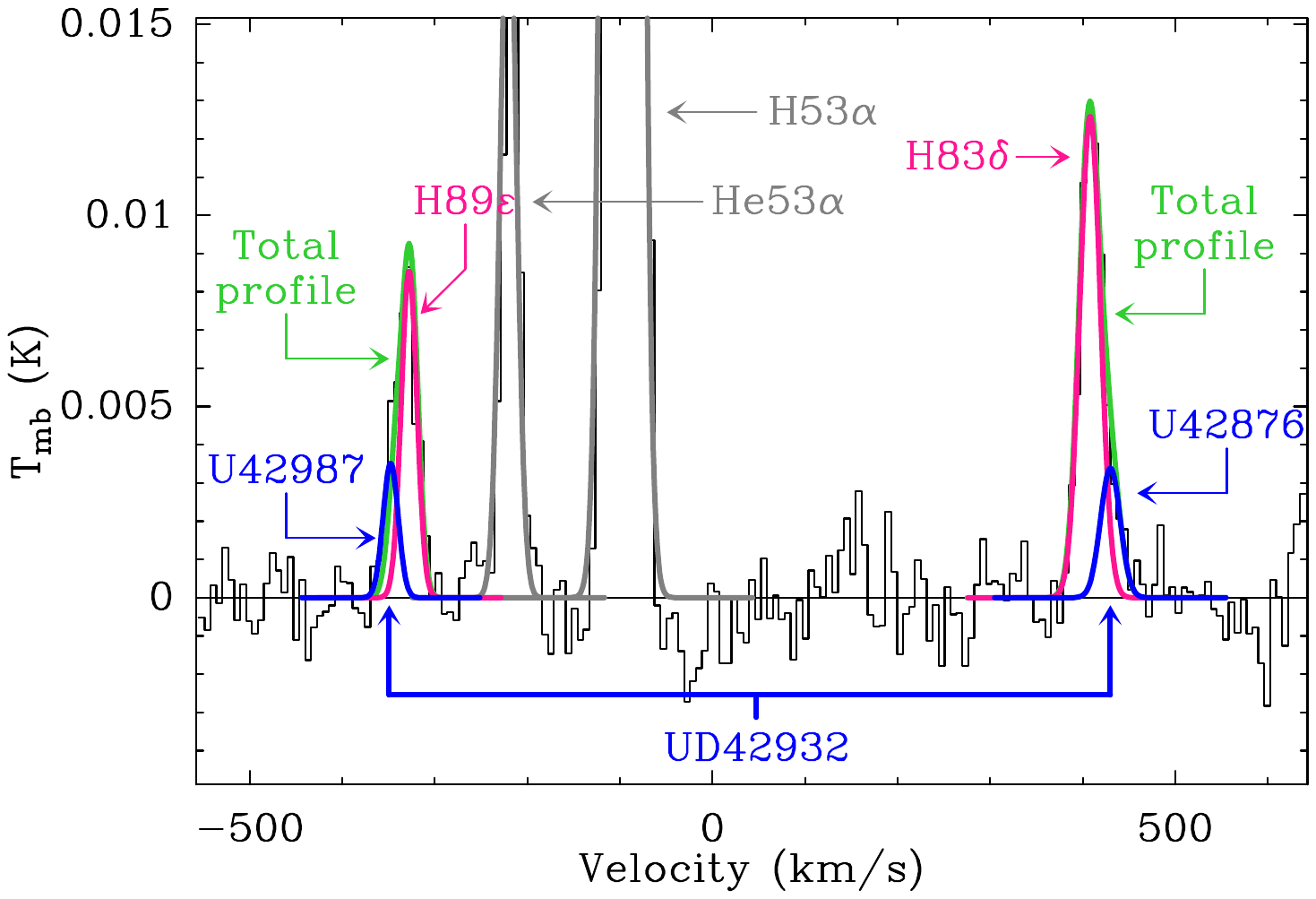}
		\caption{}\label{fig:ufs_UD42932}
	\end{subfigure}
	\hspace{-6mm}
	\begin{subfigure}[]{0.49\linewidth}
		\includegraphics[width=\linewidth]{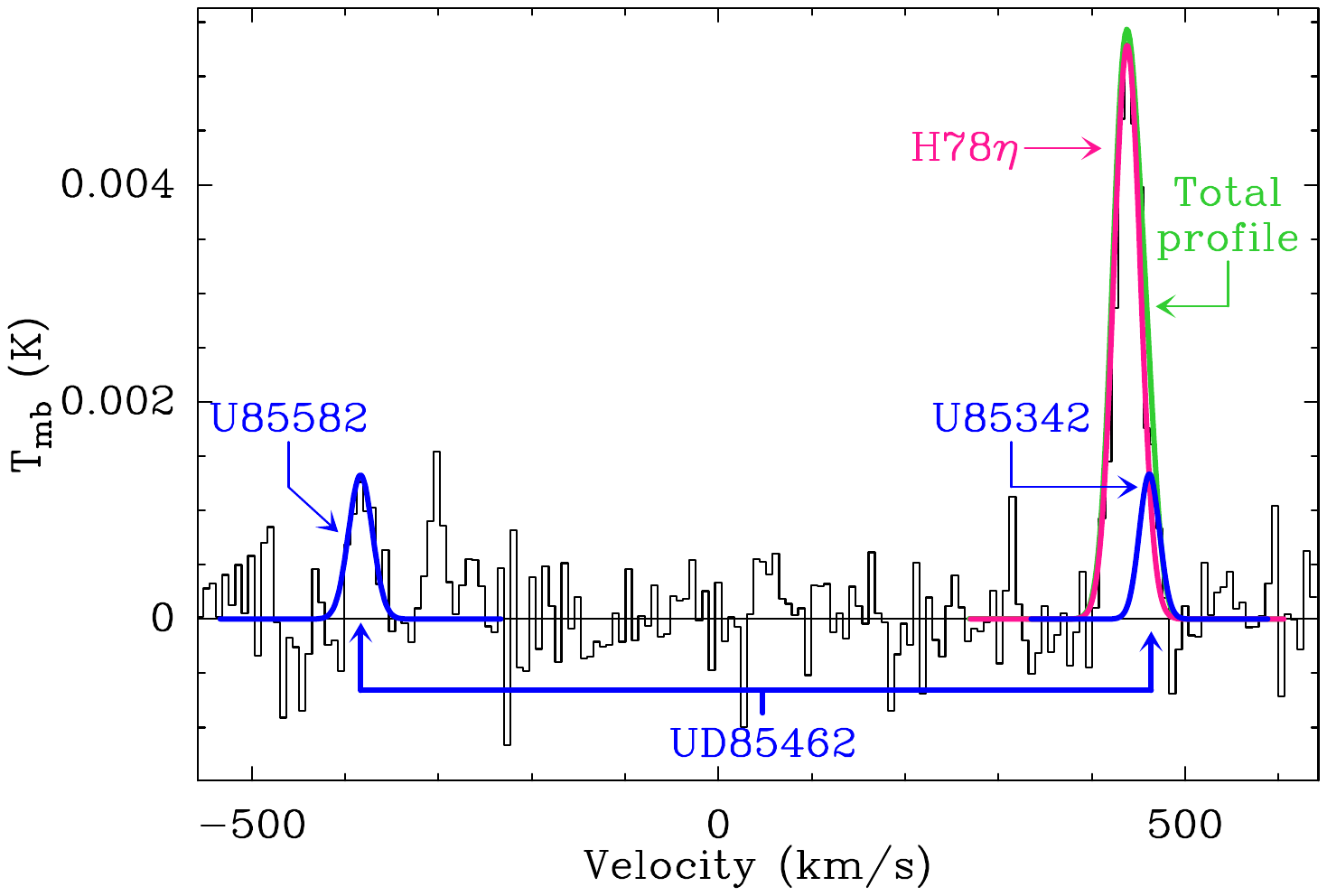}
		\caption{}\label{fig:ufs_UD85462}
	\end{subfigure}
	\hspace{-6mm}
	\begin{subfigure}[]{0.49\linewidth}
		\includegraphics[width=\linewidth]{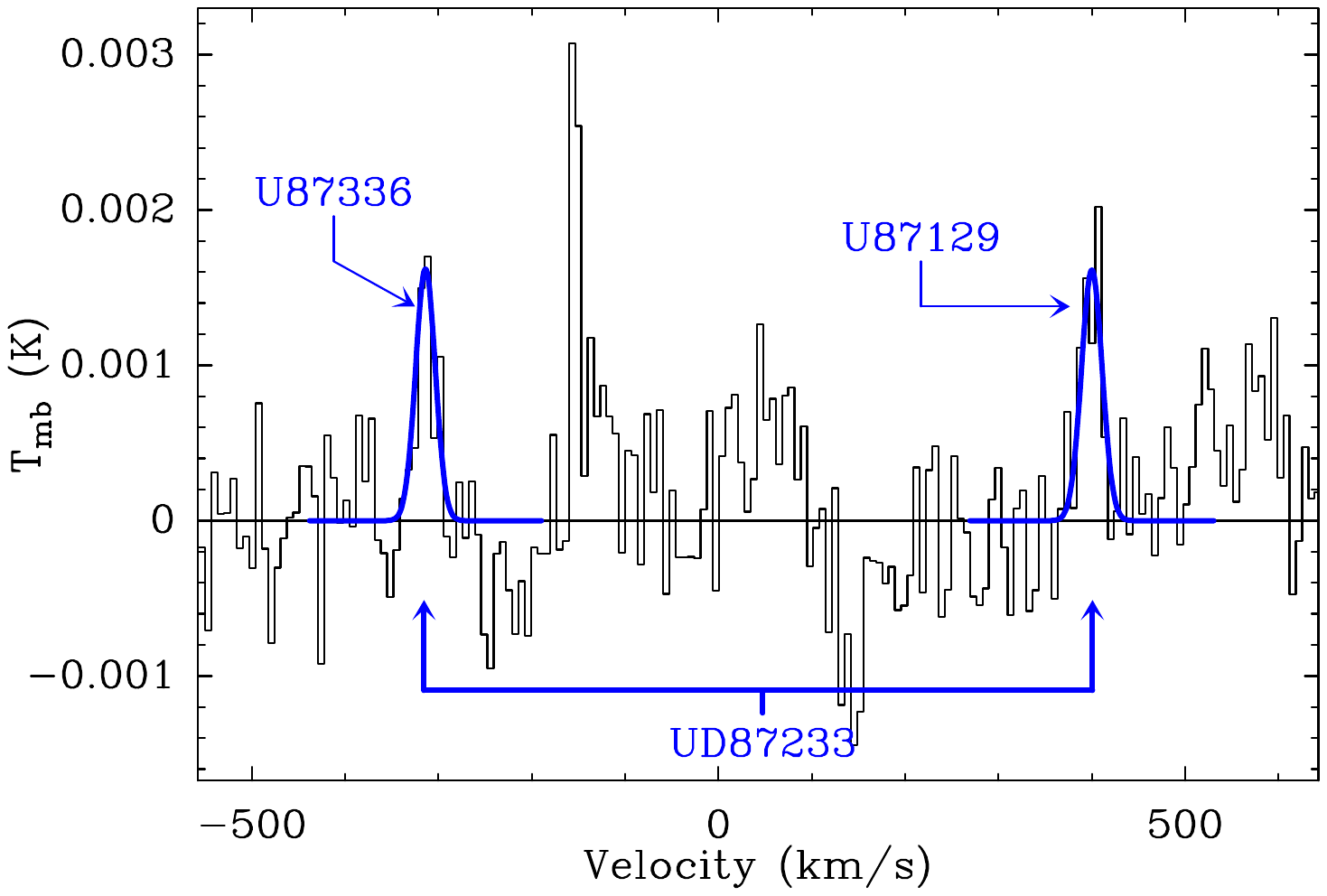}
		\caption{}\label{fig:ufs_UD87233}
	\end{subfigure}
	\hspace{-6mm}
	\begin{subfigure}[]{0.49\linewidth}
		\includegraphics[width=\linewidth]{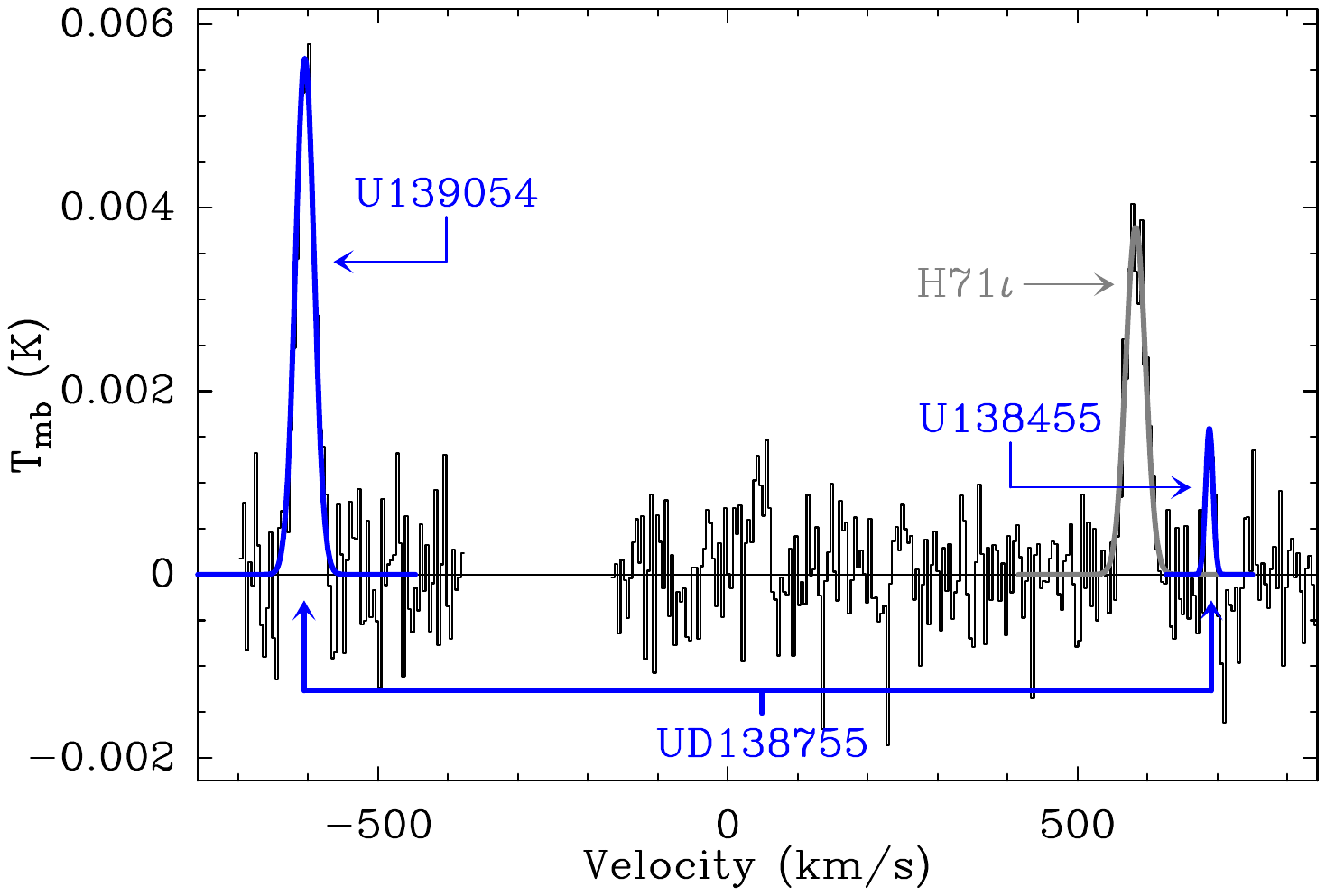}
		\caption{}\label{fig:ufs_UD138755}
	\end{subfigure}
	\caption{Spectra around the doublets formed by UFs from Table \ref{table:ufs_frequencies} that have been identified. The UFs that form the doublets are represented as Gaussian fits in blue. Pink curves represent those RRLs blended with UFs, while single RRLs that do not influence the doublet are represented with gray Gaussians. Green curves show the total profile of blendings.}
	\label{fig:ufs_doublets}
\end{figure*}

Fig.~\ref{fig:ufs_doublets} shows the spectra of the doublets that have been identified in the spectra (see their frequencies in Table \ref{table:ufs_doublets}). Each subplot corresponds to one doublet in which the two UFs that form it (see Tables \ref{table:ufs_frequencies} and \ref{table:ufs_doublets}) are indicated.

\section{Fitting results of the doublets}
\label{appen:fit_plots_doublets}

\begin{figure*}[ht]
	\captionsetup[sub]{skip=0mm, belowskip=0pt}
	\centering
	\hspace{-2mm}
	\begin{subfigure}[]{0.33\linewidth}
		\includegraphics[width=\linewidth]{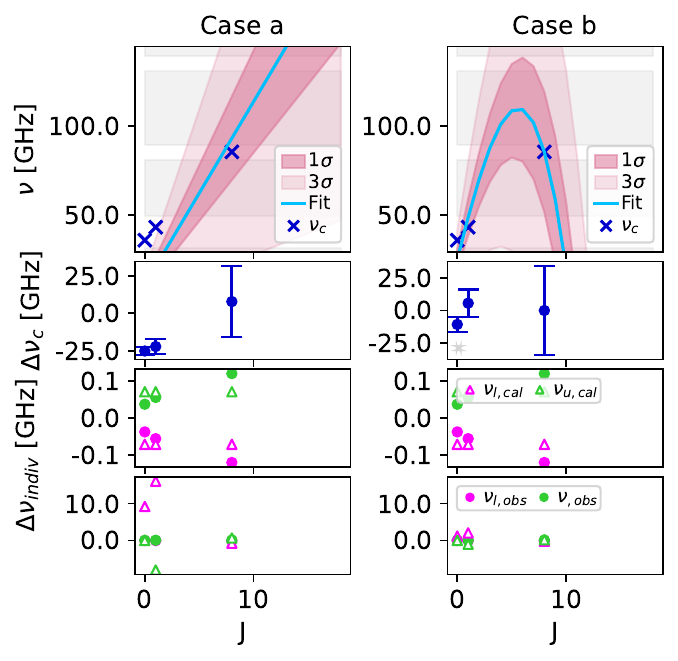}
		\caption{$J_\mathrm{init} = 0$} \label{fig:fit_UD35UD42UD85_J0-0}
	\end{subfigure}
	\hspace{-2mm}
	\begin{subfigure}[]{0.33\linewidth}
		\includegraphics[width=\linewidth]{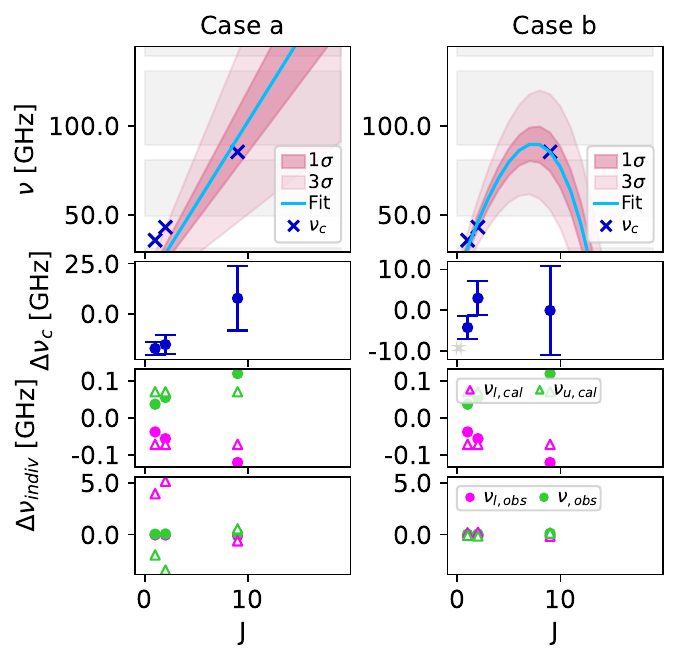}
		\caption{$J_\mathrm{init} = 1$} \label{fig:fit_UD35UD42UD85_J0-1}
	\end{subfigure}
	\hspace{-2mm}
	\begin{subfigure}[]{0.33\linewidth}
		\includegraphics[width=\linewidth]{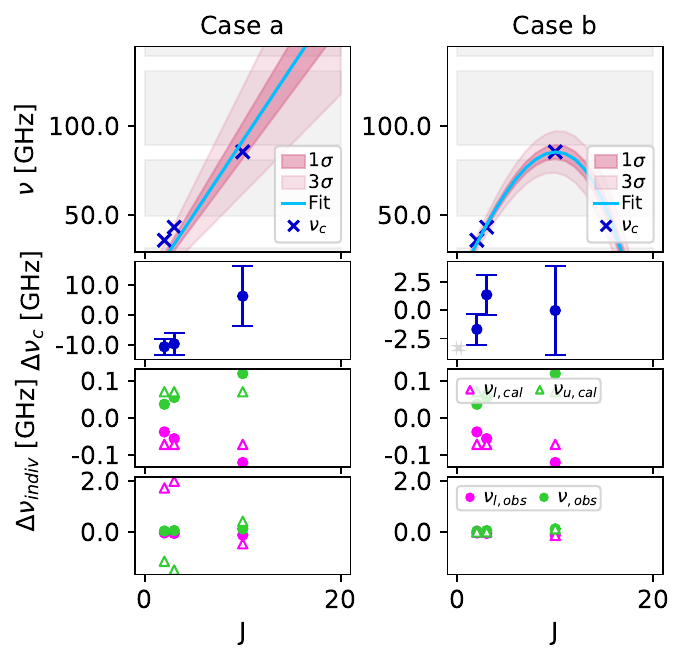}
		\caption{$J_\mathrm{init} = 2$} \label{fig:fit_UD35UD42UD85_J0-2}
	\end{subfigure}
	\hspace{-2mm}
	\begin{subfigure}[]{0.33\linewidth}
		\includegraphics[width=\linewidth]{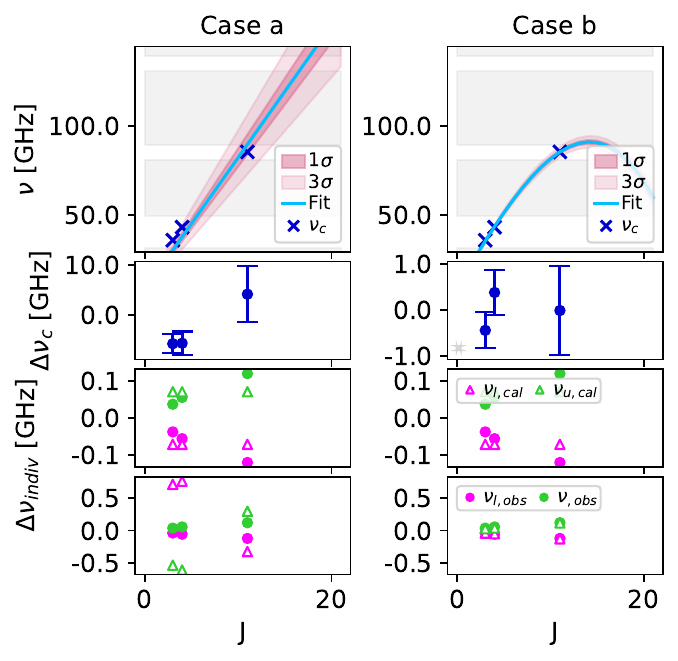}
		\caption{$J_\mathrm{init} = 3$} \label{fig:fit_UD35UD42UD85_J0-3}
	\end{subfigure}
	\hspace{-2mm}
	\begin{subfigure}[]{0.33\linewidth}
		\includegraphics[width=\linewidth]{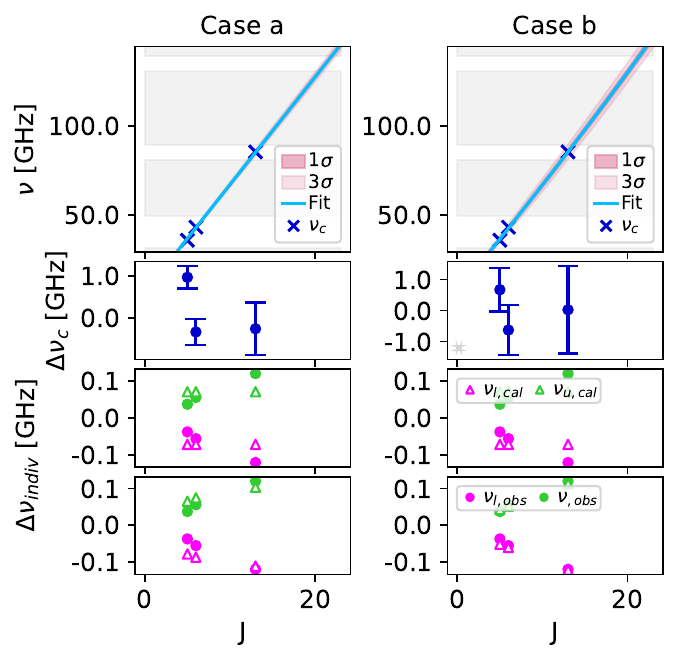}
		\caption{$J_\mathrm{init} = 5$} \label{fig:fit_UD35UD42UD85_J0-5}
	\end{subfigure}
	\hspace{-2mm}
	\begin{subfigure}[]{0.33\linewidth}
		\includegraphics[width=\linewidth]{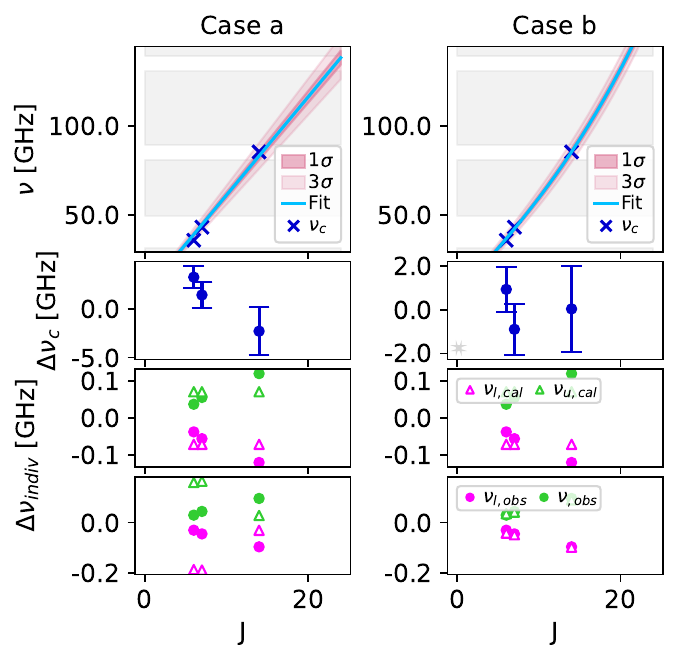}
		\caption{$J_\mathrm{init} = 6$} \label{fig:fit_UD35UD42UD85_J0-6}
	\end{subfigure}
	\hspace{-2mm}
	\begin{subfigure}[]{0.33\linewidth}
		\includegraphics[width=\linewidth]{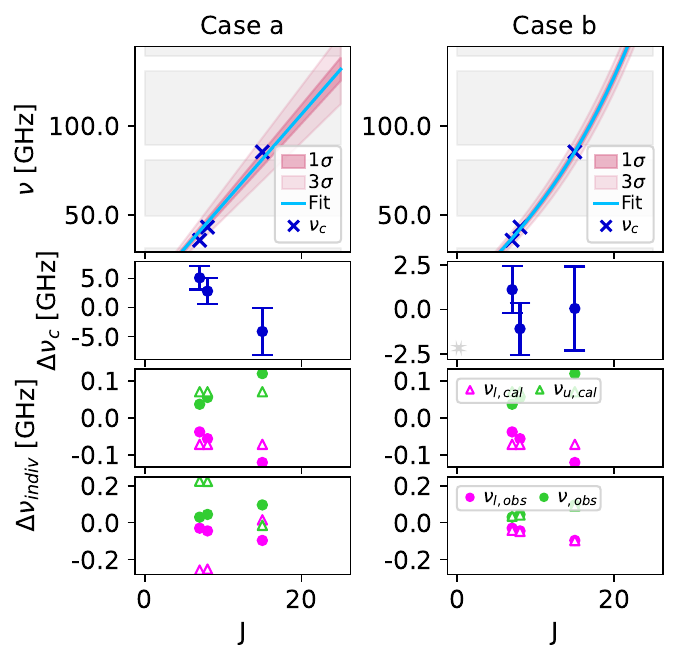}
		\caption{$J_\mathrm{init} = 7$} \label{fig:fit_UD35UD42UD85_J0-7}
	\end{subfigure}
	\hspace{-2mm}
	\begin{subfigure}[]{0.33\linewidth}
		\includegraphics[width=\linewidth]{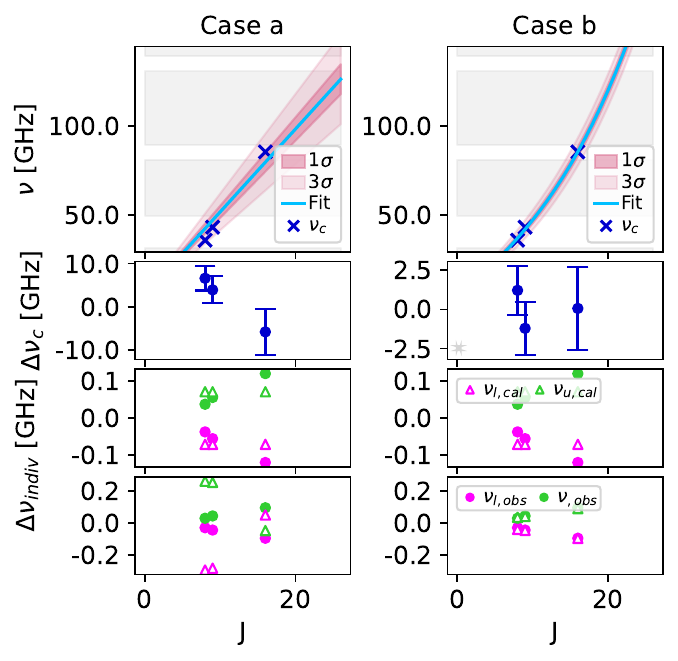}
		\caption{$J_\mathrm{init} = 8$} \label{fig:fit_UD35UD42UD85_J0-8}
	\end{subfigure}
	\hspace{-2mm}
	\begin{subfigure}[]{0.33\linewidth}
		\includegraphics[width=\linewidth]{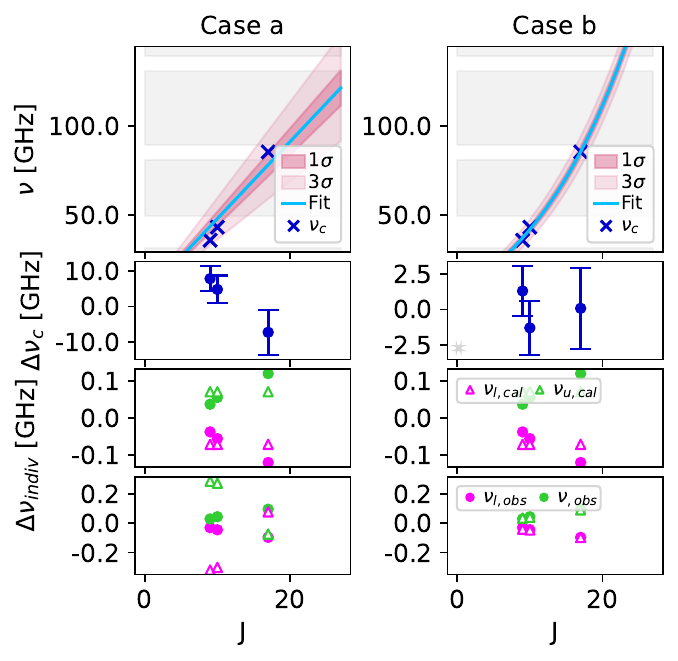}
		\caption{$J_\mathrm{init} = 9$} \label{fig:fit_UD35UD42UD85_J0-9}
	\end{subfigure}
	\hspace{-2mm}
	\begin{subfigure}[]{0.33\linewidth}
		\includegraphics[width=\linewidth]{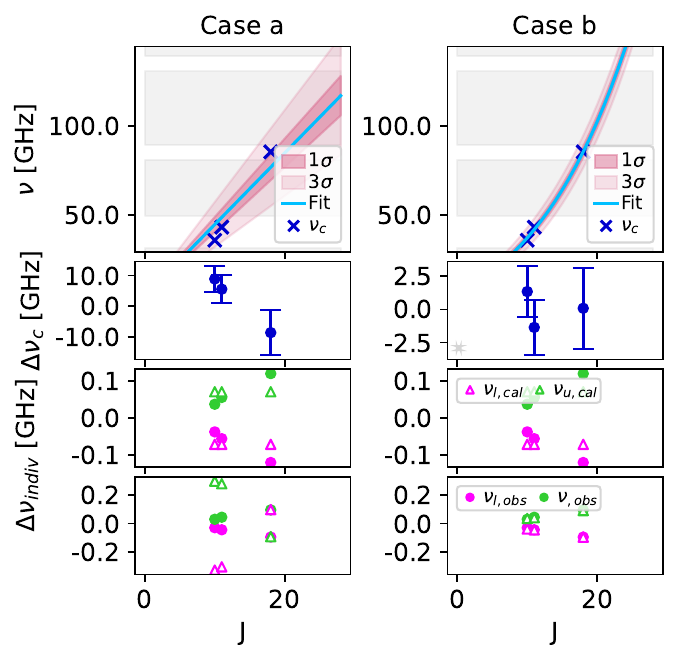}
		\caption{$J_\mathrm{init} = 10$} \label{fig:fit_UD35UD42UD85_J0-10}
	\end{subfigure}
	\hspace{-2mm}
	\begin{subfigure}[]{0.33\linewidth}
		\includegraphics[width=\linewidth]{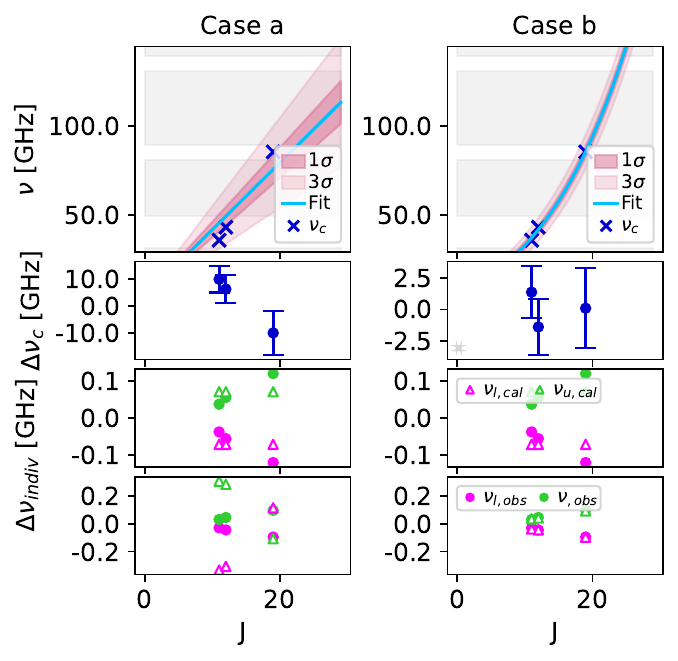}
		\caption{$J_\mathrm{init} = 11$} \label{fig:fit_UD35UD42UD85_J0-11}
	\end{subfigure}
	\caption{Fitting results of UD35556, UD42932, and UD85462 using Eqs.~\ref{eq:freq_RotTrans} and \ref{eq:freq_RotTrans_G}. Three modeling cases have been considered, depending on the number of rotational constants used. Blue crosses show the central frequency of the observational pairs. Blue dots represent the difference between the observed central frequency and the predicted value. Finally, green and pink markers show the difference between individual lines and the central frequency of the doublet.The light gray areas in all plots indicate the unobserved ranges.}
	\label{fig:FitSummary-fit_UD35UD42UD85_1}
\end{figure*}

\begin{figure*}[ht]
	\captionsetup[sub]{skip=0mm, belowskip=0pt}
	\centering
	\hspace{-2mm}
	\begin{subfigure}[]{0.33\linewidth}
		\includegraphics[width=\linewidth]{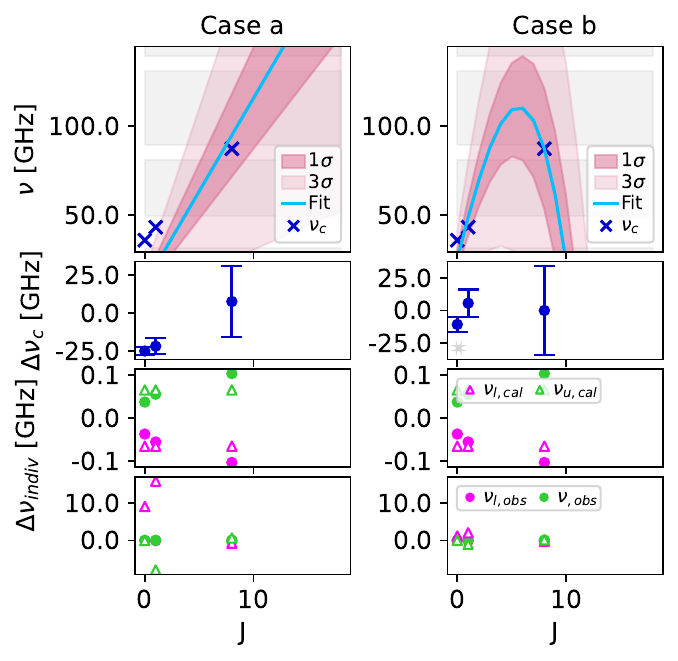}
		\caption{$J_\mathrm{init} = 0$} \label{fig:fit_UD35UD42UD87_J0-0}
	\end{subfigure}
	\hspace{-2mm}
	\begin{subfigure}[]{0.33\linewidth}
		\includegraphics[width=\linewidth]{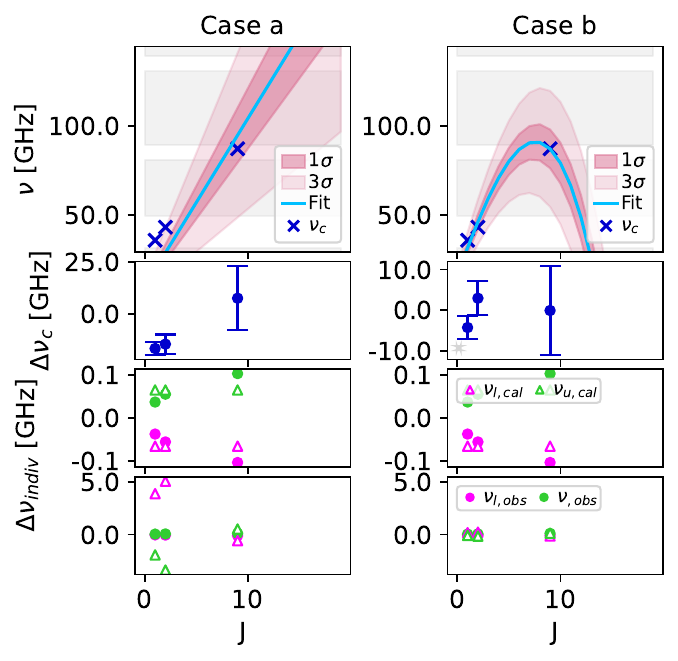}
		\caption{$J_\mathrm{init} = 1$} \label{fig:fit_UD35UD42UD87_J0-1}
	\end{subfigure}
	\hspace{-2mm}
	\begin{subfigure}[]{0.33\linewidth}
		\includegraphics[width=\linewidth]{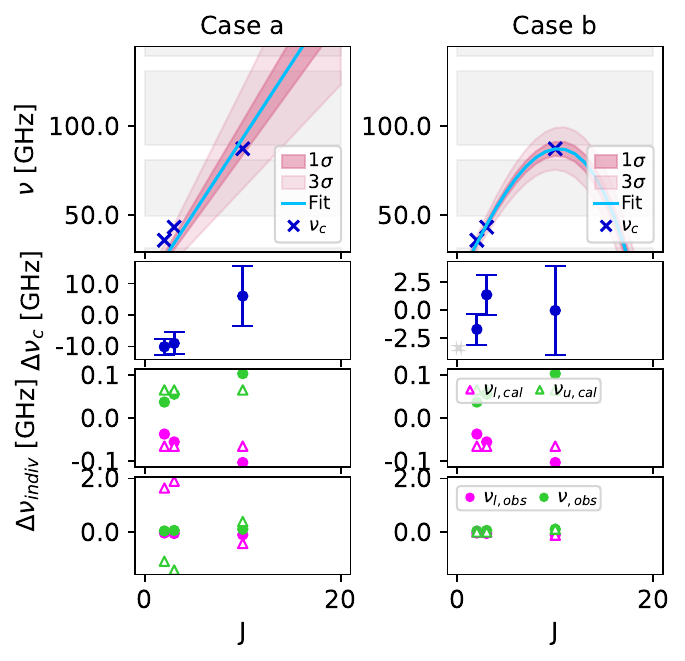}
		\caption{$J_\mathrm{init} = 2$} \label{fig:fit_UD35UD42UD87_J0-2}
	\end{subfigure}
	\hspace{-2mm}
	\begin{subfigure}[]{0.33\linewidth}
		\includegraphics[width=\linewidth]{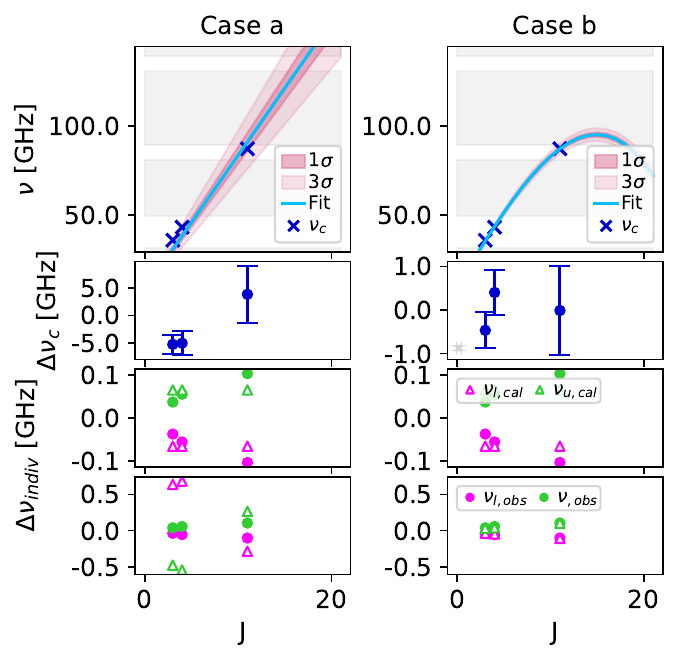}
		\caption{$J_\mathrm{init} = 3$} \label{fig:fit_UD35UD42UD87_J0-3}
	\end{subfigure}
	\hspace{-2mm}
	\begin{subfigure}[]{0.33\linewidth}
		\includegraphics[width=\linewidth]{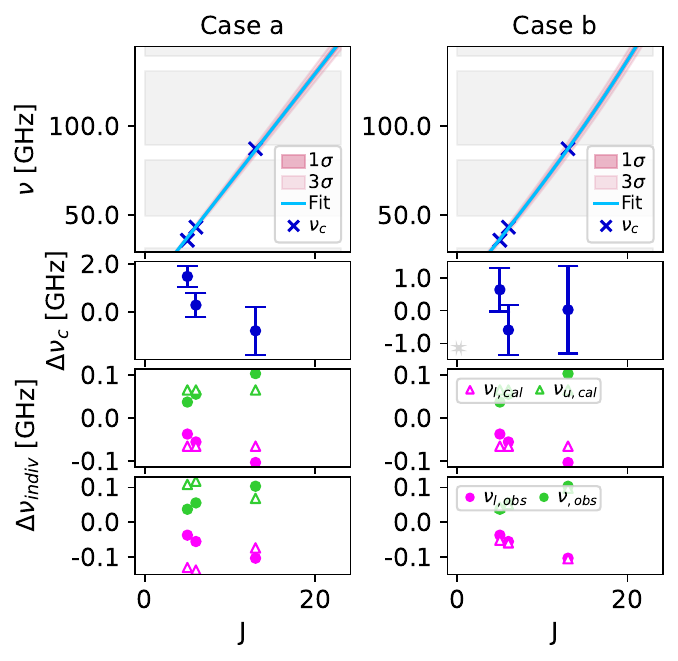}
		\caption{$J_\mathrm{init} = 5$} \label{fig:fit_UD35UD42UD87_J0-5}
	\end{subfigure}
	\hspace{-2mm}
	\begin{subfigure}[]{0.33\linewidth}
		\includegraphics[width=\linewidth]{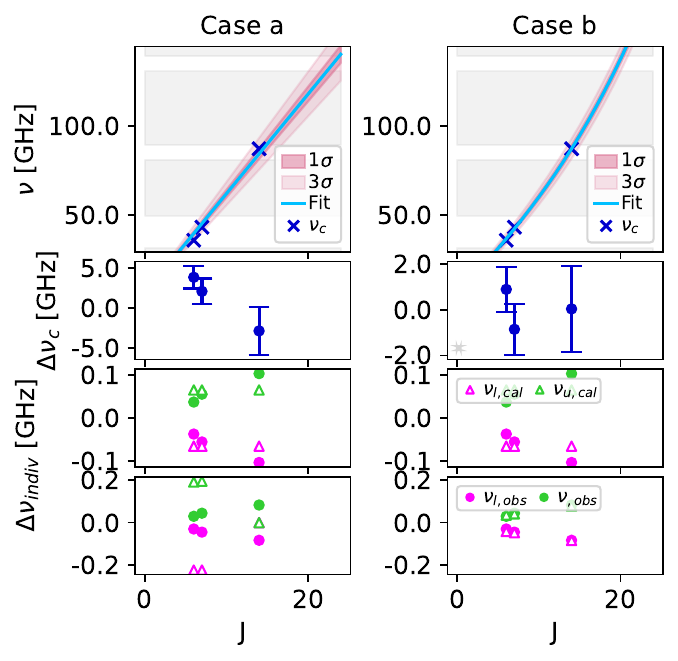}
		\caption{$J_\mathrm{init} = 6$} \label{fig:fit_UD35UD42UD87_J0-6}
	\end{subfigure}\hspace{-2mm}
	\begin{subfigure}[]{0.33\linewidth}
	\includegraphics[width=\linewidth]{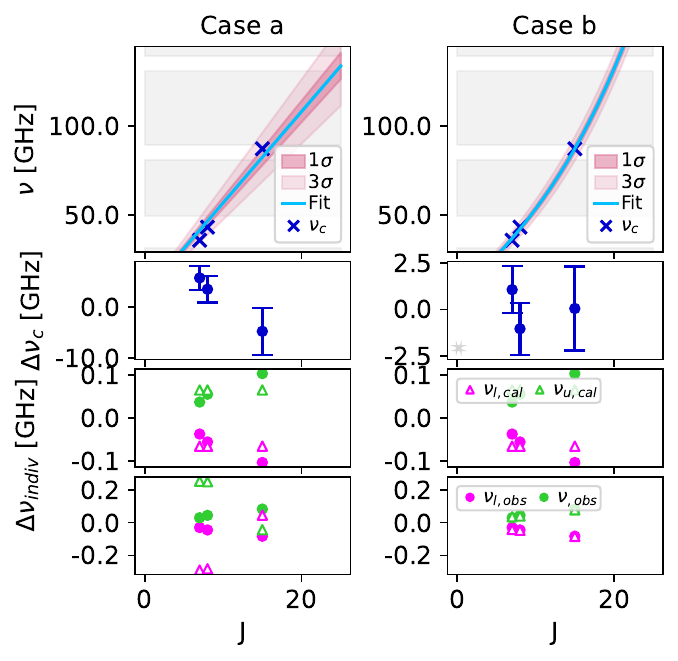}
	\caption{$J_\mathrm{init} = 7$} \label{fig:fit_UD35UD42UD87_J0-7}
	\end{subfigure}
	\hspace{-2mm}
	\begin{subfigure}[]{0.33\linewidth}
	\includegraphics[width=\linewidth]{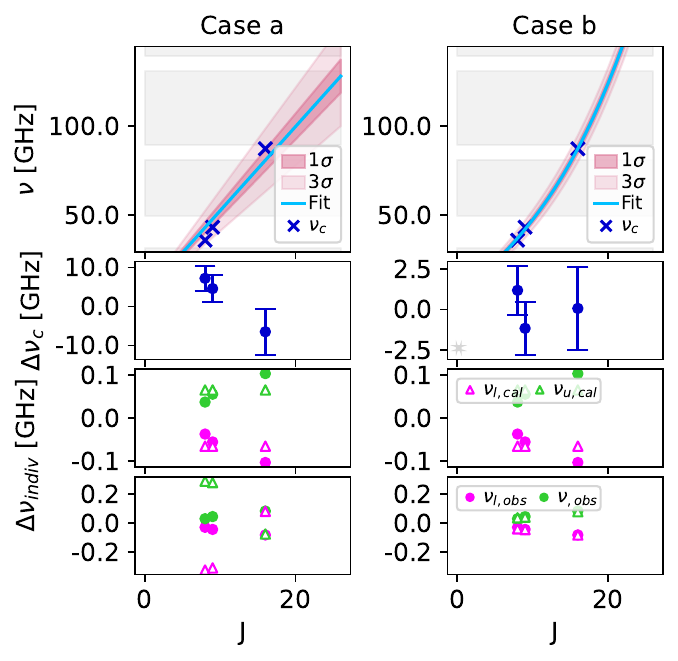}
	\caption{$J_\mathrm{init} = 8$} \label{fig:fit_UD35UD42UD87_J0-8}
	\end{subfigure}
	\hspace{-2mm}
	\begin{subfigure}[]{0.33\linewidth}
	\includegraphics[width=\linewidth]{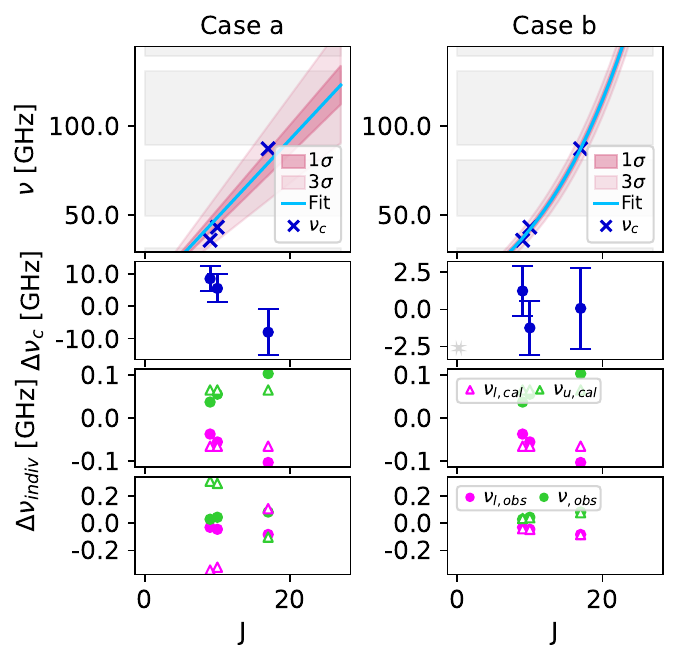}
	\caption{$J_\mathrm{init} = 9$} \label{fig:fit_UD35UD42UD87_J0-9}
	\end{subfigure}
	\hspace{-2mm}
	\begin{subfigure}[]{0.33\linewidth}
	\includegraphics[width=\linewidth]{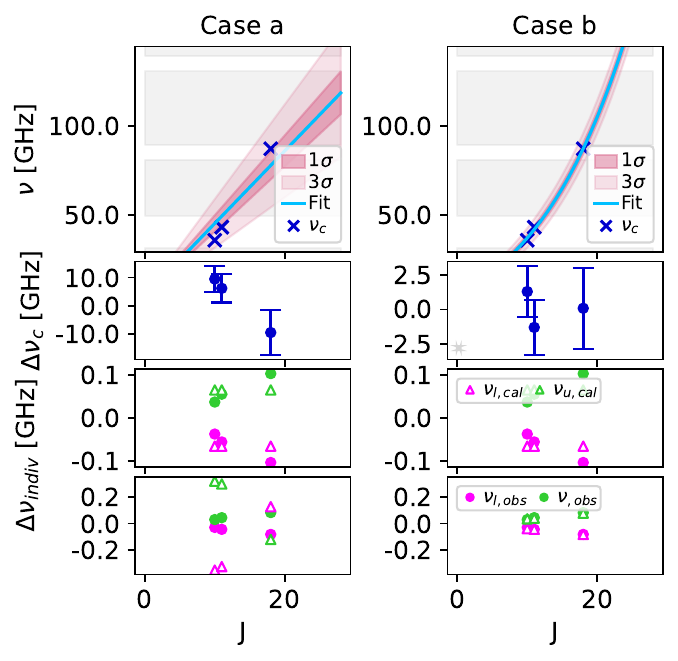}
	\caption{$J_\mathrm{init} = 10$} \label{fig:fit_UD35UD42UD87_J0-10}
	\end{subfigure}
	\hspace{-2mm}
	\begin{subfigure}[]{0.33\linewidth}
	\includegraphics[width=\linewidth]{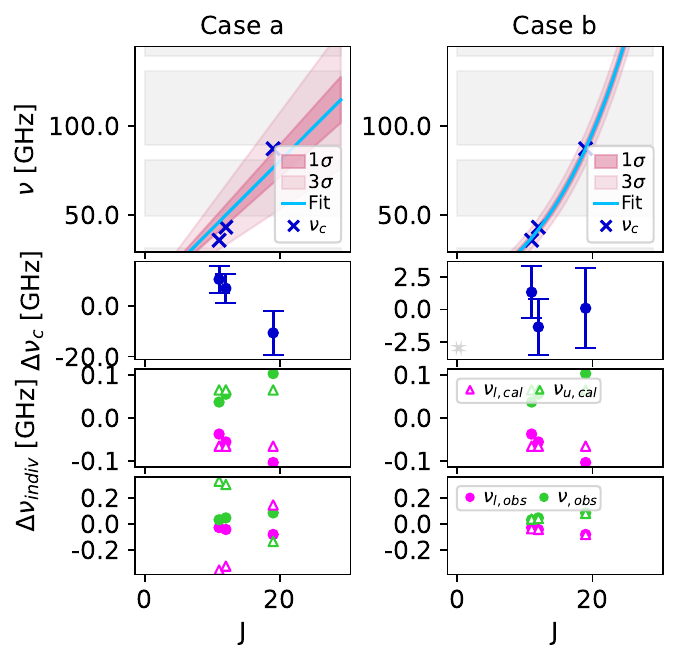}
	\caption{$J_\mathrm{init} = 11$} \label{fig:fit_UD35UD42UD87_J0-11}
	\end{subfigure}
	\caption{Fitting results of UD35556, UD42932, and UD87233 using Eqs.~\ref{eq:freq_RotTrans} and \ref{eq:freq_RotTrans_G}.}
	\label{fig:FitSummary-fit_UD35UD42UD87_1}
\end{figure*}

\begin{figure*}[ht]
	\captionsetup[sub]{skip=0mm, belowskip=0pt}
	\centering
	\hspace{-2mm}
	\begin{subfigure}[]{0.33\linewidth}
		\includegraphics[width=\linewidth]{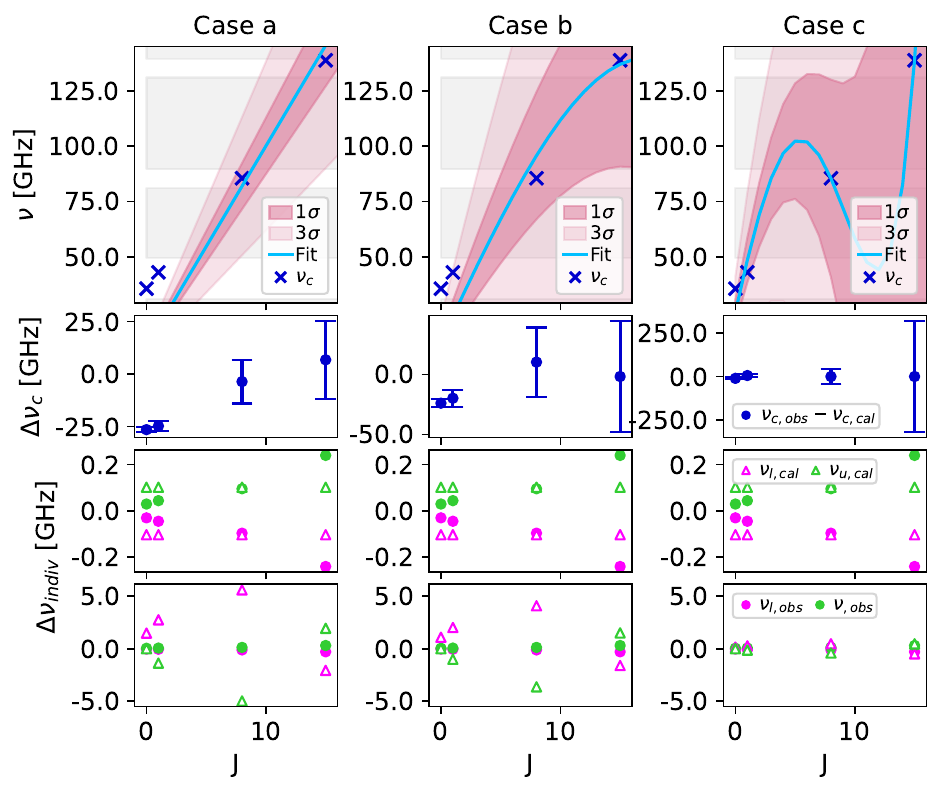}
		\caption{$J_\mathrm{init} = 0$} \label{fig:fit_UD35UD42UD85UD138_J0-0}
	\end{subfigure}
	\hspace{-2mm}
	\begin{subfigure}[]{0.33\linewidth}
		\includegraphics[width=\linewidth]{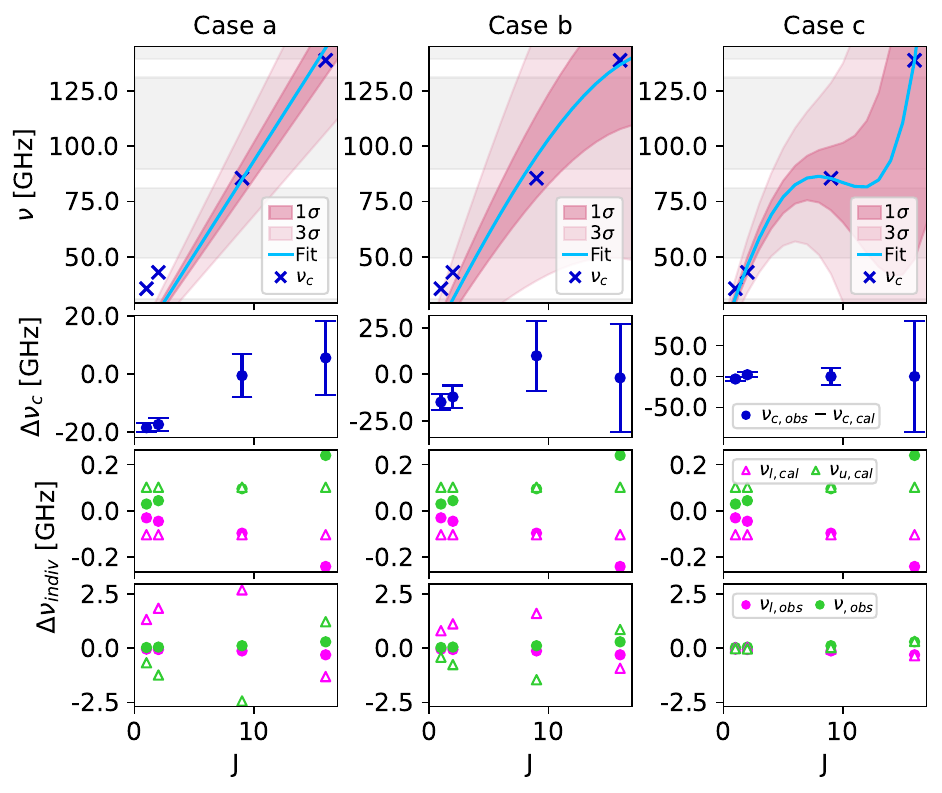}
		\caption{$J_\mathrm{init} = 1$} \label{fig:fit_UD35UD42UD85UD138_J0-1}
	\end{subfigure}
	\hspace{-2mm}
	\begin{subfigure}[]{0.33\linewidth}
		\includegraphics[width=\linewidth]{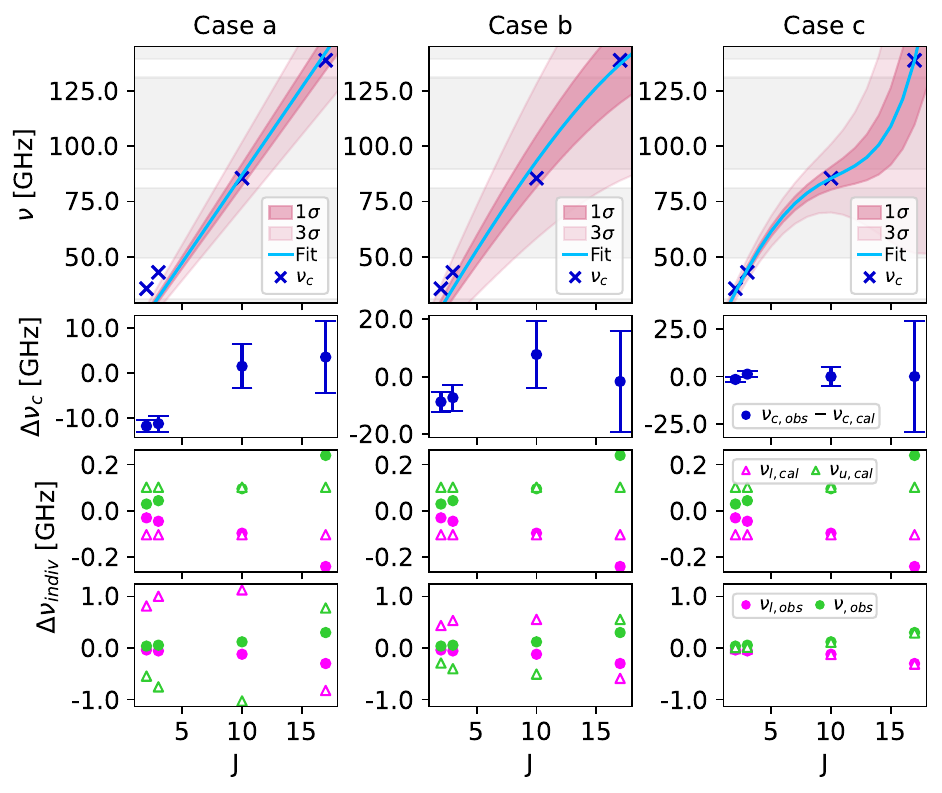}
		\caption{$J_\mathrm{init} = 2$} \label{fig:fit_UD35UD42UD85UD138_J0-2}
	\end{subfigure}
	\hspace{-2mm}
	\begin{subfigure}[]{0.33\linewidth}
		\includegraphics[width=\linewidth]{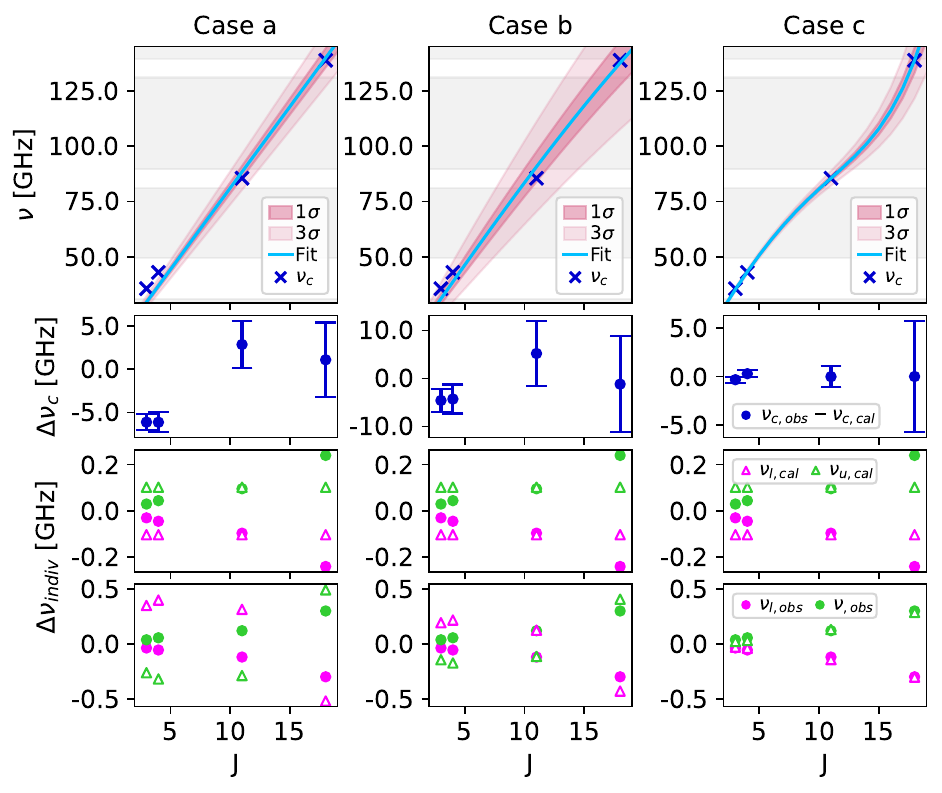}
		\caption{$J_\mathrm{init} = 3$} \label{fig:fit_UD35UD42UD85UD138_J0-3}
	\end{subfigure}
	\hspace{-2mm}
	\begin{subfigure}[]{0.33\linewidth}
		\includegraphics[width=\linewidth]{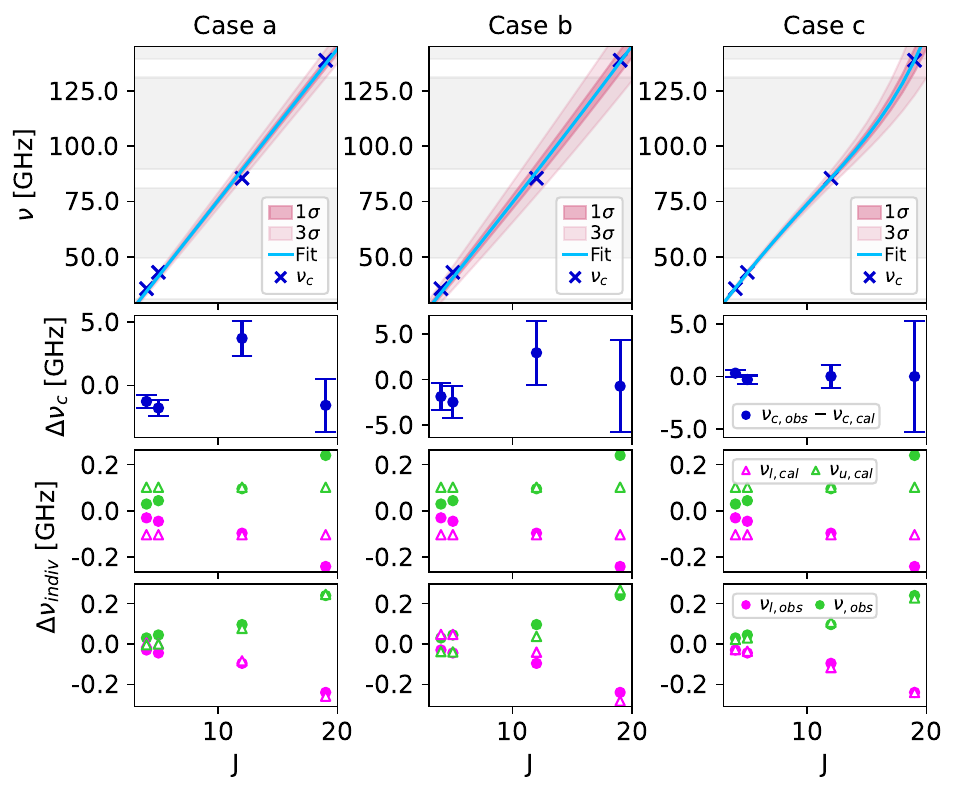}
		\caption{$J_\mathrm{init} = 4$} \label{fig:fit_UD35UD42UD85UD138_J0-4}
	\end{subfigure}
	\hspace{-2mm}
	\begin{subfigure}[]{0.33\linewidth}
		\includegraphics[width=\linewidth]{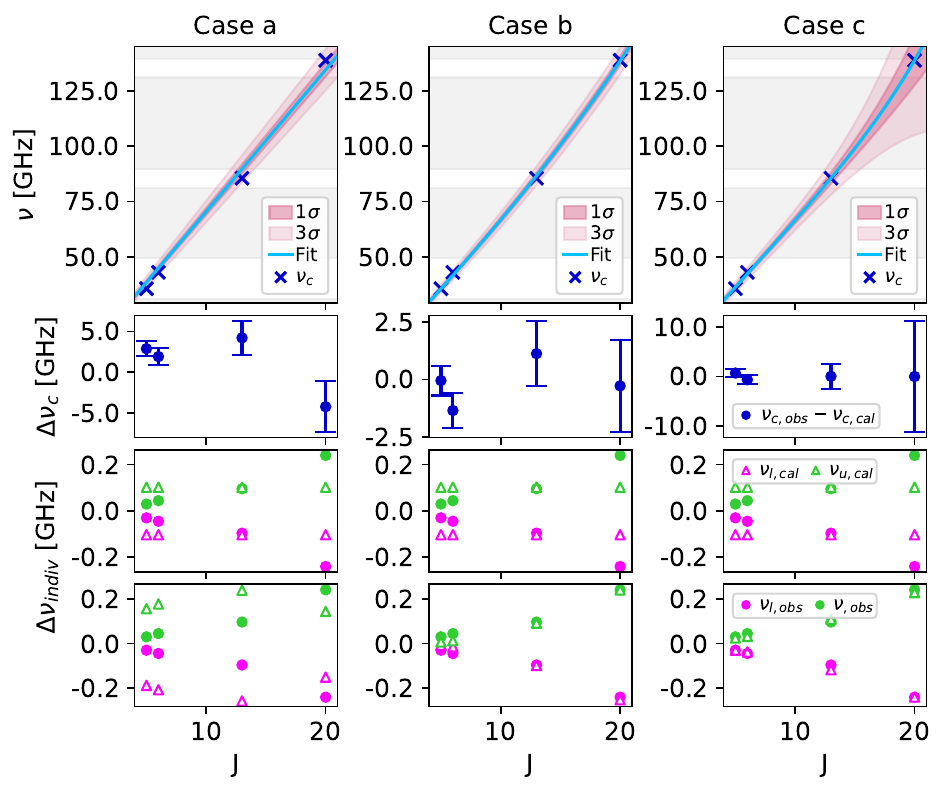}
		\caption{$J_\mathrm{init} = 5$} \label{fig:fit_UD35UD42UD85UD138_J0-5}
	\end{subfigure}\hspace{-2mm}
	\begin{subfigure}[]{0.33\linewidth}
	\includegraphics[width=\linewidth]{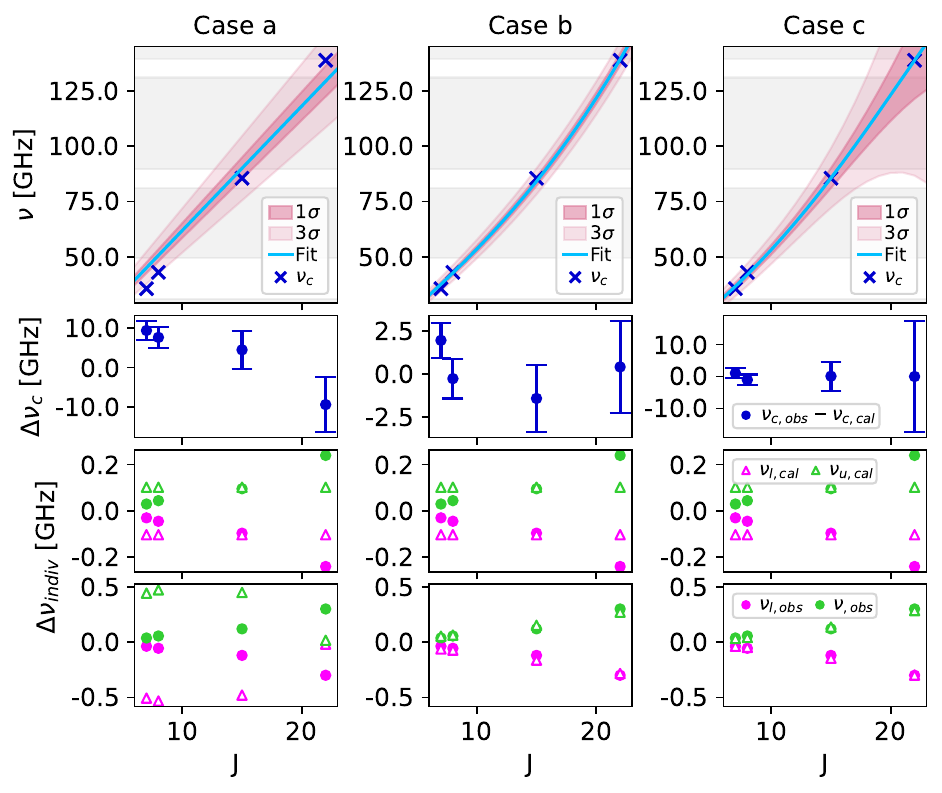}
	\caption{$J_\mathrm{init} = 7$} \label{fig:fit_UD35UD42UD85UD138_J0-7}
	\end{subfigure}
	\hspace{-2mm}
	\begin{subfigure}[]{0.33\linewidth}
	\includegraphics[width=\linewidth]{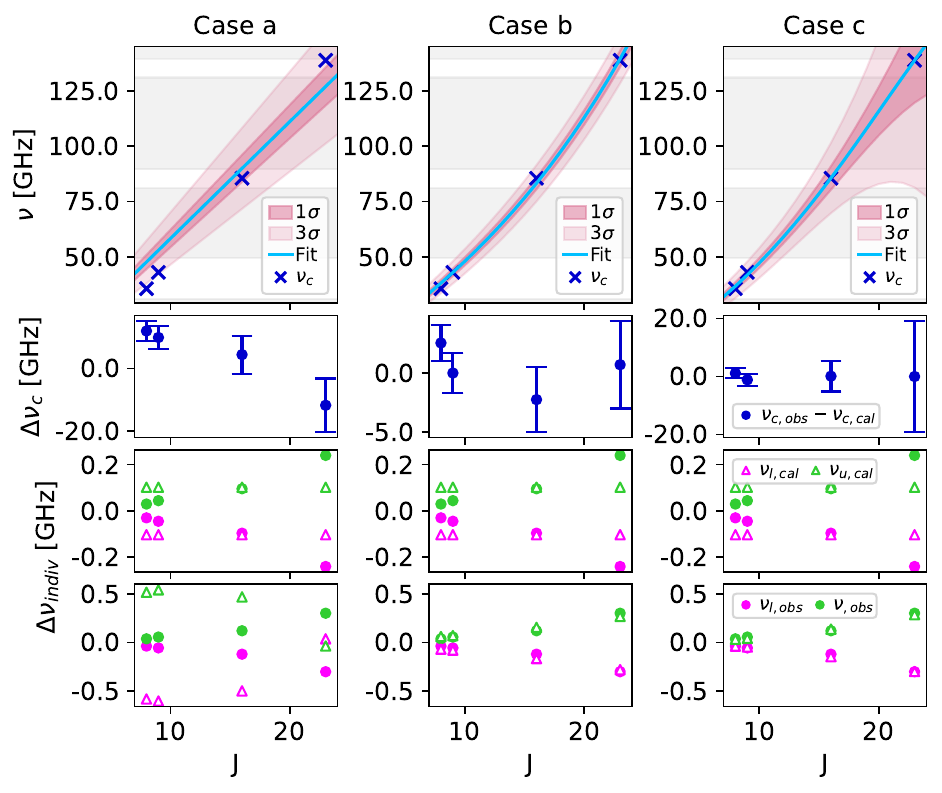}
	\caption{$J_\mathrm{init} = 8$} \label{fig:fit_UD35UD42UD85UD138_J0-8}
	\end{subfigure}
	\hspace{-2mm}
	\begin{subfigure}[]{0.33\linewidth}
	\includegraphics[width=\linewidth]{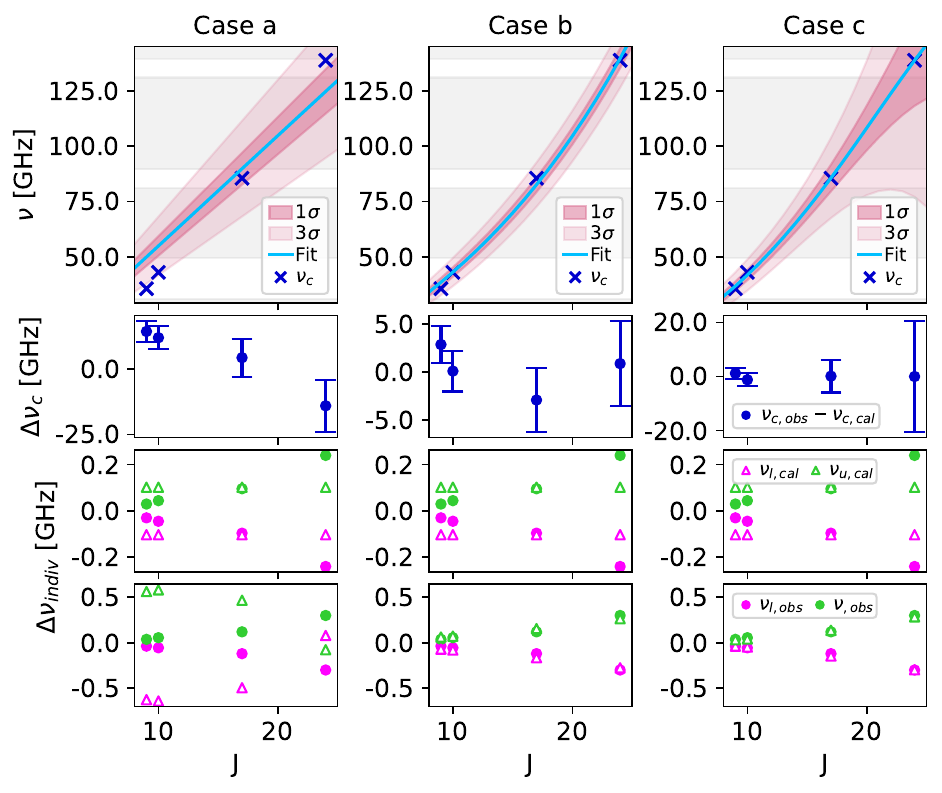}
	\caption{$J_\mathrm{init} = 9$} \label{fig:fit_UD35UD42UD85UD138_J0-9}
	\end{subfigure}
	\hspace{-2mm}
	\begin{subfigure}[]{0.33\linewidth}
	\includegraphics[width=\linewidth]{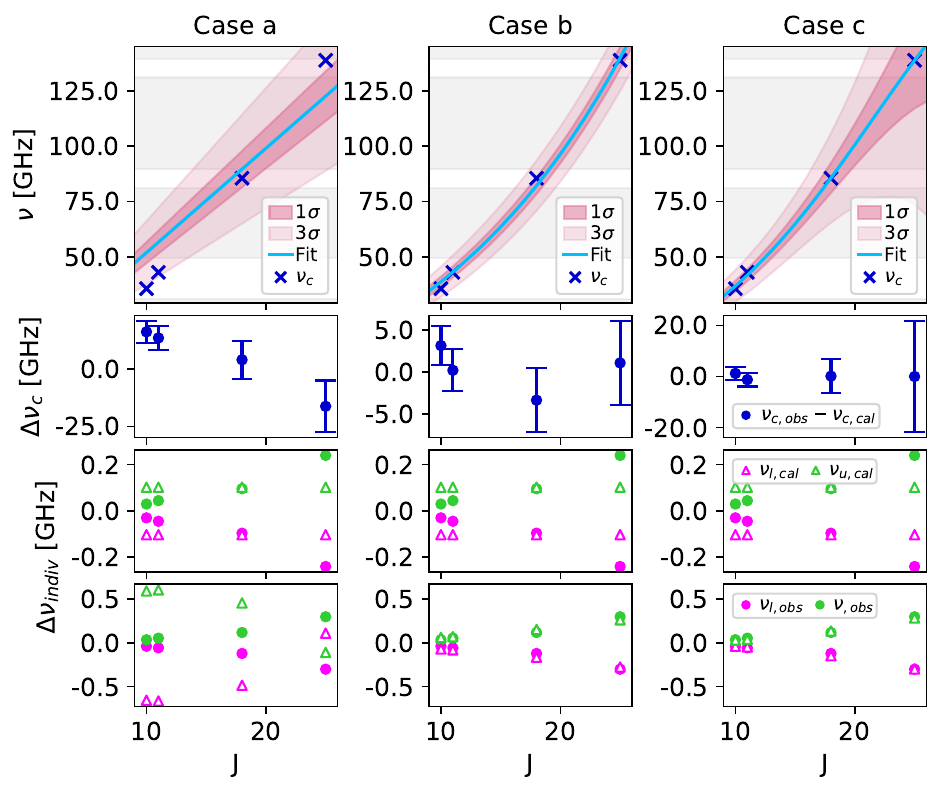}
	\caption{$J_\mathrm{init} = 10$} \label{fig:fit_UD35UD42UD85UD138_J0-10}
	\end{subfigure}
	\hspace{-2mm}
	\begin{subfigure}[]{0.33\linewidth}
	\includegraphics[width=\linewidth]{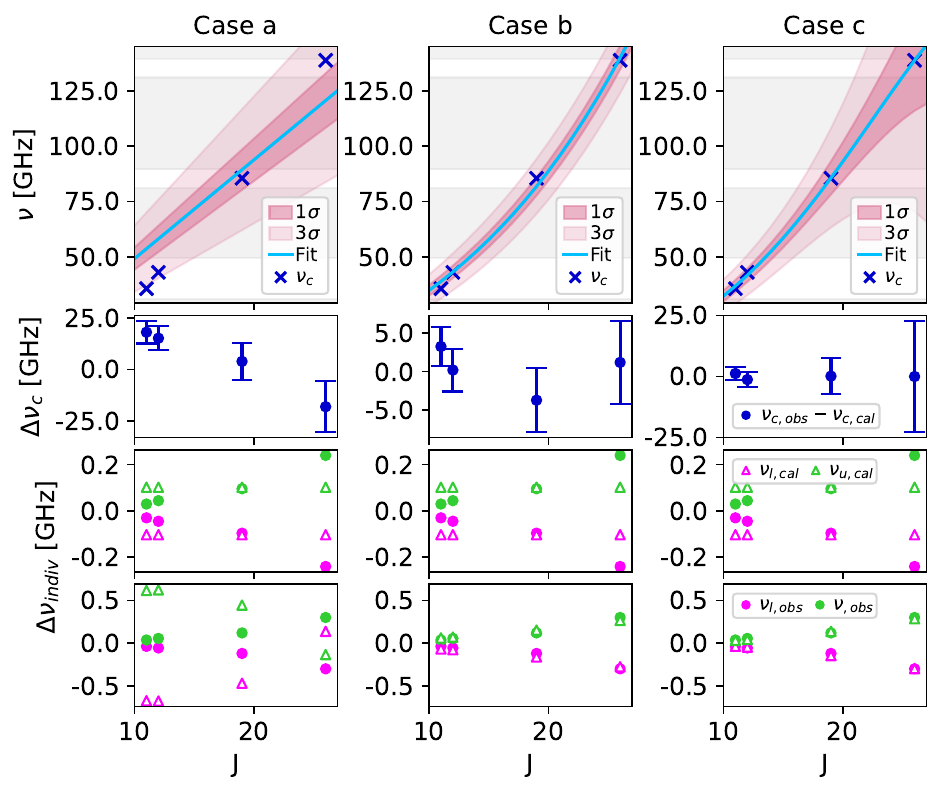}
	\caption{$J_\mathrm{init} = 11$} \label{fig:fit_UD35UD42UD85UD138_J0-11}
	\end{subfigure}
	\caption{Fitting results of UD35556, UD42932, UD85462, and UD138755 using Eqs.~\ref{eq:freq_RotTrans} and \ref{eq:freq_RotTrans_G}.}
	\label{fig:FitSummary-fit_UD35UD42UD85UD138_1}
\end{figure*}

\begin{figure*}[ht]
	\captionsetup[sub]{skip=0mm, belowskip=0pt}
	\centering
	\hspace{-2mm}
	\begin{subfigure}[]{0.33\linewidth}
		\includegraphics[width=\linewidth]{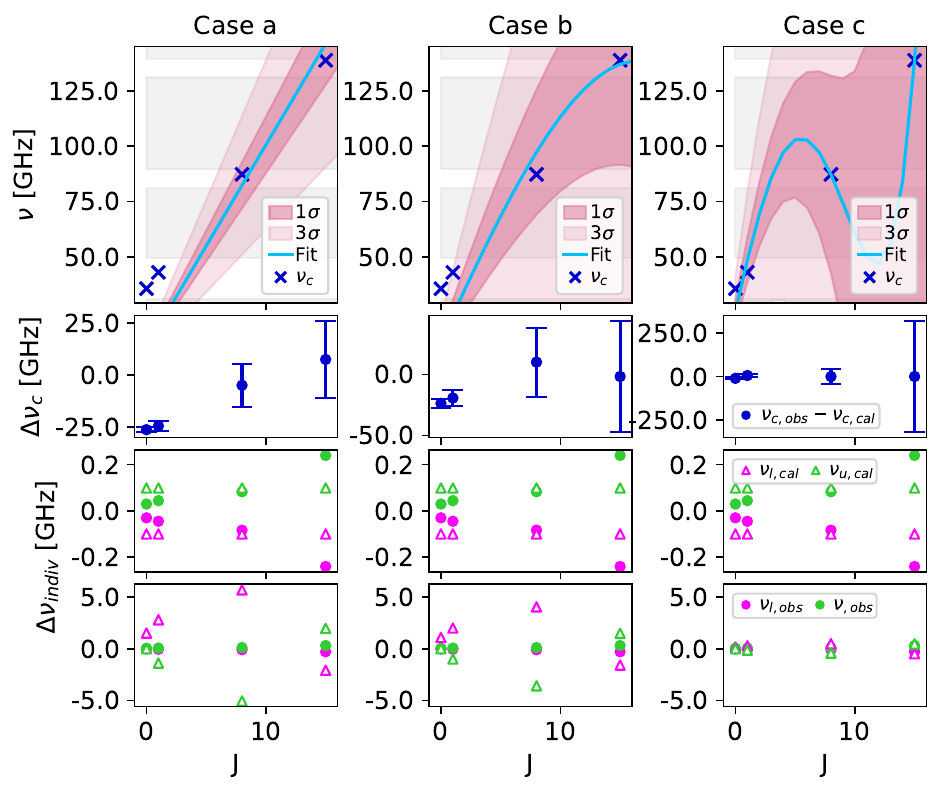}
		\caption{$J_\mathrm{init} = 0$} \label{fig:fit_UD35UD42UD87UD138_J0-0}
	\end{subfigure}
	\hspace{-2mm}
	\begin{subfigure}[]{0.33\linewidth}
		\includegraphics[width=\linewidth]{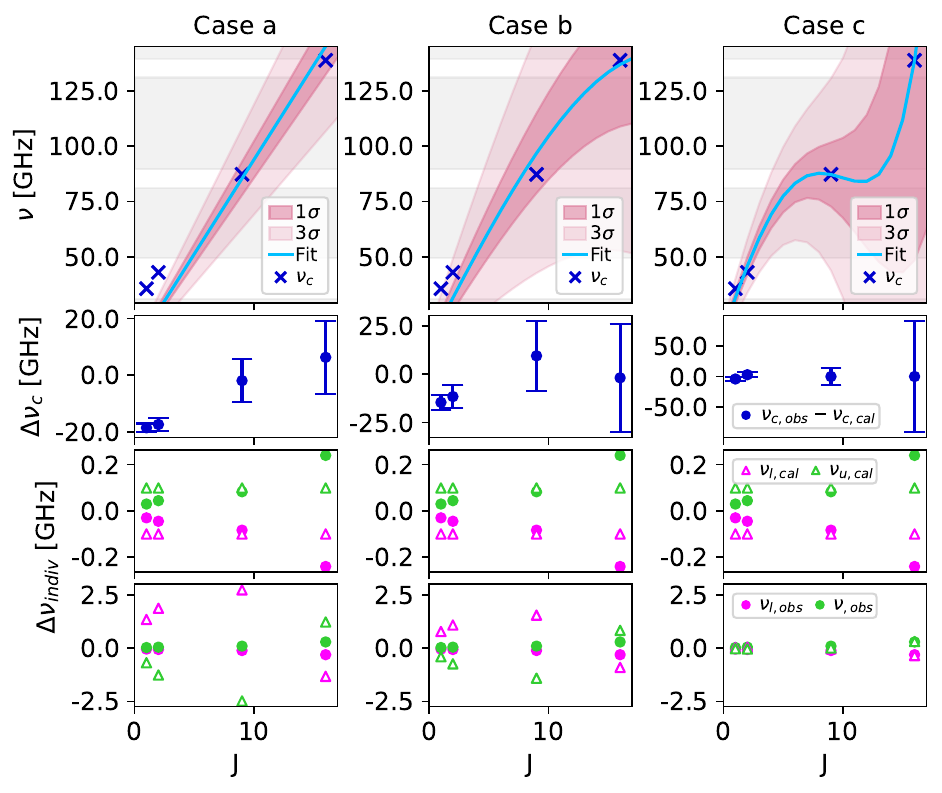}
		\caption{$J_\mathrm{init} = 1$} \label{fig:fit_UD35UD42UD87UD138_J0-1}
	\end{subfigure}
	\hspace{-2mm}
	\begin{subfigure}[]{0.33\linewidth}
		\includegraphics[width=\linewidth]{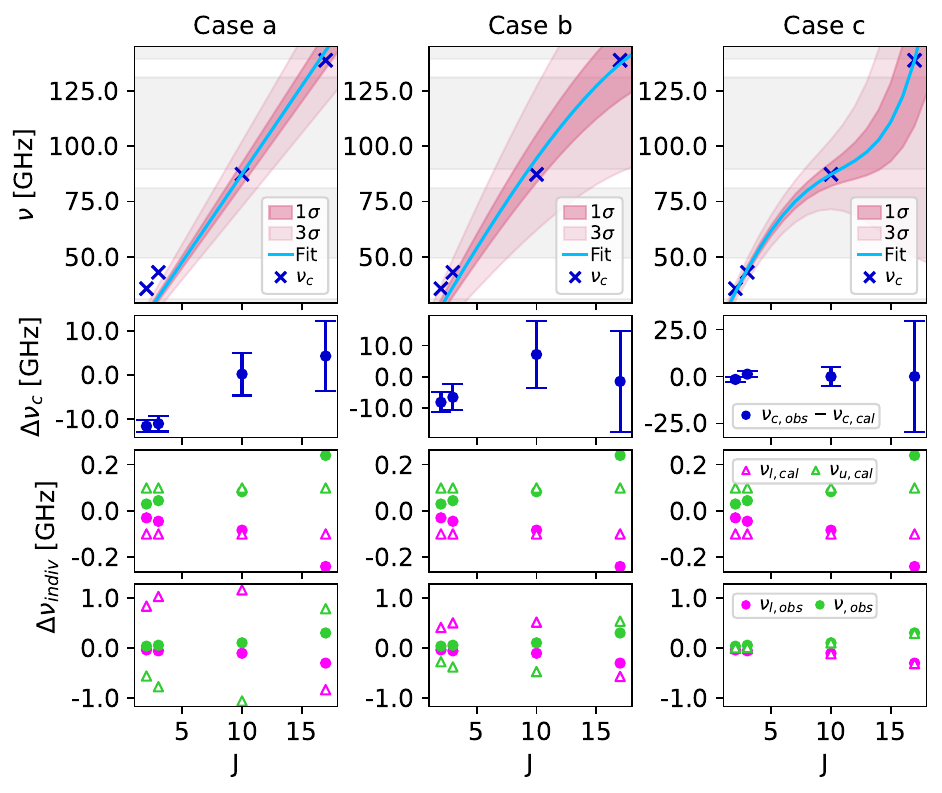}
		\caption{$J_\mathrm{init} = 2$} \label{fig:fit_UD35UD42UD87UD138_J0-2}
	\end{subfigure}
	\hspace{-2mm}
	\begin{subfigure}[]{0.33\linewidth}
		\includegraphics[width=\linewidth]{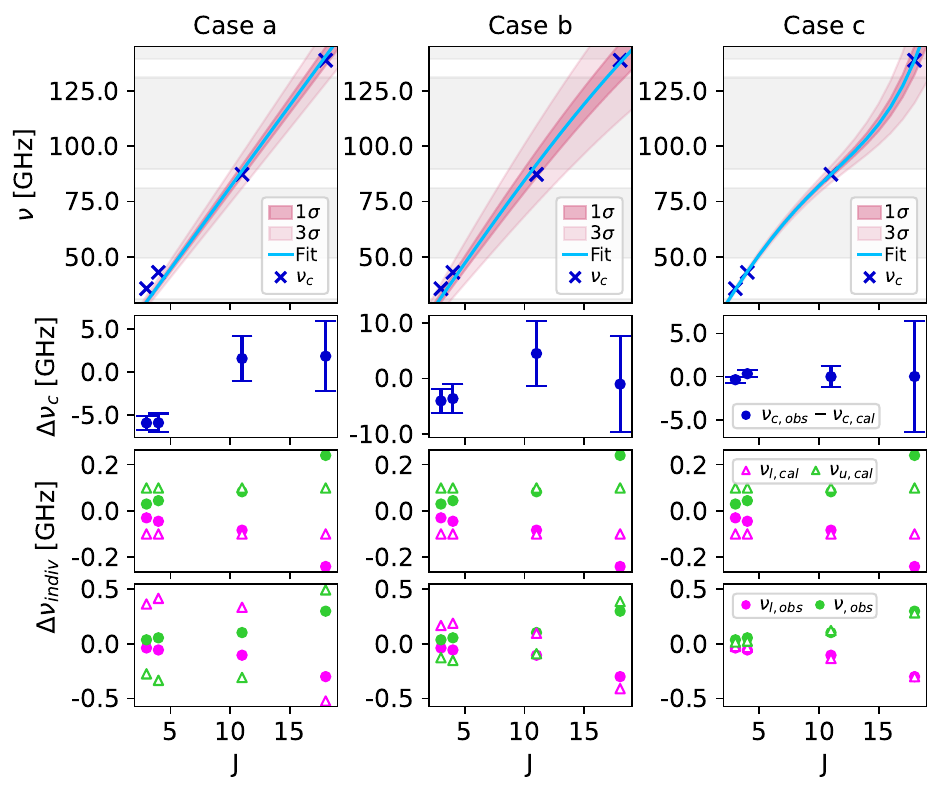}
		\caption{$J_\mathrm{init} = 3$} \label{fig:fit_UD35UD42UD87UD138_J0-3}
	\end{subfigure}
	\hspace{-2mm}
	\begin{subfigure}[]{0.33\linewidth}
		\includegraphics[width=\linewidth]{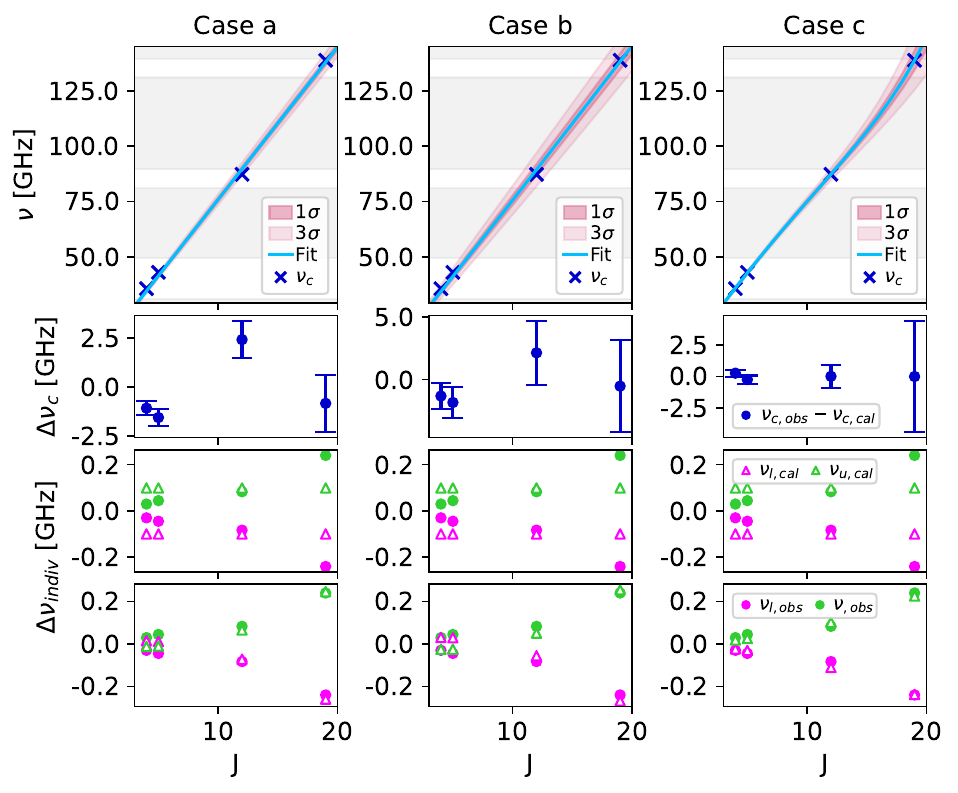}
		\caption{$J_\mathrm{init} = 4$} \label{fig:fit_UD35UD42UD87UD138_J0-4}
	\end{subfigure}
	\hspace{-2mm}
	\begin{subfigure}[]{0.33\linewidth}
		\includegraphics[width=\linewidth]{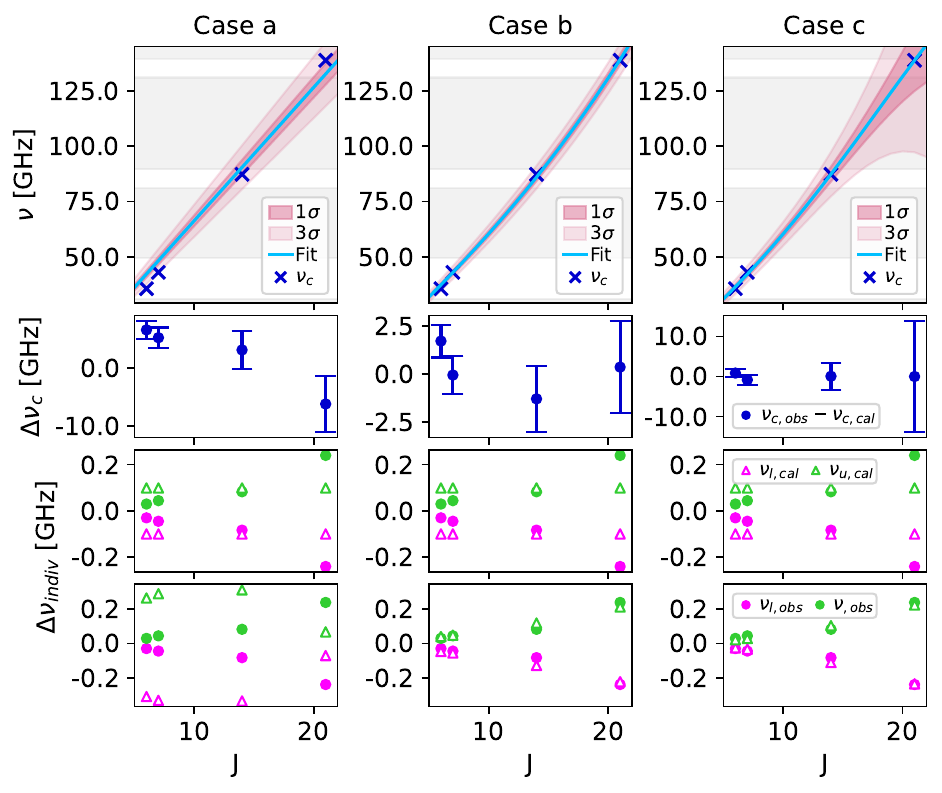}
		\caption{$J_\mathrm{init} = 6$} \label{fig:fit_UD35UD42UD87UD138_J0-6}
	\end{subfigure}\hspace{-2mm}
	\begin{subfigure}[]{0.33\linewidth}
	\includegraphics[width=\linewidth]{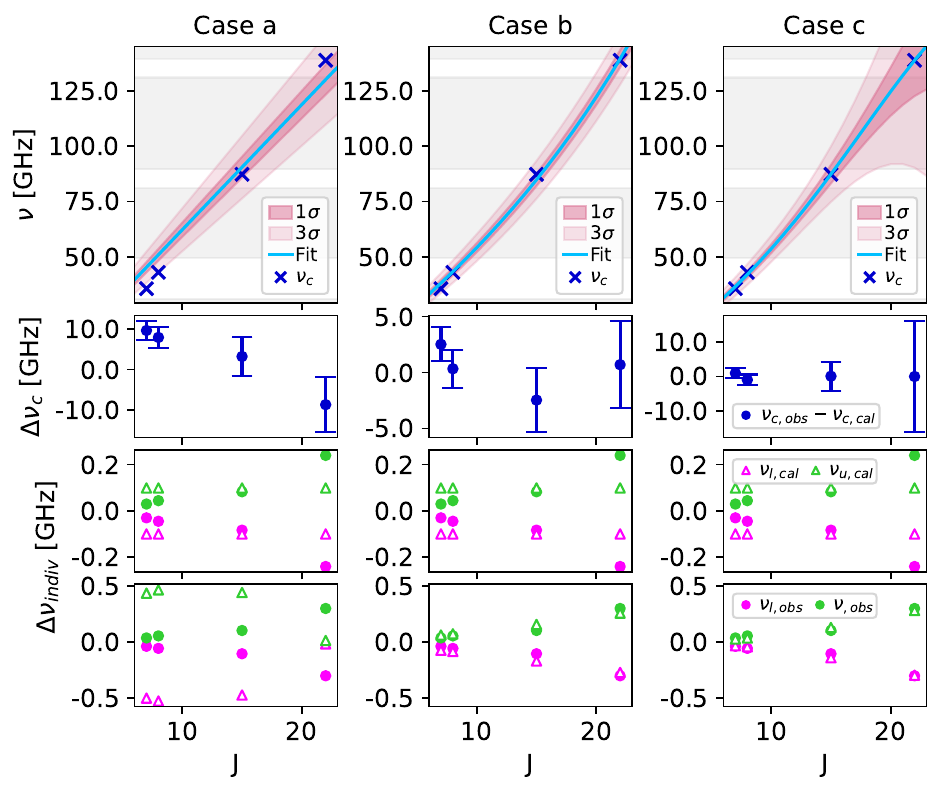}
	\caption{$J_\mathrm{init} = 7$} \label{fig:fit_UD35UD42UD87UD138_J0-7}
	\end{subfigure}
	\hspace{-2mm}
	\begin{subfigure}[]{0.33\linewidth}
	\includegraphics[width=\linewidth]{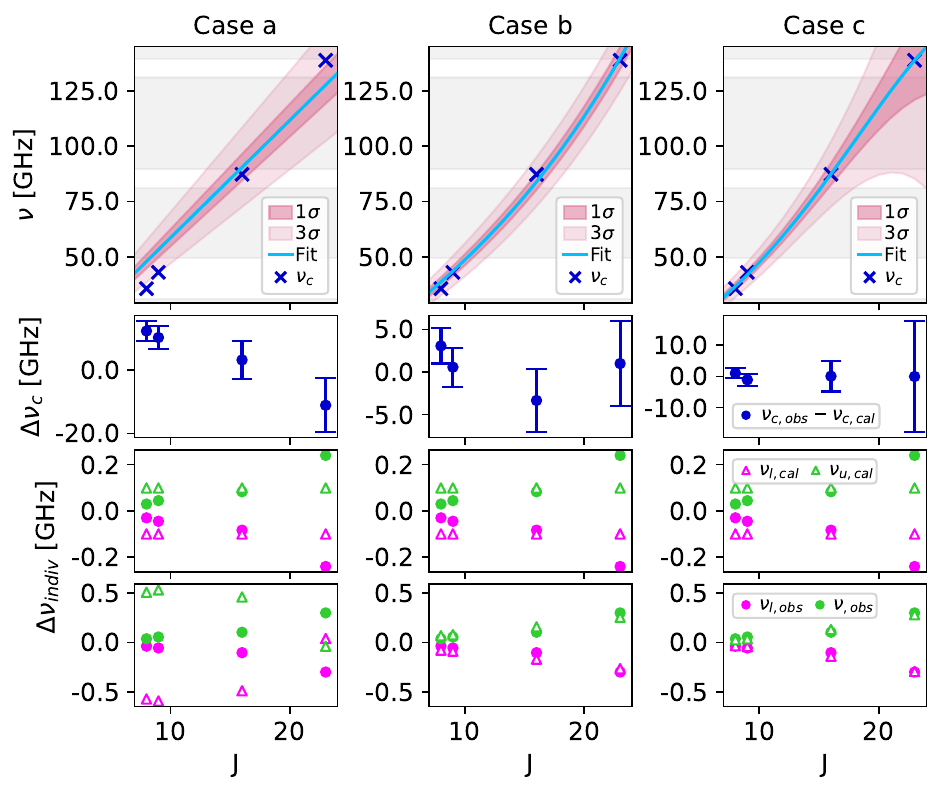}
	\caption{$J_\mathrm{init} = 8$} \label{fig:fit_UD35UD42UD87UD138_J0-8}
	\end{subfigure}
	\hspace{-2mm}
	\begin{subfigure}[]{0.33\linewidth}
	\includegraphics[width=\linewidth]{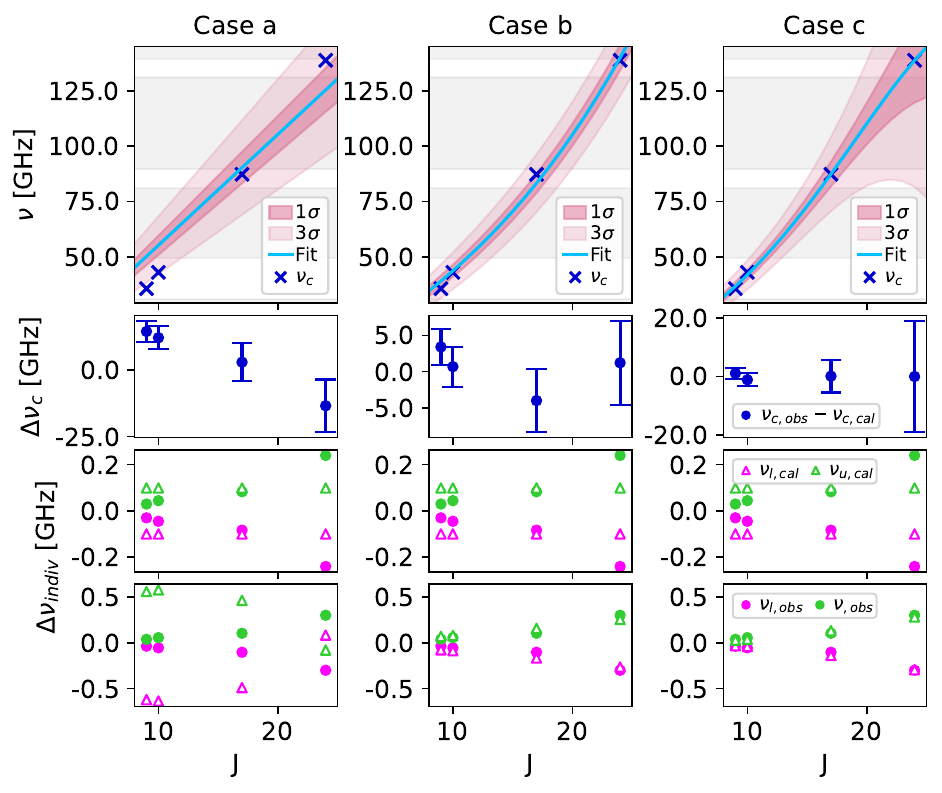}
	\caption{$J_\mathrm{init} = 9$} \label{fig:fit_UD35UD42UD87UD138_J0-9}
	\end{subfigure}
	\hspace{-2mm}
	\begin{subfigure}[]{0.33\linewidth}
	\includegraphics[width=\linewidth]{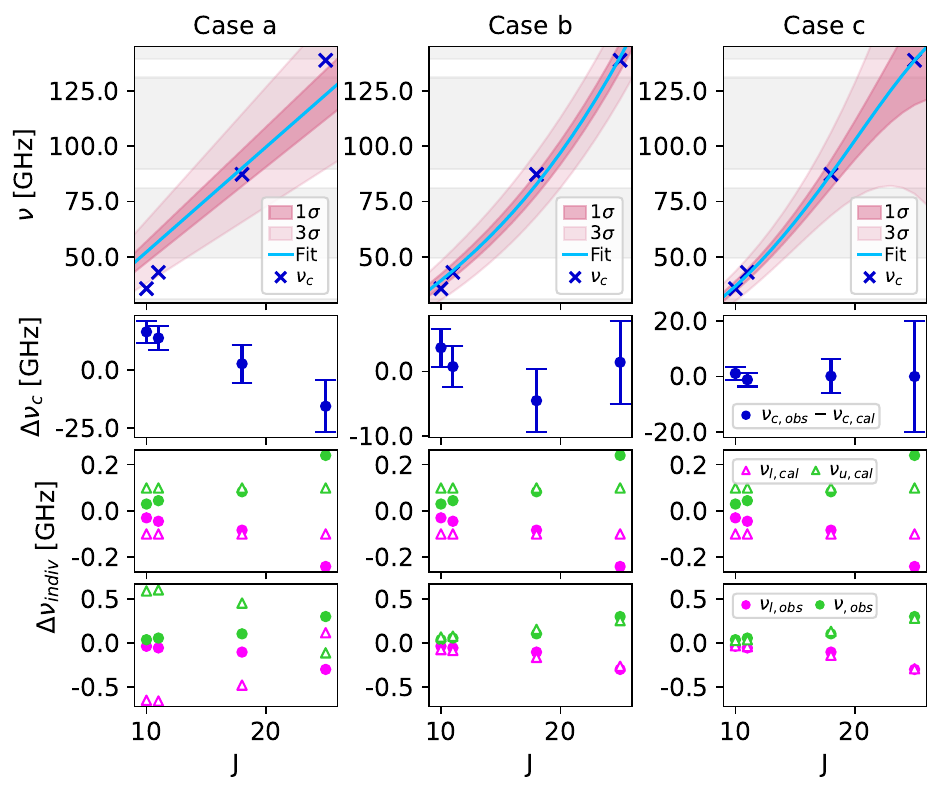}
	\caption{$J_\mathrm{init} = 10$} \label{fig:fit_UD35UD42UD87UD138_J0-10}
	\end{subfigure}
	\hspace{-2mm}
	\begin{subfigure}[]{0.33\linewidth}
	\includegraphics[width=\linewidth]{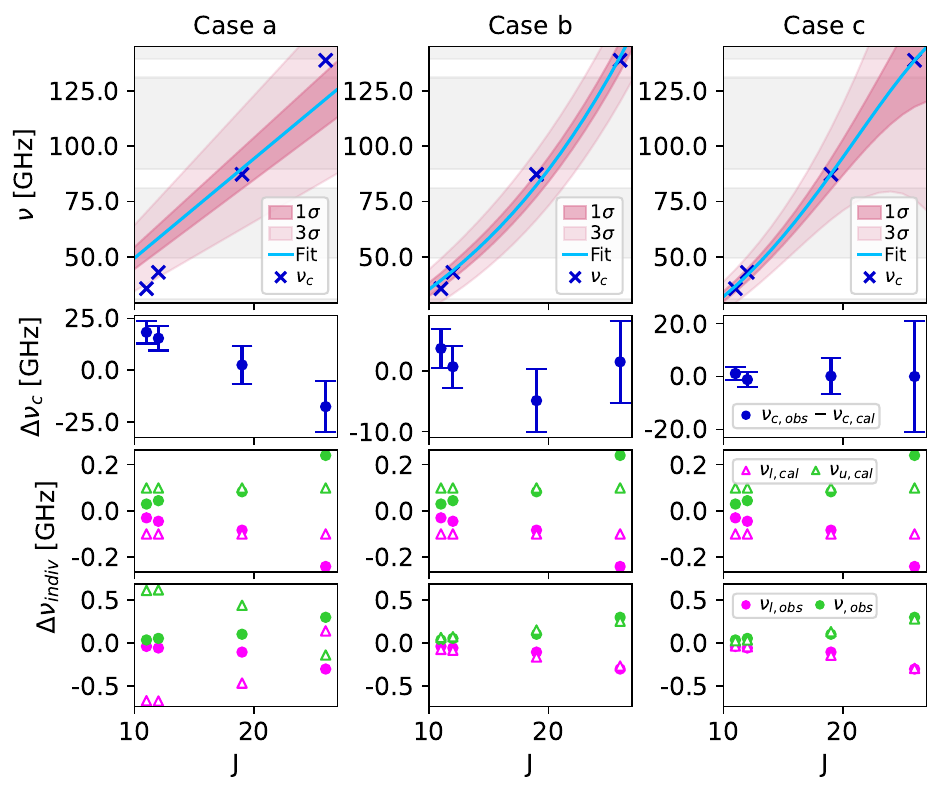}
	\caption{$J_\mathrm{init} = 11$} \label{fig:fit_UD35UD42UD87UD138_J0-11}
	\end{subfigure}
	\caption{Fitting results of UD35556, UD42932, UD87233, and UD138755 using Eqs.~\ref{eq:freq_RotTrans} and \ref{eq:freq_RotTrans_G}.}
	\label{fig:FitSummary-fit_UD35UD42UD87UD138_1}
\end{figure*}

The whole fits of the four combinations of doublets explained in Sec.~\ref{sec:linear_rotor_model} are plotted in Figs.~\ref{fig:FitSummary-fit_UD35UD42UD85_1}, \ref{fig:FitSummary-fit_UD35UD42UD87_1}, \ref{fig:FitSummary-fit_UD35UD42UD85UD138_1}, and \ref{fig:FitSummary-fit_UD35UD42UD87UD138_1}. The parameters can be found in Tables \ref{tab:pattern1_params}, \ref{tab:pattern2_params}, \ref{tab:pattern3_params}, and \ref{tab:pattern4_params}.

\input{pattern1234_table.tex}

\section{Loomis-Woods diagrams of the UFs}
\label{sec:LW_diagrams}

\begin{figure*}[ht]
	\centering
	\includegraphics[width=0.7\linewidth]{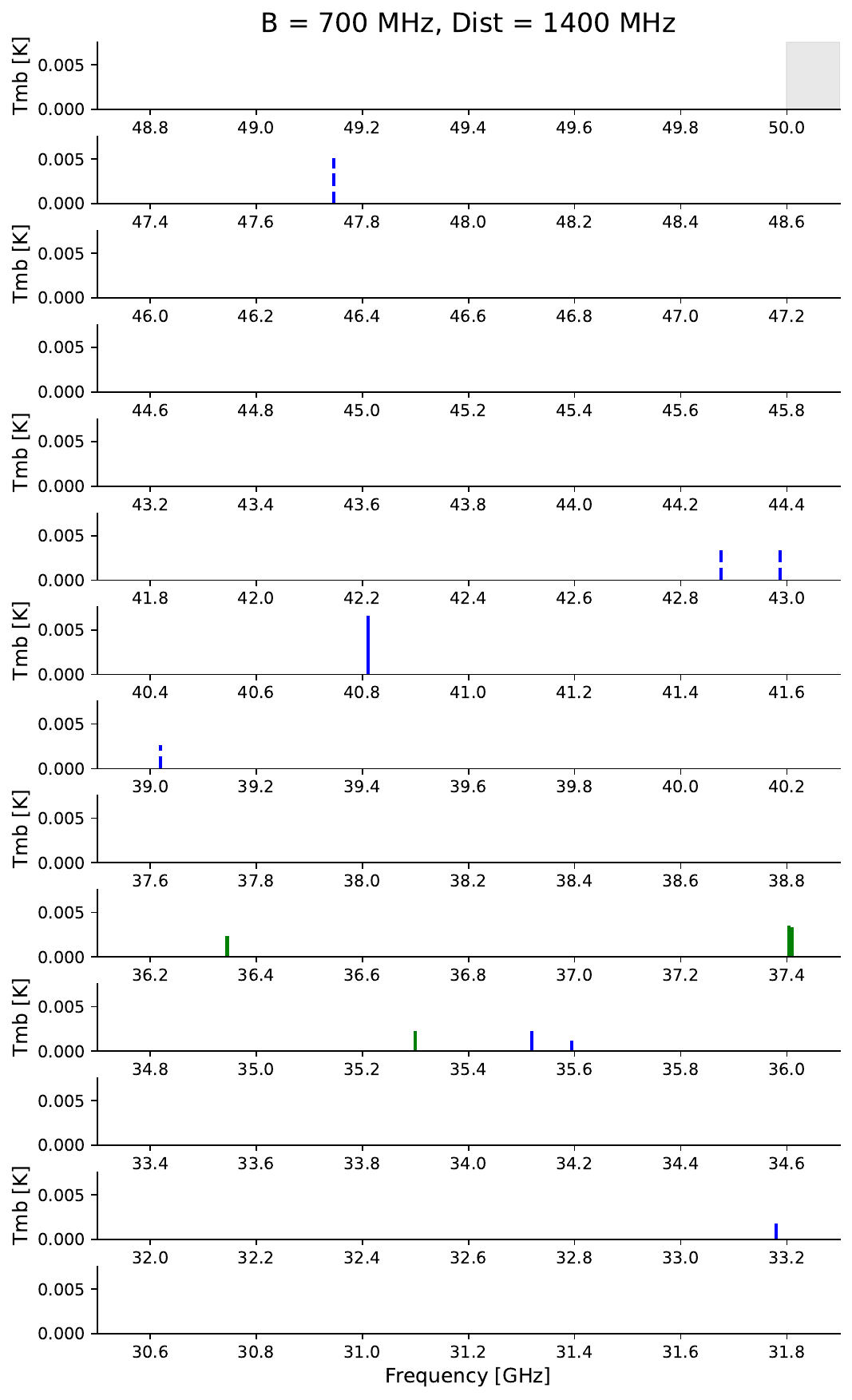}
	\caption{Loomis-Woods diagram assuming $B \sim 700$\mhz. The UFs are plotted with vertical lines to point their frequencies. Blue lines are blended UFs, while green are isolated UFs. Solid lines show detected UFs, otherwise they are tentative.} \label{fig:LW_700MHz}
\end{figure*}

\begin{figure*}[ht]
	\centering
	\includegraphics[width=0.7\linewidth]{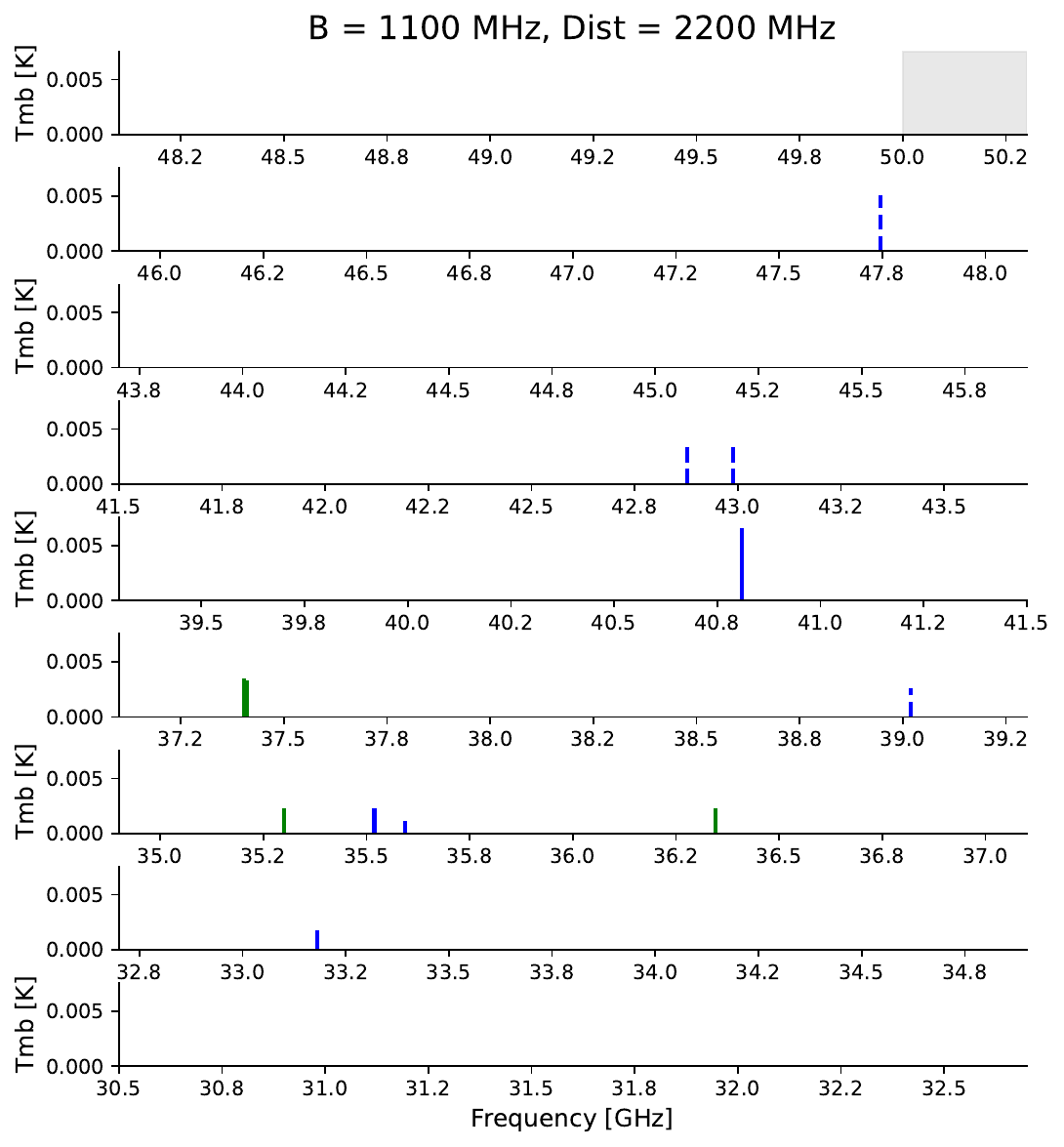}
	\caption{Loomis-Woods diagram assuming $B \sim 1100$\mhz. The same color code of Fig.~\ref{fig:LW_700MHz} is used.} \label{fig:LW_1100MHz}
\end{figure*}

\begin{figure*}[ht]
	\centering
	\includegraphics[width=0.7\linewidth]{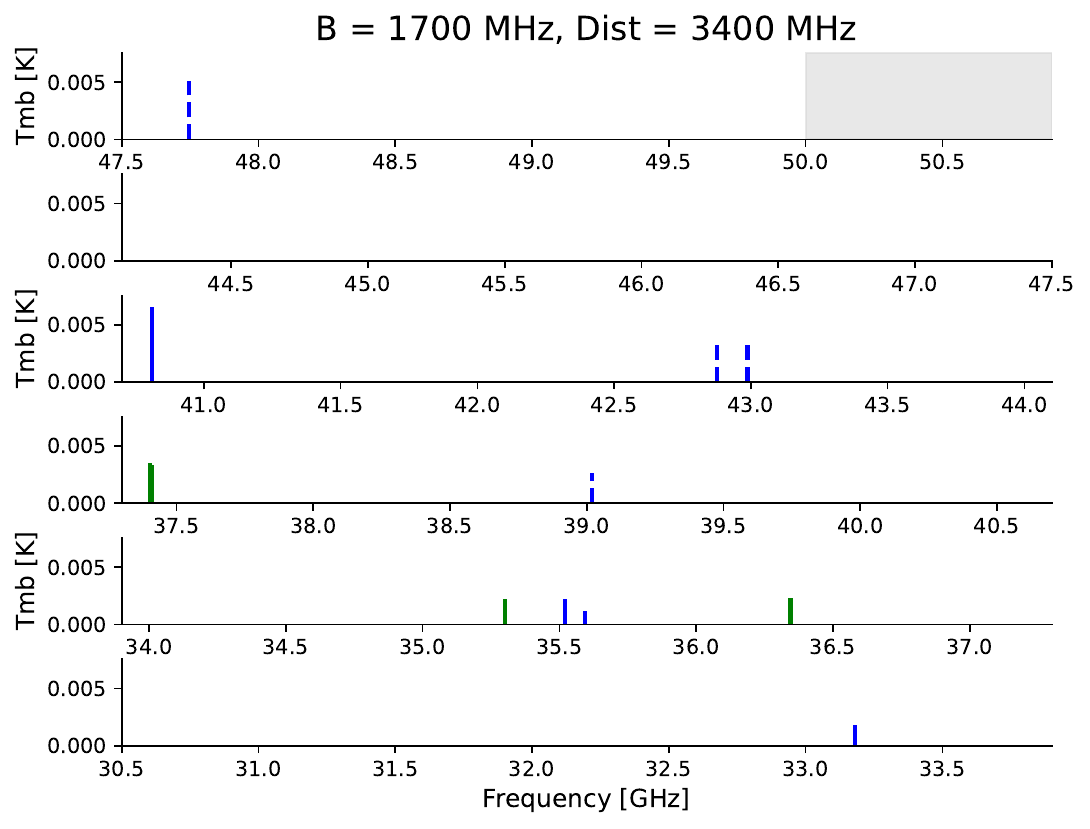}
	\caption{Loomis-Woods diagram assuming $B \sim 1700$\mhz. The same color code of Fig.~\ref{fig:LW_700MHz} is used.} \label{fig:LW_1700MHz}
\end{figure*}

\begin{figure*}[ht]
	\centering
	\includegraphics[width=0.7\linewidth]{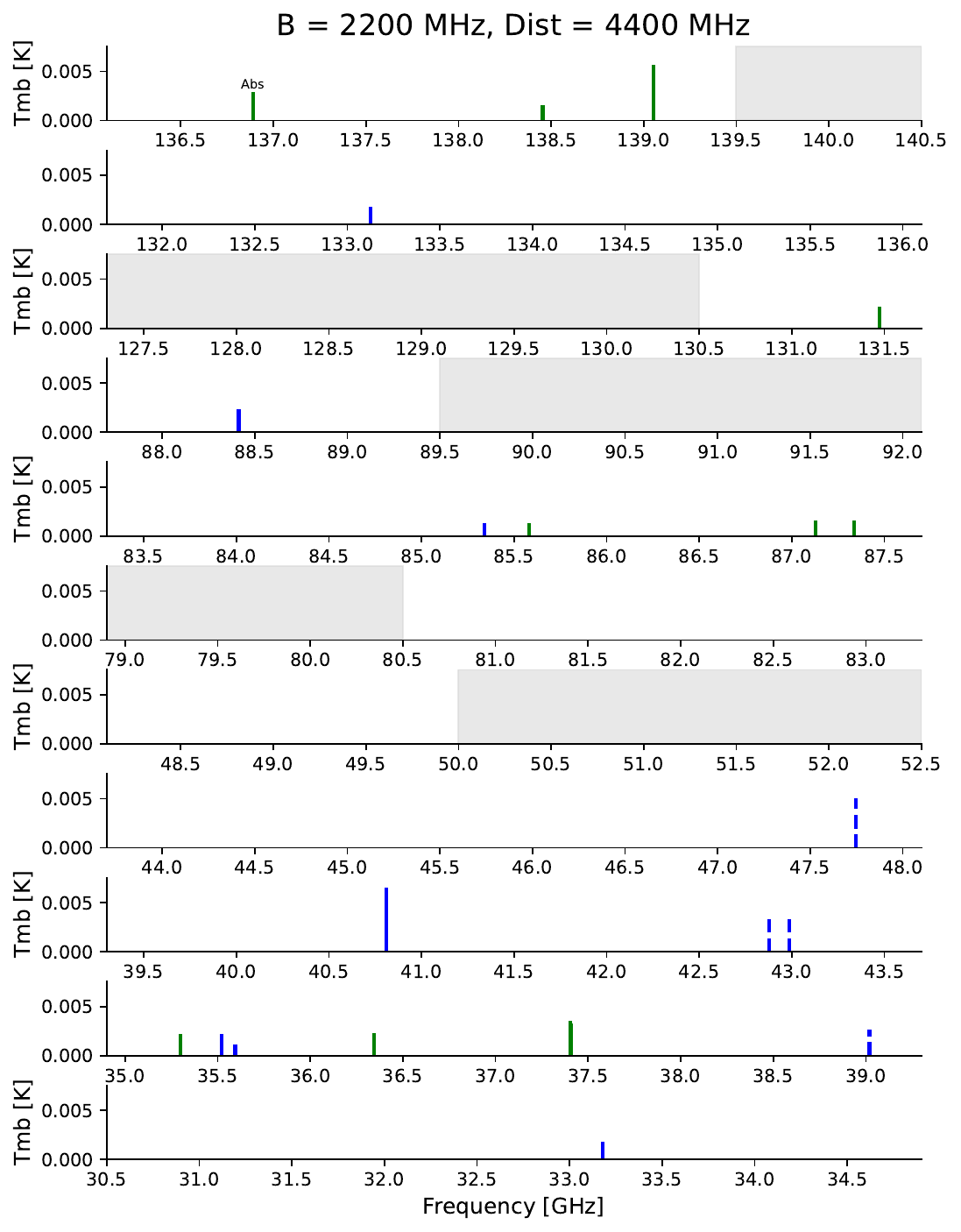}
	\caption{Loomis-Woods diagram assuming $B \sim 2200$\mhz. The same color code of Fig.~\ref{fig:LW_700MHz} is used.} \label{fig:LW_2200MHz}
\end{figure*}

\begin{figure*}[ht]
	\centering
	\includegraphics[width=0.7\linewidth]{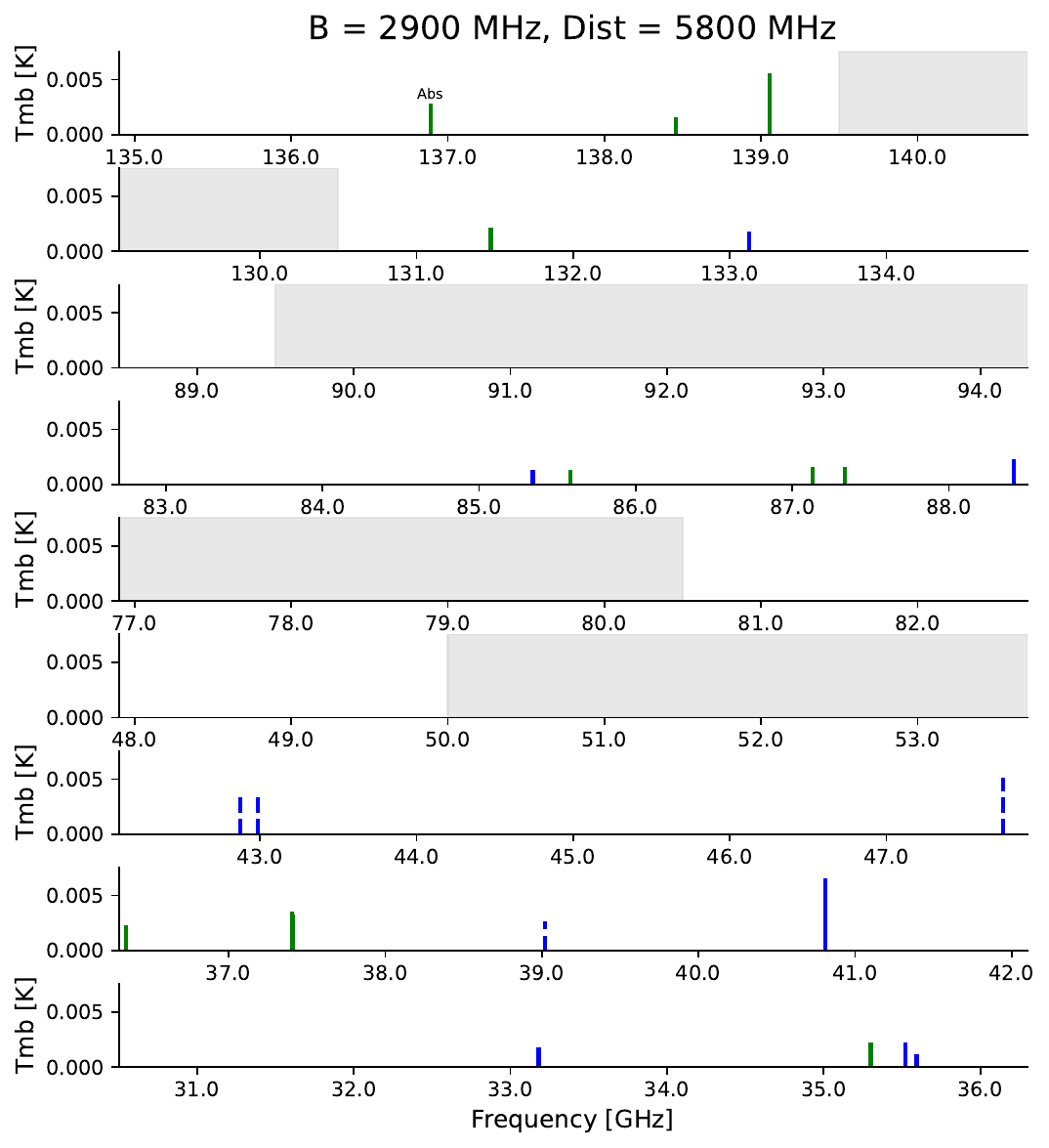}
	\caption{Loomis-Woods diagram assuming $B \sim 2900$\mhz. The same color code of Fig.~\ref{fig:LW_700MHz} is used.} \label{fig:LW_2900MHz}
\end{figure*}

\begin{figure*}[ht]
	\centering
	\includegraphics[width=0.7\linewidth]{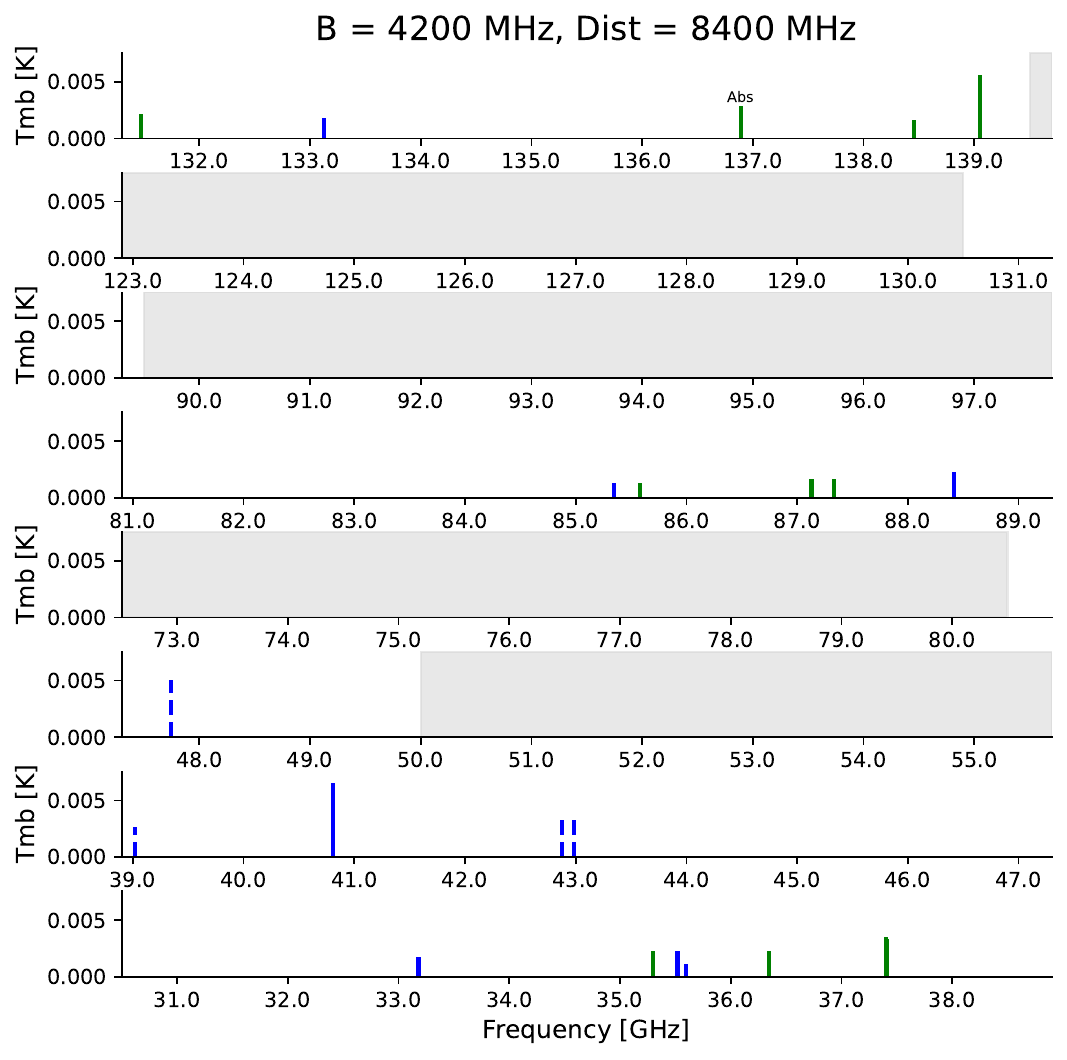}
	\caption{Loomis-Woods diagram assuming $B \sim 4200$\mhz. The same color code of Fig.~\ref{fig:LW_700MHz} is used.} \label{fig:LW_420MHz}
\end{figure*}

\normalsize
Figs.~\ref{fig:LW_700MHz}, \ref{fig:LW_1100MHz}, \ref{fig:LW_1700MHz}, \ref{fig:LW_2200MHz}, \ref{fig:LW_2900MHz}, and \ref{fig:LW_420MHz} are some examples of the Loomis-Woods diagrams that we used to search for correlations among the UFs of Table \ref{table:ufs_frequencies}.

\end{appendix}

\end{document}

%% file: pattern1234_table.tex
\onecolumn
\scriptsize
\begin{landscape}
\begin{longtable}{c | cccc | ccccc}
	\caption{Fitting results for pattern 1 of the doublets}\label{tab:pattern1_params}\\
	\hline\hline
 	& \multicolumn{4}{c}{Case a} & \multicolumn{5}{c}{Case b} \\
	\cline{2-5}\cline{6-10}
	$J_0$ & $B$ [MHz] & $\gamma$ [MHz]$^a$ & $\gamma$ [MHz]$^b$ & $\gamma_l$ [MHz] & $B$ [MHz] & $D$ [MHz] & $\gamma$ [MHz]$^a$ & $\gamma$ [MHz] & $\gamma_l$ [MHz] \\
	\hline
	\endfirsthead

	\caption[]{(continued)}\\
	\hline\hline
	 & \multicolumn{4}{c}{Case a} & \multicolumn{5}{c}{Case b} \\
	\cline{2-5}\cline{6-10}
	$J_0$ & $B$ [MHz] & $\gamma$ [MHz]$^a$ & $\gamma$ [MHz]$^b$ & $\gamma_l$ [MHz] & $B$ [MHz] & $D$ [MHz] & $\gamma$ [MHz]$^a$ & $\gamma$ [MHz] & $\gamma_l$ [MHz] \\
	\hline
	\endhead

	\hline
	\endfoot

	\hline\hline
	\endlastfoot

		0 & $5177.8 \pm 1314.5$ & 141.9 $\pm$ 17804.4 & -20761.7 $\pm$ 22726.9 & 2325.3 $\pm$ 2574.6 & $12509.8 \pm 2821.7$ & $47.93 \pm 17.95$ & 141.9 $\pm$ 6243.4 & -2653.4 $\pm$ 8660.4 & 300.0 $\pm$ 981.1 \\
		1 & $4666.1 \pm 794.2$ & 141.9 $\pm$ 12331.0 & -4995.1 $\pm$ 10666.9 & 511.8 $\pm$ 1114.3 & $7983.7 \pm 772.4$ & $18.56 \pm 4.10$ & 141.9 $\pm$ 2661.9 & -222.9 $\pm$ 2362.3 & 25.3 $\pm$ 246.8 \\
		2 & $4172.9 \pm 455.2$ & 141.9 $\pm$ 8034.3 & -1611.1 $\pm$ 5325.3 & 154.1 $\pm$ 519.1 & $5786.8 \pm 243.1$ & $7.87 \pm 1.10$ & 141.9 $\pm$ 1115.2 & -10.2 $\pm$ 747.5 & 3.1 $\pm$ 72.9 \\
		3 & $3736.3 \pm 233.7$ & 141.9 $\pm$ 4643.1 & -556.7 $\pm$ 2570.5 & 50.9 $\pm$ 235.1 & $4493.1 \pm 53.5$ & $3.24 \pm 0.21$ & 141.9 $\pm$ 302.1 & 18.3 $\pm$ 167.5 & 0.2 $\pm$ 15.3 \\
		4 & $3361.7 \pm 88.0$ & 141.9 $\pm$ 1949.7 & -147.3 $\pm$ 950.7 & 13.8 $\pm$ 81.8 & $3632.0 \pm 27.0$ & $1.02 \pm 0.09$ & 141.9 $\pm$ 179.4 & 20.0 $\pm$ 86.3 & -0.1 $\pm$ 7.4 \\
		5 & $3043.3 \pm 22.1$ & 141.9 $\pm$ 541.5 & 33.8 $\pm$ 240.1 & -1.3 $\pm$ 19.4 & $3011.4 \pm 64.7$ & $-0.11 \pm 0.20$ & 141.9 $\pm$ 475.8 & 17.0 $\pm$ 211.2 & 0.0 $\pm$ 17.1 \\
		6 & $2772.6 \pm 81.3$ & 141.9 $\pm$ 2184.2 & 120.1 $\pm$ 903.4 & -7.7 $\pm$ 69.0 & $2538.8 \pm 83.1$ & $-0.69 \pm 0.22$ & 141.9 $\pm$ 669.6 & 13.6 $\pm$ 277.3 & 0.2 $\pm$ 21.2 \\
		7 & $2541.4 \pm 126.2$ & 141.9 $\pm$ 3691.5 & 162.7 $\pm$ 1447.2 & -10.3 $\pm$ 104.4 & $2164.5 \pm 92.0$ & $-0.99 \pm 0.22$ & 141.9 $\pm$ 799.3 & 10.4 $\pm$ 313.6 & 0.3 $\pm$ 22.6 \\
		8 & $2342.8 \pm 157.2$ & 141.9 $\pm$ 4976.3 & 183.4 $\pm$ 1870.8 & -11.2 $\pm$ 127.6 & $1859.1 \pm 96.1$ & $-1.14 \pm 0.21$ & 141.9 $\pm$ 887.2 & 7.7 $\pm$ 333.7 & 0.4 $\pm$ 22.8 \\
		9 & $2170.8 \pm 178.3$ & 141.9 $\pm$ 6078.2 & 192.7 $\pm$ 2210.5 & -11.3 $\pm$ 142.7 & $1604.3 \pm 97.6$ & $-1.19 \pm 0.19$ & 141.9 $\pm$ 947.1 & 5.4 $\pm$ 344.6 & 0.4 $\pm$ 22.2 \\
		10 & $2020.8 \pm 192.4$ & 141.9 $\pm$ 7031.8 & 195.6 $\pm$ 2490.1 & -11.0 $\pm$ 152.5 & $1387.8 \pm 97.6$ & $-1.20 \pm 0.17$ & 141.9 $\pm$ 987.7 & 3.5 $\pm$ 349.9 & 0.5 $\pm$ 21.4 \\
		11 & $1889.2 \pm 201.6$ & 141.9 $\pm$ 7863.9 & 194.9 $\pm$ 2725.4 & -10.5 $\pm$ 158.6 & $1201.1 \pm 96.8$ & $-1.17 \pm 0.15$ & 141.9 $\pm$ 1014.7 & 1.8 $\pm$ 351.7 & 0.5 $\pm$ 20.5 \\
\end{longtable}

\begin{longtable}{c | cccc | ccccc}
	\caption{Fitting results for pattern 2 of the doublets}\label{tab:pattern2_params}\\
	\hline\hline
 	& \multicolumn{4}{c}{Case a} & \multicolumn{5}{c}{Case b} \\
	\cline{2-5}\cline{6-10}
	$J_0$ & $B$ [MHz] & $\gamma$ [MHz]$^a$ & $\gamma$ [MHz]$^b$ & $\gamma_l$ [MHz] & $B$ [MHz] & $D$ [MHz] & $\gamma$ [MHz]$^a$ & $\gamma$ [MHz] & $\gamma_l$ [MHz] \\
	\hline
	\endfirsthead

	\caption[]{(continued)}\\
	\hline\hline
	 & \multicolumn{4}{c}{Case a} & \multicolumn{5}{c}{Case b} \\
	\cline{2-5}\cline{6-10}
	$J_0$ & $B$ [MHz] & $\gamma$ [MHz]$^a$ & $\gamma$ [MHz]$^b$ & $\gamma_l$ [MHz] & $B$ [MHz] & $D$ [MHz] & $\gamma$ [MHz]$^a$ & $\gamma$ [MHz] & $\gamma_l$ [MHz] \\
	\hline
	\endhead

	\hline
	\endfoot

	\hline\hline
	\endlastfoot

		0 & $5270.5 \pm 1299.3$ & 131.0 $\pm$ 17599.4 & -20522.8 $\pm$ 22465.1 & 2298.2 $\pm$ 2545.0 & $12505.5 \pm 2823.3$ & $47.30 \pm 17.96$ & 131.0 $\pm$ 6246.8 & -2653.9 $\pm$ 8665.4 & 299.6 $\pm$ 981.7 \\
		1 & $4744.4 \pm 774.8$ & 131.0 $\pm$ 12030.2 & -4871.8 $\pm$ 10406.9 & 498.9 $\pm$ 1087.1 & $7976.6 \pm 774.8$ & $18.09 \pm 4.12$ & 131.0 $\pm$ 2670.0 & -222.5 $\pm$ 2369.6 & 25.0 $\pm$ 247.5 \\
		2 & $4239.6 \pm 434.1$ & 131.0 $\pm$ 7661.8 & -1534.0 $\pm$ 5078.6 & 146.5 $\pm$ 495.0 & $5776.8 \pm 245.9$ & $7.49 \pm 1.11$ & 131.0 $\pm$ 1128.0 & -9.2 $\pm$ 756.2 & 2.8 $\pm$ 73.7 \\
		3 & $3793.7 \pm 212.2$ & 131.0 $\pm$ 4215.1 & -501.9 $\pm$ 2333.8 & 45.8 $\pm$ 213.4 & $4480.3 \pm 56.6$ & $2.94 \pm 0.22$ & 131.0 $\pm$ 318.9 & 19.7 $\pm$ 177.3 & -0.1 $\pm$ 16.2 \\
		4 & $3411.8 \pm 66.8$ & 131.0 $\pm$ 1479.3 & -105.0 $\pm$ 721.6 & 10.1 $\pm$ 62.1 & $3616.6 \pm 23.7$ & $0.77 \pm 0.08$ & 131.0 $\pm$ 157.0 & 21.7 $\pm$ 75.7 & -0.4 $\pm$ 6.5 \\
		5 & $3087.4 \pm 36.1$ & 131.0 $\pm$ 884.1 & 68.2 $\pm$ 391.9 & -4.1 $\pm$ 31.7 & $2993.5 \pm 61.2$ & $-0.31 \pm 0.19$ & 131.0 $\pm$ 450.1 & 19.0 $\pm$ 199.9 & -0.3 $\pm$ 16.2 \\
		6 & $2811.9 \pm 100.0$ & 131.0 $\pm$ 2684.6 & 149.3 $\pm$ 1110.4 & -10.0 $\pm$ 84.8 & $2518.7 \pm 79.5$ & $-0.87 \pm 0.21$ & 131.0 $\pm$ 640.6 & 15.7 $\pm$ 265.3 & -0.1 $\pm$ 20.3 \\
		7 & $2576.8 \pm 144.4$ & 131.0 $\pm$ 4223.3 & 188.1 $\pm$ 1655.7 & -12.2 $\pm$ 119.4 & $2142.4 \pm 88.4$ & $-1.14 \pm 0.21$ & 131.0 $\pm$ 767.2 & 12.7 $\pm$ 301.1 & 0.0 $\pm$ 21.7 \\
		8 & $2374.8 \pm 174.6$ & 131.0 $\pm$ 5529.1 & 206.1 $\pm$ 2078.6 & -12.8 $\pm$ 141.7 & $1835.2 \pm 92.4$ & $-1.27 \pm 0.20$ & 131.0 $\pm$ 852.4 & 10.1 $\pm$ 320.7 & 0.1 $\pm$ 21.9 \\
		9 & $2200.0 \pm 195.0$ & 131.0 $\pm$ 6647.7 & 213.3 $\pm$ 2417.5 & -12.7 $\pm$ 156.1 & $1578.7 \pm 93.7$ & $-1.31 \pm 0.18$ & 131.0 $\pm$ 909.9 & 7.9 $\pm$ 331.1 & 0.2 $\pm$ 21.4 \\
		10 & $2047.7 \pm 208.4$ & 131.0 $\pm$ 7615.0 & 214.6 $\pm$ 2696.6 & -12.2 $\pm$ 165.2 & $1360.7 \pm 93.7$ & $-1.30 \pm 0.16$ & 131.0 $\pm$ 948.2 & 6.1 $\pm$ 335.9 & 0.3 $\pm$ 20.6 \\
		11 & $1914.1 \pm 216.9$ & 131.0 $\pm$ 8458.9 & 212.6 $\pm$ 2931.6 & -11.5 $\pm$ 170.6 & $1172.6 \pm 92.8$ & $-1.26 \pm 0.15$ & 131.0 $\pm$ 973.1 & 4.5 $\pm$ 337.4 & 0.3 $\pm$ 19.6 \\
\end{longtable}
\begin{longtable}{c | cccc | ccccc}
	\caption{Fitting results for pattern 3 of the doublets}\label{tab:pattern3_params}\\
	\hline\hline
	& \multicolumn{4}{c}{Case a} & \multicolumn{5}{c}{Case b} \\
	\cline{2-5}\cline{6-10}
	$J_0$ & $B$ [MHz] & $\gamma$ [MHz]$^a$ & $\gamma$ [MHz]$^b$ & $\gamma_l$ [MHz] & $B$ [MHz] & $D$ [MHz] & $\gamma$ [MHz]$^a$ & $\gamma$ [MHz] & $\gamma_l$ [MHz] \\
	\hline
	\endfirsthead
	
	\caption[]{(continued)}\\
	\hline\hline
	& \multicolumn{4}{c}{Case a} & \multicolumn{5}{c}{Case b} \\
	\cline{2-5}\cline{6-10}
	$J_0$ & $B$ [MHz] & $\gamma$ [MHz]$^a$ & $\gamma$ [MHz]$^b$ & $\gamma_l$ [MHz] & $B$ [MHz] & $D$ [MHz] & $\gamma$ [MHz]$^a$ & $\gamma$ [MHz] & $\gamma_l$ [MHz] \\
	\hline
	\endhead
	
	\hline
	\endfoot
	
	\hline\hline
	\endlastfoot
	
	0 & $4547.7 \pm 578.0$ & 256.4 $\pm$ 13996.0 & -3182.7 $\pm$ 7702.6 & 215.1 $\pm$ 527.2 & $5795.8 \pm 1799.7$ & $2.96 \pm 4.00$ & 256.4 $\pm$ 12401.4 & -2337.2 $\pm$ 6855.4 & 158.6 $\pm$ 469.2 \\
	1 & $4245.5 \pm 375.8$ & 256.4 $\pm$ 9866.7 & -1534.0 $\pm$ 4731.8 & 99.3 $\pm$ 307.9 & $5163.2 \pm 1053.9$ & $1.97 \pm 2.11$ & 256.4 $\pm$ 8230.5 & -939.4 $\pm$ 3963.0 & 61.7 $\pm$ 257.9 \\
	2 & $3953.3 \pm 224.0$ & 256.4 $\pm$ 6357.9 & -670.8 $\pm$ 2662.9 & 42.3 $\pm$ 165.5 & $4489.6 \pm 593.6$ & $1.05 \pm 1.08$ & 256.4 $\pm$ 5231.4 & -362.0 $\pm$ 2197.6 & 23.7 $\pm$ 136.5 \\
	3 & $3680.2 \pm 113.5$ & 256.4 $\pm$ 3472.5 & -236.2 $\pm$ 1292.5 & 15.3 $\pm$ 76.9 & $3882.2 \pm 316.6$ & $0.36 \pm 0.52$ & 256.4 $\pm$ 3117.3 & -134.1 $\pm$ 1162.2 & 9.4 $\pm$ 69.2 \\
	4 & $3429.7 \pm 52.7$ & 256.4 $\pm$ 1732.2 & -14.0 $\pm$ 583.9 & 2.3 $\pm$ 33.3 & $3364.1 \pm 150.9$ & $-0.11 \pm 0.23$ & 256.4 $\pm$ 1641.9 & -43.3 $\pm$ 553.2 & 3.9 $\pm$ 31.6 \\
	5 & $3202.3 \pm 74.2$ & 256.4 $\pm$ 2610.1 & 101.8 $\pm$ 808.1 & -4.0 $\pm$ 44.4 & $2928.6 \pm 56.9$ & $-0.42 \pm 0.08$ & 256.4 $\pm$ 679.6 & -7.2 $\pm$ 209.4 & 1.8 $\pm$ 11.5 \\
	6 & $2996.9 \pm 116.3$ & 256.4 $\pm$ 4366.0 & 162.4 $\pm$ 1259.8 & -7.0 $\pm$ 66.5 & $2562.3 \pm 40.8$ & $-0.61 \pm 0.05$ & 256.4 $\pm$ 528.2 & 6.4 $\pm$ 151.0 & 0.9 $\pm$ 8.0 \\
	7 & $2811.6 \pm 151.7$ & 256.4 $\pm$ 6058.9 & 193.3 $\pm$ 1647.9 & -8.3 $\pm$ 83.8 & $2251.7 \pm 68.4$ & $-0.73 \pm 0.08$ & 256.4 $\pm$ 940.7 & 10.4 $\pm$ 255.3 & 0.6 $\pm$ 13.0 \\
	8 & $2644.5 \pm 179.3$ & 256.4 $\pm$ 7594.8 & 207.8 $\pm$ 1964.7 & -8.8 $\pm$ 96.2 & $1985.7 \pm 89.6$ & $-0.80 \pm 0.10$ & 256.4 $\pm$ 1307.3 & 10.3 $\pm$ 337.9 & 0.5 $\pm$ 16.5 \\
	9 & $2493.6 \pm 200.4$ & 256.4 $\pm$ 8975.3 & 213.0 $\pm$ 2224.6 & -8.8 $\pm$ 105.0 & $1755.7 \pm 102.9$ & $-0.83 \pm 0.10$ & 256.4 $\pm$ 1582.6 & 8.6 $\pm$ 392.1 & 0.5 $\pm$ 18.5 \\
	10 & $2357.0 \pm 216.2$ & 256.4 $\pm$ 10216.0 & 212.8 $\pm$ 2440.9 & -8.5 $\pm$ 111.2 & $1554.8 \pm 110.5$ & $-0.84 \pm 0.11$ & 256.4 $\pm$ 1782.7 & 6.3 $\pm$ 425.8 & 0.6 $\pm$ 19.4 \\
	11 & $2233.2 \pm 228.0$ & 256.4 $\pm$ 11334.2 & 209.7 $\pm$ 2623.6 & -8.1 $\pm$ 115.3 & $1377.7 \pm 114.4$ & $-0.83 \pm 0.10$ & 256.4 $\pm$ 1924.6 & 3.8 $\pm$ 445.4 & 0.6 $\pm$ 19.6 \\
\end{longtable}

\begin{longtable}{c | cccccc}
	\hline\hline
	& \multicolumn{6}{c}{Case c} \\
	\cline{2-7}
	$J_0$ & $B$ [MHz] & $D$ [MHz] & $H$ [MHz] & $\gamma$ [MHz]$^a$ & $\gamma$ [MHz]$^b$ & $\gamma_l$ [MHz] \\
	\hline
	\endfirsthead
	
	\hline\hline
	& \multicolumn{6}{c}{Case c} \\
	\cline{2-7}
	$J_0$ & $B$ [MHz] & $D$ [MHz] & $H$ [MHz] & $\gamma$ [MHz]$^a$ & $\gamma$ [MHz]$^b$ & $\gamma_l$ [MHz] \\
	\hline
	\endhead
	
	\hline
	\endfoot
	
	\hline\hline
	\endlastfoot
	
	0 & $12624.4 \pm 2822.0$ & $63.73 \pm 23.81$ & $0.1236 \pm 0.0483$ & 256.4 $\pm$ 4510.6 & -315.1 $\pm$ 2513.9 & 23.5 $\pm$ 172.1 \\
	
	1 & $8084.4 \pm 761.9$ & $25.52 \pm 5.54$ & $0.0428 \pm 0.0100$ & 256.4 $\pm$ 1873.6 & -40.1 $\pm$ 905.5 & 4.7 $\pm$ 58.9 \\
	
	2 & $5870.0 \pm 232.2$ & $11.26 \pm 1.49$ & $0.0168 \pm 0.0024$ & 256.4 $\pm$ 750.3 & 3.7 $\pm$ 314.3 & 1.7 $\pm$ 19.5 \\
	
	3 & $4555.5 \pm 43.2$ & $4.92 \pm 0.25$ & $0.0068 \pm 0.0004$ & 256.4 $\pm$ 183.9 & 9.9 $\pm$ 62.3 & 1.2 $\pm$ 3.7 \\
	
	4 & $3671.6 \pm 37.4$ & $1.78 \pm 0.19$ & $0.0025 \pm 0.0003$ & 256.4 $\pm$ 181.1 & 8.6 $\pm$ 55.2 & 1.1 $\pm$ 3.2 \\
	
	5 & $3027.0 \pm 75.8$ & $0.13 \pm 0.35$ & $0.0007 \pm 0.0004$ & 256.4 $\pm$ 369.8 & 5.8 $\pm$ 112.1 & 1.1 $\pm$ 6.2 \\
	
	6 & $2530.3 \pm 95.6$ & $-0.77 \pm 0.40$ & $-0.0002 \pm 0.0004$ & 256.4 $\pm$ 490.7 & 3.0 $\pm$ 140.0 & 1.1 $\pm$ 7.4 \\
	
	7 & $2132.3 \pm 106.5$ & $-1.26 \pm 0.41$ & $-0.0005 \pm 0.0004$ & 256.4 $\pm$ 567.3 & 0.5 $\pm$ 153.0 & 1.1 $\pm$ 7.8 \\
	
	8 & $1804.0 \pm 113.0$ & $-1.53 \pm 0.39$ & $-0.0007 \pm 0.0004$ & 256.4 $\pm$ 615.8 & -1.7 $\pm$ 158.2 & 1.1 $\pm$ 7.7 \\
	
	9 & $1527.4 \pm 117.2$ & $-1.65 \pm 0.37$ & $-0.0007 \pm 0.0003$ & 256.4 $\pm$ 645.8 & -3.6 $\pm$ 159.0 & 1.1 $\pm$ 7.5 \\
	
	10 & $1290.2 \pm 120.2$ & $-1.69 \pm 0.34$ & $-0.0007 \pm 0.0003$ & 256.4 $\pm$ 663.6 & -5.3 $\pm$ 157.6 & 1.1 $\pm$ 7.2 \\
	
	11 & $1084.0 \pm 122.6$ & $-1.69 \pm 0.32$ & $-0.0006 \pm 0.0002$ & 256.4 $\pm$ 673.2 & -6.8 $\pm$ 154.9 & 1.1 $\pm$ 6.8 \\
\end{longtable}
\begin{longtable}{c | cccc | ccccc}
	\caption{Fitting results for pattern 4 of the doublets}\label{tab:pattern4_params}\\
	\hline\hline
	& \multicolumn{4}{c}{Case a} & \multicolumn{5}{c}{Case b} \\
	\cline{2-5}\cline{6-10}
	$J_0$ & $B$ [MHz] & $\gamma$ [MHz]$^a$ & $\gamma$ [MHz]$^b$ & $\gamma_l$ [MHz] & $B$ [MHz] & $D$ [MHz] & $\gamma$ [MHz]$^a$ & $\gamma$ [MHz] & $\gamma_l$ [MHz] \\
	\hline
	\endfirsthead
	
	\caption[]{(continued)}\\
	\hline\hline
	& \multicolumn{4}{c}{Case a} & \multicolumn{5}{c}{Case b} \\
	\cline{2-5}\cline{6-10}
	$J_0$ & $B$ [MHz] & $\gamma$ [MHz]$^a$ & $\gamma$ [MHz]$^b$ & $\gamma_l$ [MHz] & $B$ [MHz] & $D$ [MHz] & $\gamma$ [MHz]$^a$ & $\gamma$ [MHz] & $\gamma_l$ [MHz] \\
	\hline
	\endhead
	
	\hline
	\endfoot
	
	\hline\hline
	\endlastfoot
	
	0 & $4571.1 \pm 581.2$ & 248.2 $\pm$ 14073.1 & -3221.2 $\pm$ 7743.6 & 217.6 $\pm$ 530.0 & $5922.8 \pm 1769.4$ & $3.21 \pm 3.93$ & 248.2 $\pm$ 12193.0 & -2305.4 $\pm$ 6739.7 & 156.5 $\pm$ 461.3 \\
	1 & $4267.5 \pm 377.0$ & 248.2 $\pm$ 9897.9 & -1557.8 $\pm$ 4745.7 & 100.8 $\pm$ 308.8 & $5267.4 \pm 1014.9$ & $2.15 \pm 2.03$ & 248.2 $\pm$ 7926.0 & -910.1 $\pm$ 3816.1 & 59.8 $\pm$ 248.3 \\
	2 & $3974.0 \pm 222.8$ & 248.2 $\pm$ 6323.9 & -685.8 $\pm$ 2647.9 & 43.2 $\pm$ 164.5 & $4574.8 \pm 551.3$ & $1.18 \pm 1.00$ & 248.2 $\pm$ 4858.8 & -339.9 $\pm$ 2041.0 & 22.4 $\pm$ 126.8 \\
	3 & $3699.7 \pm 108.1$ & 248.2 $\pm$ 3307.5 & -246.1 $\pm$ 1230.4 & 15.8 $\pm$ 73.2 & $3952.8 \pm 273.7$ & $0.45 \pm 0.45$ & 248.2 $\pm$ 2695.5 & -118.1 $\pm$ 1004.7 & 8.5 $\pm$ 59.8 \\
	4 & $3448.0 \pm 36.6$ & 248.2 $\pm$ 1204.4 & -20.8 $\pm$ 405.5 & 2.7 $\pm$ 23.2 & $3423.4 \pm 109.0$ & $-0.04 \pm 0.16$ & 248.2 $\pm$ 1186.5 & -31.8 $\pm$ 399.3 & 3.3 $\pm$ 22.8 \\
	5 & $3219.5 \pm 63.6$ & 248.2 $\pm$ 2238.8 & 96.9 $\pm$ 692.8 & -3.8 $\pm$ 38.0 & $2979.4 \pm 29.0$ & $-0.37 \pm 0.04$ & 248.2 $\pm$ 352.3 & 1.3 $\pm$ 106.6 & 1.3 $\pm$ 5.9 \\
	6 & $3013.1 \pm 110.3$ & 248.2 $\pm$ 4139.8 & 158.7 $\pm$ 1194.4 & -6.8 $\pm$ 63.1 & $2606.3 \pm 64.5$ & $-0.57 \pm 0.08$ & 248.2 $\pm$ 829.5 & 12.7 $\pm$ 238.6 & 0.6 $\pm$ 12.6 \\
	7 & $2826.8 \pm 147.6$ & 248.2 $\pm$ 5895.2 & 190.5 $\pm$ 1603.3 & -8.2 $\pm$ 81.5 & $2290.4 \pm 99.5$ & $-0.70 \pm 0.12$ & 248.2 $\pm$ 1367.0 & 15.3 $\pm$ 371.6 & 0.4 $\pm$ 18.9 \\
	8 & $2658.9 \pm 176.3$ & 248.2 $\pm$ 7466.8 & 205.6 $\pm$ 1931.5 & -8.7 $\pm$ 94.6 & $2020.1 \pm 121.3$ & $-0.77 \pm 0.13$ & 248.2 $\pm$ 1769.1 & 14.2 $\pm$ 457.6 & 0.3 $\pm$ 22.4 \\
	9 & $2507.2 \pm 198.1$ & 248.2 $\pm$ 8871.3 & 211.2 $\pm$ 2198.8 & -8.7 $\pm$ 103.8 & $1786.6 \pm 134.1$ & $-0.81 \pm 0.14$ & 248.2 $\pm$ 2062.5 & 11.7 $\pm$ 511.2 & 0.4 $\pm$ 24.1 \\
	10 & $2370.0 \pm 214.4$ & 248.2 $\pm$ 10129.8 & 211.5 $\pm$ 2420.3 & -8.5 $\pm$ 110.2 & $1582.7 \pm 141.1$ & $-0.82 \pm 0.13$ & 248.2 $\pm$ 2274.0 & 8.8 $\pm$ 543.3 & 0.4 $\pm$ 24.7 \\
	11 & $2245.4 \pm 226.5$ & 248.2 $\pm$ 11262.0 & 208.6 $\pm$ 2606.9 & -8.1 $\pm$ 114.6 & $1403.1 \pm 144.1$ & $-0.82 \pm 0.13$ & 248.2 $\pm$ 2423.8 & 5.9 $\pm$ 561.1 & 0.5 $\pm$ 24.7 \\
\end{longtable}

\begin{longtable}{c | cccccc}
	\hline\hline
	& \multicolumn{6}{c}{Case c} \\
	\cline{2-7}
	$J_0$ & $B$ [MHz] & $D$ [MHz] & $H$ [MHz] & $\gamma$ [MHz]$^a$ & $\gamma$ [MHz]$^b$ & $\gamma_l$ [MHz] \\
	\hline
	\endfirsthead
	
	\hline\hline
	& \multicolumn{6}{c}{Case c} \\
	\cline{2-7}
	$J_0$ & $B$ [MHz] & $D$ [MHz] & $H$ [MHz] & $\gamma$ [MHz]$^a$ & $\gamma$ [MHz]$^b$ & $\gamma_l$ [MHz] \\
	\hline
	\endhead
	
	\hline
	\endfoot
	
	\hline\hline
	\endlastfoot
	
	0 & $12618.2 \pm 2824.4$ & $62.79 \pm 23.83$ & $0.1212 \pm 0.0483$ & 248.2 $\pm$ 4514.4 & -322.8 $\pm$ 2515.8 & 24.0 $\pm$ 172.2 \\
	1 & $8073.6 \pm 765.6$ & $24.77 \pm 5.57$ & $0.0412 \pm 0.0101$ & 248.2 $\pm$ 1882.6 & -46.1 $\pm$ 909.8 & 5.1 $\pm$ 59.2 \\
	2 & $5854.1 \pm 236.8$ & $10.64 \pm 1.52$ & $0.0155 \pm 0.0025$ & 248.2 $\pm$ 764.9 & -0.9 $\pm$ 320.5 & 2.0 $\pm$ 19.9 \\
	3 & $4534.3 \pm 48.4$ & $4.39 \pm 0.28$ & $0.0058 \pm 0.0004$ & 248.2 $\pm$ 202.4 & 6.2 $\pm$ 69.9 & 1.3 $\pm$ 4.2 \\
	4 & $3645.0 \pm 31.6$ & $1.32 \pm 0.16$ & $0.0018 \pm 0.0002$ & 248.2 $\pm$ 158.9 & 5.6 $\pm$ 46.8 & 1.2 $\pm$ 2.7 \\
	5 & $2995.0 \pm 69.4$ & $-0.28 \pm 0.32$ & $0.0001 \pm 0.0004$ & 248.2 $\pm$ 340.5 & 3.3 $\pm$ 102.8 & 1.2 $\pm$ 5.6 \\
	6 & $2493.0 \pm 88.7$ & $-1.14 \pm 0.37$ & $-0.0006 \pm 0.0004$ & 248.2 $\pm$ 456.6 & 0.9 $\pm$ 130.0 & 1.2 $\pm$ 6.9 \\
	7 & $2089.8 \pm 99.2$ & $-1.60 \pm 0.38$ & $-0.0009 \pm 0.0004$ & 248.2 $\pm$ 528.9 & -1.3 $\pm$ 142.5 & 1.2 $\pm$ 7.2 \\
	8 & $1756.4 \pm 105.1$ & $-1.83 \pm 0.36$ & $-0.0010 \pm 0.0003$ & 248.2 $\pm$ 573.5 & -3.3 $\pm$ 147.1 & 1.2 $\pm$ 7.2 \\
	9 & $1474.7 \pm 108.8$ & $-1.93 \pm 0.34$ & $-0.0009 \pm 0.0003$ & 248.2 $\pm$ 599.9 & -5.0 $\pm$ 147.6 & 1.1 $\pm$ 7.0 \\
	10 & $1232.5 \pm 111.2$ & $-1.95 \pm 0.32$ & $-0.0009 \pm 0.0002$ & 248.2 $\pm$ 614.5 & -6.5 $\pm$ 145.7 & 1.1 $\pm$ 6.6 \\
	11 & $1021.5 \pm 112.9$ & $-1.93 \pm 0.30$ & $-0.0008 \pm 0.0002$ & 248.2 $\pm$ 621.0 & -7.9 $\pm$ 142.7 & 1.1 $\pm$ 6.3 \\
	
\end{longtable}
\tablefoot{
	\tablefoottext{a}{Value obtained using Eq.~\ref{eq:freq_RotTrans_G}.}
	\tablefoottext{b}{Value obtained using Eq.~\ref{eq:freq_RotTrans_JG}.}
}
\end{landscape}